\documentclass{article}

    \PassOptionsToPackage{numbers, compress}{natbib}

    \usepackage[preprint]{neurips_2024}

\usepackage[utf8]{inputenc} 
\usepackage[T1]{fontenc}    
\usepackage{hyperref}       
\usepackage{url}            
\usepackage{booktabs}       
\usepackage{amsfonts}       
\usepackage{nicefrac}       
\usepackage{microtype}      
\usepackage{xcolor}         

\usepackage{wrapfig}
\usepackage{caption}
\usepackage{xspace}
\usepackage{subcaption}
\usepackage{graphicx}
\usepackage{amsmath}
\usepackage{amssymb}
\usepackage{float}
\usepackage{algorithm}
\usepackage{algorithmic}
\usepackage{comment}
\usepackage{enumitem}
\usepackage{multirow}
\usepackage[normalem]{ulem}
\useunder{\uline}{\ul}{}

\def\eg{\emph{e.g.}\xspace} 

\def\ie{\emph{i.e.}\xspace}

\def\etc{\emph{etc}\xspace} 

\def\wrt{\emph{w.r.t.}\xspace}

\usepackage{xcolor}

\newcommand{\nameMetric}{NWC\xspace}
\newcommand{\nameMetrics}{NWCs\xspace}
\newcommand{\nameFramework}{TSBD\xspace}

\title{Unveiling and Mitigating Backdoor Vulnerabilities based on Unlearning Weight Changes and Backdoor Activeness}

\definecolor{darkred}{rgb}{0.7,0,0}
\definecolor{darkgreen}{rgb}{0,0.46,0}
\definecolor{purple}{rgb}{0.6,0,0.5}
\definecolor{cholocate}{HTML}{d2691e}
\definecolor{slateblue}{HTML}{6a5acd}

\newcommand{\fin}{\color{black}}

\author{%
    Weilin Lin$^1$ \quad Li Liu$^{1}$\thanks{Corresponds to Li Liu (\href{avrillliu@hkust-gz.edu.cn}{avrillliu@hkust-gz.edu.cn})} \quad Shaokui Wei$^2$ \quad Jianze Li$^{3,2}$ \quad Hui Xiong$^1$\\
    $^1$The Hong Kong University of Science and Technology (Guangzhou)\\
    $^2$The Chinese University of Hong Kong, Shenzhen\\
    $^3$Shenzhen Research Institute of Big Data
}

\begin{document}

\maketitle

\begin{abstract}

The security threat of backdoor attacks is a central concern for deep neural networks (DNNs). Recently, without poisoned data, unlearning models with clean data and then learning a pruning mask have contributed to backdoor defense. Additionally, vanilla fine-tuning with those clean data can help recover the lost clean accuracy. However, the behavior of clean unlearning is still under-explored, and vanilla fine-tuning unintentionally induces back the backdoor effect. In this work, we first investigate model unlearning from the perspective of weight changes and gradient norms, and find two interesting observations in the backdoored model: 1) the weight changes between poison and clean unlearning are positively correlated, making it possible for us to identify the backdoored-related neurons without using poisoned data; 2) the neurons of the backdoored model are more active (\ie, larger changes in gradient norm) than those in the clean model, suggesting the need to suppress the gradient norm during fine-tuning. Then, we propose an effective two-stage defense method. In the first stage, an efficient \textit{Neuron Weight Change (NWC)-based Backdoor Reinitialization} is proposed based on observation 1). In the second stage, based on observation 2), we design an \textit{Activeness-Aware Fine-Tuning} to replace the vanilla fine-tuning. Extensive experiments, involving eight backdoor attacks on three benchmark datasets, demonstrate the superior performance of our proposed method compared to recent state-of-the-art backdoor defense approaches. 
\end{abstract}

\section{Introduction}
\label{sec:intro}


Over the past few years, \emph{deep neural networks} (DNNs) have achieved surprising success in several real-world applications, such as \emph{face recognition}~\cite{taigman2014deepface, parmar2014face,ibrahim2011study}, \emph{medical image processing}~\cite{chen2021uscl,chen2023metalr}, and \emph{autonomous driving}~\cite{yurtsever2020survey, caesar2020nuscenes}, \etc. However, DNNs are susceptible to malicious attacks that can compromise their security and reliability.
One typical example is the \emph{backdoor attack}~\cite{gu2019badnets,li2021invisible,wu2023adversarial,zheng2022pre}, where the adversary maliciously manipulates the training dataset or training process to produce a backdoored model, which performs normally on clean data while predicting any sample with a particular trigger pattern to a pre-defined target label. 
In this work, we focus on the \emph{post-training defense} scenario where, given a backdoored model and a small set of clean training samples, one aims to mitigate the backdoor effect while maintaining the performance on clean data, thereby obtaining a benign model.

\begin{figure}
    \centering
    \includegraphics[width=\linewidth]{ 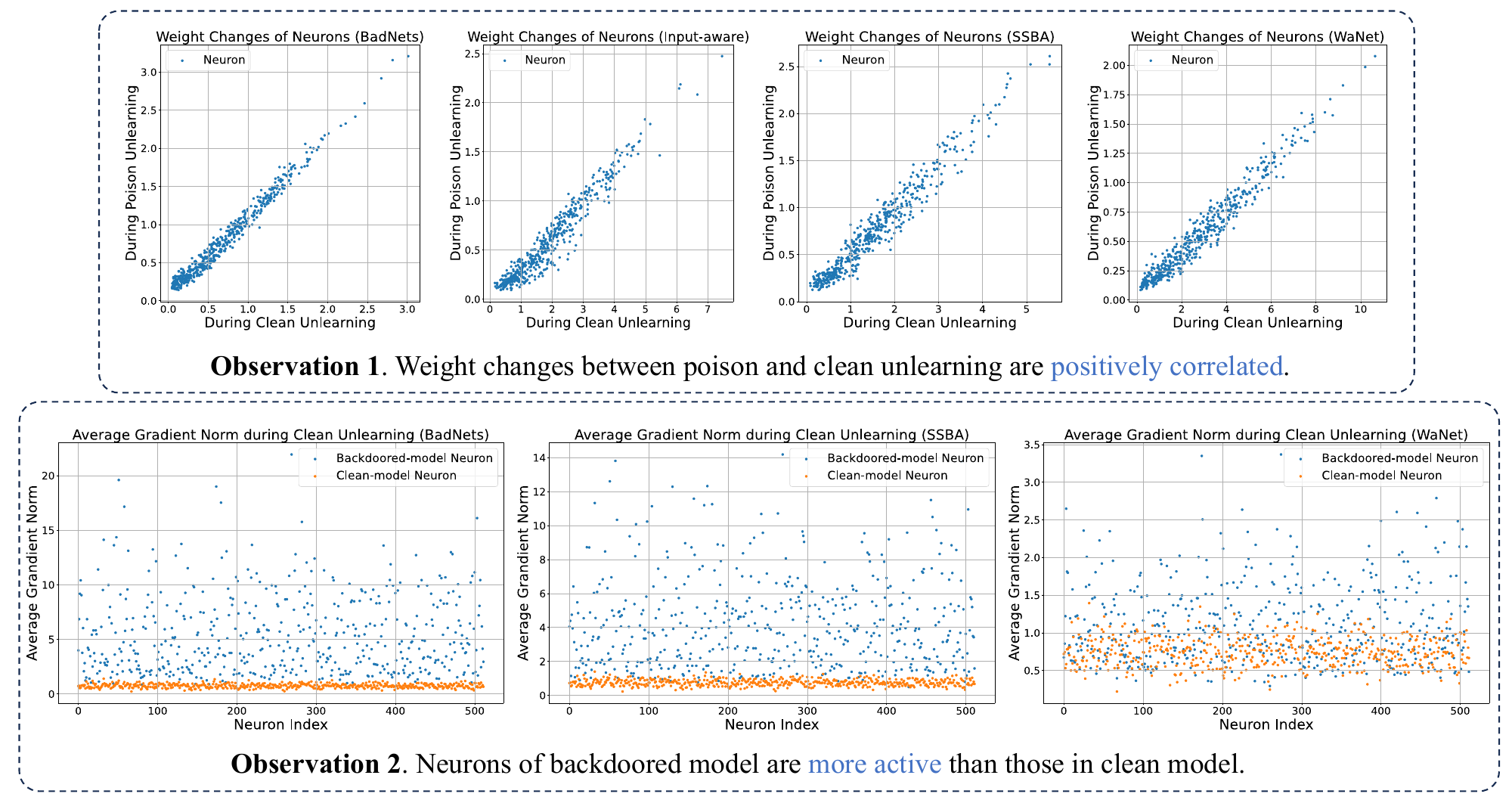}
    \vspace{-5mm}
    \caption{Illustration of two observations. Figures for Observation 1 show distributions of neuron weight changes during clean unlearning and poison unlearning. Figures for Observation 2 compare the average gradient norm for each neuron on the backdoored model and clean model, which are calculated with one-epoch clean unlearning. More active means a larger change in the gradient norm. Experiments are conducted on PreAct-ResNet18~\cite{he2016identity} on CIFAR-10~\cite{krizhevsky2009learning} for the clean model and additional attacks with 10\% poisoning ratio for the backdoored model. The last convolutional layers are chosen for illustration.}
    \label{fig:facts}
    \vspace{-5mm}
\end{figure}

Up to now, several important methods have been developed for \emph{backdoor defense}~\cite{li2021anti, chen2018detecting, tran2018spectral, qi2023towards, liu2023beating}. One promising approach is \emph{poison unlearning}, which involves updating a backdoored model by unlearning from poisoned data. This technique has been utilized in various backdoor defenses such as ABL~\cite{li2021anti}, D-BR~\cite{chen2022effective}, \emph{Neural Cleanse} (NC)~\cite{wang2019neural}, and i-BAU~\cite{zeng2021adversarial}, \etc. 
To avoid approximating poisoned data, another approach called \emph{clean unlearning} was conducted by RNP~\cite{li2023reconstructive}. This technique only uses clean data for unlearning and then prunes the backdoored model, which has been proven to be effective. 
Through relevant experiments, we find an interesting connection between poison unlearning and clean unlearning, as illustrated in \textbf{Observation 1} of Figure~\ref{fig:facts}. Specifically, by calculating the weight changes of each neuron during the two unlearning processes on the backdoored models\footnote{Four attacked models on BadNets~\cite{gu2019badnets}, Input-aware~\cite{nguyen2020input}, SSBA~\cite{li2021invisible}, and WaNet~\cite{nguyen2021wanet}, are used for illustration.}, we find that they exhibit a strong positive correlation, \ie, the neurons exhibiting significant weight changes during clean unlearning also tend to play crucial roles in poison unlearning, indicating a stronger association with backdoor-related activities. 
Moreover, we further investigate the \textbf{backdoor activeness} during learning processes\footnote{Unlearning, as the opposite process of model learning, can also be considered as a kind of learning process.}, \ie, comparing the average gradient norm for each neuron in both the backdoored and clean models. \fin
The results are shown in \textbf{Observation 2} of Figure~\ref{fig:facts}, revealing that neurons in the backdoored model are always more active compared to those in the clean model.


%

Inspired by the above two observations regarding the backdoored model
, we propose \textbf{T}wo-\textbf{S}tage \textbf{B}ackdoor \textbf{D}efense (\textbf{\nameFramework}), consisting of stage 1) \textit{Neuron Weight Change-based Backdoor Reinitialization} and stage 2)\textit{ Activeness-Aware Fine-tuning}. In the first stage, we first conduct clean unlearning on the backdoored model, followed by the neuron weight change calculation, where both the changes of each subweight\footnote{A subweight represents one learnable weight in a neuron weight matrix.}  and neuron are recorded. Then, we conduct zero reinitialization to mitigate the backdoor effect by reinitializing the most-changed subweights among the top-$n\%$ most-changed neurons as $0$ in the original backdoored model.
In the second stage, we adopt activeness-aware fine-tuning with gradient-norm regulation to recover clean accuracy and suppress the reactivation of the backdoor effect. 
Extensive experiments demonstrate the superior defense performance of the proposed method compared to state-of-the-art (SOTA) backdoor defense methods.


To summarize, our main contributions are three-fold. \textbf{(1) Novel Insight:} We are the first to uncover the strong positive correlation between neuron weight changes in clean unlearning and poison unlearning. We also reveal the high backdoor activeness in the backdoored model during the learning process. \textbf{(2) Effective Defense Method:} We further develop an effective two-stage defense method based on unlearning weight changes and backdoor activeness, considering both backdoor mitigation and clean-accuracy recovery, respectively. \textbf{(3) SOTA Performance:} Experimental results and analysis show that our proposed method achieves SOTA performance in backdoor defense.

\begin{figure*}[h]
    \centering
    \includegraphics[width=\linewidth]{ 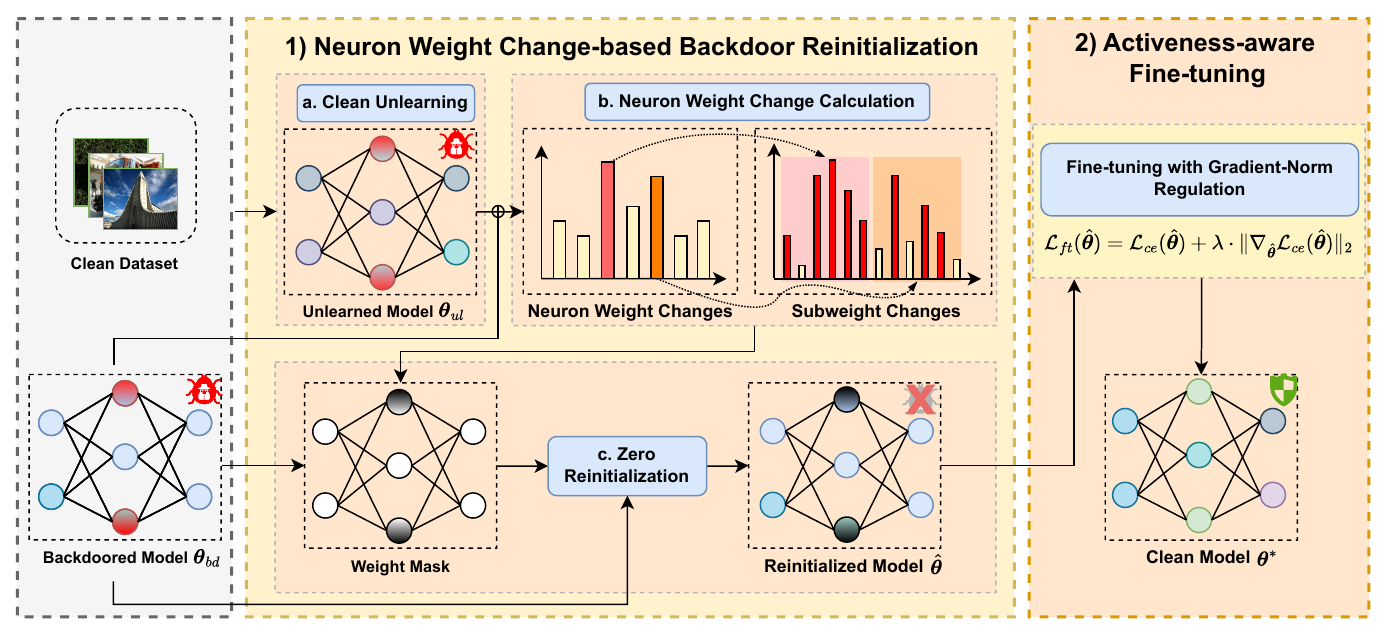}
    \vspace{-6mm}
    \caption{Overview of the proposed Two-Stage Backdoor Defense framework.}
    \label{fig:framework}
    \vspace{-4mm}
\end{figure*}

\section{Related Work}
\label{sec: related work}

\subsection{Backdoor Attack}
In the literature, various backdoor attacks on DNNs have been proposed, which can be categorized into \emph{data poisoning attacks} and \emph{training-controllable attacks}. BadNets~\cite{gu2019badnets} is one of the earliest data poisoning attacks in this field. In this attack, a small proportion of the original data is selected and patched with a pre-defined pattern, known as a \textit{trigger}. The labels of these patched data points are then modified to a target label. The mixed dataset, containing both clean and poisoned data, is used to train the DNNs, resulting in the implantation of the backdoor.
Under a similar procedure, Blended~\cite{chen2017targeted} was proposed as a stronger attack by blending an entire pre-defined image into the original clean data with controllable transparency. Recently, more advanced and stealthy attacks have been proposed to enhance the trigger, such as SIG~\cite{barni2019new}, label consistent attacks~\cite{shafahi2018poison, zhao2020clean}, SSBA~\cite{li2021invisible}, \etc. Another category is training-controllable attacks~\cite{nguyen2020input,nguyen2021wanet,doan2021lira, bagdasaryan2021blind,doan2021backdoor}, where the attackers design triggers with permission to control the training process. 
Two significant examples are WaNet~\cite{nguyen2021wanet} and Input-aware~\cite{nguyen2020input}, which generate unique triggers for different input data by incorporating an injection function into the model training process. This approach makes these attacks more difficult to detect compared to previous attacks with fixed triggers. 

\subsection{Backdoor Defense}
According to the different stages of model training, backdoor defense methods can be classified into two types: \emph{training-stage defenses} and \emph{post-training defenses}. 

\textbf{Training-stage Defenses.}
In training-stage defenses, defenders have access to a mixed training dataset containing both clean data and poisoned data with triggers. ABL~\cite{li2021anti} discovers that the loss-dropping speed of poisoned data during the early stages of model training is faster, and thus isolates them for poison unlearning. DBD~\cite{huang2022backdoor} splits the training process into three steps to separate the training of feature extraction from that of the subsequent classifier to evade the learning of trigger-label correlation. Similarly, D-ST/D-BR~\cite{chen2022effective} observes that the transformations of poisoned-data feature representations are more sensitive than clean ones, and thus proposes to modularize the training process.

\textbf{Post-training Defenses.} 
In post-training defenses~\cite{guo2021aeva, jiang2022critical,kolouri2020universal,xu2021detecting}, defenders aim to erase the backdoor effect in the learned DNNs using a small portion of clean data. 
FP~\cite{liu2018fine} is one of the earliest defense methods, which observes that poisoned data and clean data activate different neurons in a backdoored DNN, and thus keeps pruning the less-activated neurons in response to clean data until a significant drop in accuracy occurs. After that, vanilla fine-tuning is employed to recover the lost clean accuracy. Using the pruning strategy~\cite{zheng2022data, chen2022quarantine, guan2022few},  ANP~\cite{wu2021adversarial} observes that the backdoor-related neurons exhibit higher sensitivity to adversarial perturbations compared to others, and thus trains a pruning mask using minimax optimization. Continuing along this line, AWM~\cite{chai2022one} and RNP~\cite{li2023reconstructive} use a similar mask training process with main modifications in neuron perturbations to data perturbations and \textit{clean unlearning}, respectively. 
Different strategies are also proposed for defense. For example, NC~\cite{wang2019neural} proposes to recover the trigger before the subsequent backdoor removal. NAD~\cite{li2021neural}, for the first time, adopts model distillation to guide the learning of a benign student model.
Additionally, employing unlearning techniques, SAU~\cite{wei2024shared} treats backdoor triggers as a form of adversarial perturbation, and generates poisoned data through optimization on clean data, which are then used in \textit{poison unlearning}. 

However, there exist some limitations among those techniques, \eg, clean unlearning is still under-explored, and the vanilla fine-tuning unintentionally increases the attack success rate. 
In this paper, we propose a comprehensive two-stage defense method breaking through the two limitations.

\section{Methods}
\label{sec:methods}

\subsection{Problem Formulation}

\textbf{Threat Model.}
We assume that the attacker has full access to the training data. Their goal is to poison a portion of the dataset by injecting triggers into the data so that the trained model misclassifies the poisoned data to the target class while still performing normally on clean data. The \emph{poisoning ratio} (\textit{e.g.}, 10\%) is used to depict the proportion of poisoned data within the entire dataset. We denote the parameters of the backdoored model as $\boldsymbol{\theta}_{bd}=\{\boldsymbol{\theta}_{bd}^{(l)}\}_{1\leq l\leq L}$ satisfying $\boldsymbol{\theta}_{bd}^{(l)} \in \mathbb{R}^{K^{(l)} \times I^{(l)}}$, where $\boldsymbol{K}=\{K^{(l)}\}_{1\leq l\leq L}$ and $\boldsymbol{I}=\{I^{(l)}\}_{1\leq l\leq L}$ represent the neuron numbers and learnable subweight numbers, respectively. Specifically, for the $l^{th} \in \{1,\hdots,L\}$ layer, there are $K^{(l)}$ neurons in total and $I^{(l)}$ subweights for each neuron. 

\textbf{Defense Setting.} 
The defender's goal is to remove the backdoor effect, which causes poisoned data to be misclassified to the target class, from the backdoored model while minimizing the impact on the prediction accuracy for clean data. Following the previous defense setting~\cite{liu2018fine, wu2021adversarial}, we assume that the defender knows nothing about the poisoned data and possesses only 5\% of the total dataset as clean data, denoted as $\mathcal{D}_c$.



\subsection{Neuron Weight Change \& Suggestions Given by the Two Observations} 
\label{subsec:facts}
In this subsection, we provide more details on the unlearning formulation and offer suggestions based on the two observed observations. 

\textbf{Model Unlearning.} 
Model Unlearning can be defined as the reverse process of model training~\cite{li2023reconstructive}, which involves maximizing the loss value on a given dataset. Given a DNN model $f$ parameterized as $\boldsymbol{\theta}$ and a dataset $\mathcal{D}$ for unlearning, the maximization problem can be formulated as:$
\max_{\boldsymbol{\theta}} \mathbb{E}_{(\boldsymbol{x}, y) \in \mathcal{D}} \left[\mathcal{L}(f(\boldsymbol{x};\boldsymbol{\theta}),y)\right],$
where $(\boldsymbol{x}, y) \in \mathcal{D}$ represents the images and their corresponding labels, and $\mathcal{L}$ denotes the loss function used in this task, \eg, cross-entropy loss. 

Intuitively, by maximizing the loss expectation, the unlearned model, parameterized as $\boldsymbol{\theta}_{ul}$, is prone to fail at the task specified in $\mathcal{D}$. In this paper, we term the process as \textit{clean unlearning} when all the data in $\mathcal{D}$ are clean, denoted as $\mathcal{D}_c$. On the other hand, \textit{poison unlearning} refers to the scenario where all the data in $\mathcal{D}$ are poisoned with a trigger. By default, both clean and poison unlearning are terminated when the model performs poorly on the corresponding tasks, such as achieving only 10\% clean accuracy or attack success rate.

\textbf{Neuron Weight Change.} 
To comprehensively quantify the weight changes of a neuron during the entire unlearning process, we define the \textit{Neuron Weight Change} (\nameMetric), where the $L_1$ norm is calculated on every neuron's weight differences.
The \nameMetric for the $k^{th} \in \{1,\hdots,K^{(l)}\}$ neuron in layer $l \in \{1,\hdots ,L\}$ can be formulated as:
\begin{equation}
\label{equ:ucn}
\mathrm{\nameMetric}^{(l)k} = \sum_{i=0}^{I^{(l)}} \|\boldsymbol{\theta}^{(l)ki}_{ul}-\boldsymbol{\theta}^{(l)ki}_{bd}\|_1,
\end{equation}

where $\sum_{i=0}^{I^{(l)}} \left\|\cdot\right\|_1$ is to calculate the $L_1$ norm for the differences on a neuron with totally $I^{(l)}$ subweights, $\boldsymbol{\theta}^{(l)ki}_{ul}$ and $\boldsymbol{\theta}^{(l)ki}_{bd}$ denote the $i^{th} \in \{1,\hdots, I^{(l)}\}$ subweights of $k^{th}$ neuron after and before the entire unlearning process, respectively. A larger $\mathrm{\nameMetric}^k$ indicates more significant changes occurring in neuron $k$ during the unlearning process. Similarly, in Equation~(\ref{equ:ucn}), the term $\|\boldsymbol{\theta}^{(l)ki}_{ul}-\boldsymbol{\theta}^{(l)ki}_{bd}\|_1$ represents the changes in the $i^{th}$ subweight of neuron $k$ in layer $l$, \ie, defined as \textit{Subweight Change}.

\textbf{Suggestions Given by the Two Observations.} 
As demonstrated in Section~\ref{sec:intro}, we have two interesting observations regarding the backdoored model.
\textbf{Observation 1} shows that the neurons exhibiting significant weight changes during clean unlearning also tend to play crucial roles in poison unlearning. It suggests that we can employ clean unlearning to identify and eliminate backdoor-related neurons using \nameMetric, at the expense of reducing clean accuracy. 
On the other hand, \textbf{Observation 2} reveals that neurons in the backdoored model are always more active compared to those in the clean model. It suggests that we should suppress the gradient norm during the learning process if we want to recover it to a clean model.
These two suggestions act as the main supports to our proposed \nameFramework.

\begin{figure}[]
    \centering
    \begin{subfigure}{0.313\linewidth}
        \includegraphics[width=\linewidth]{ 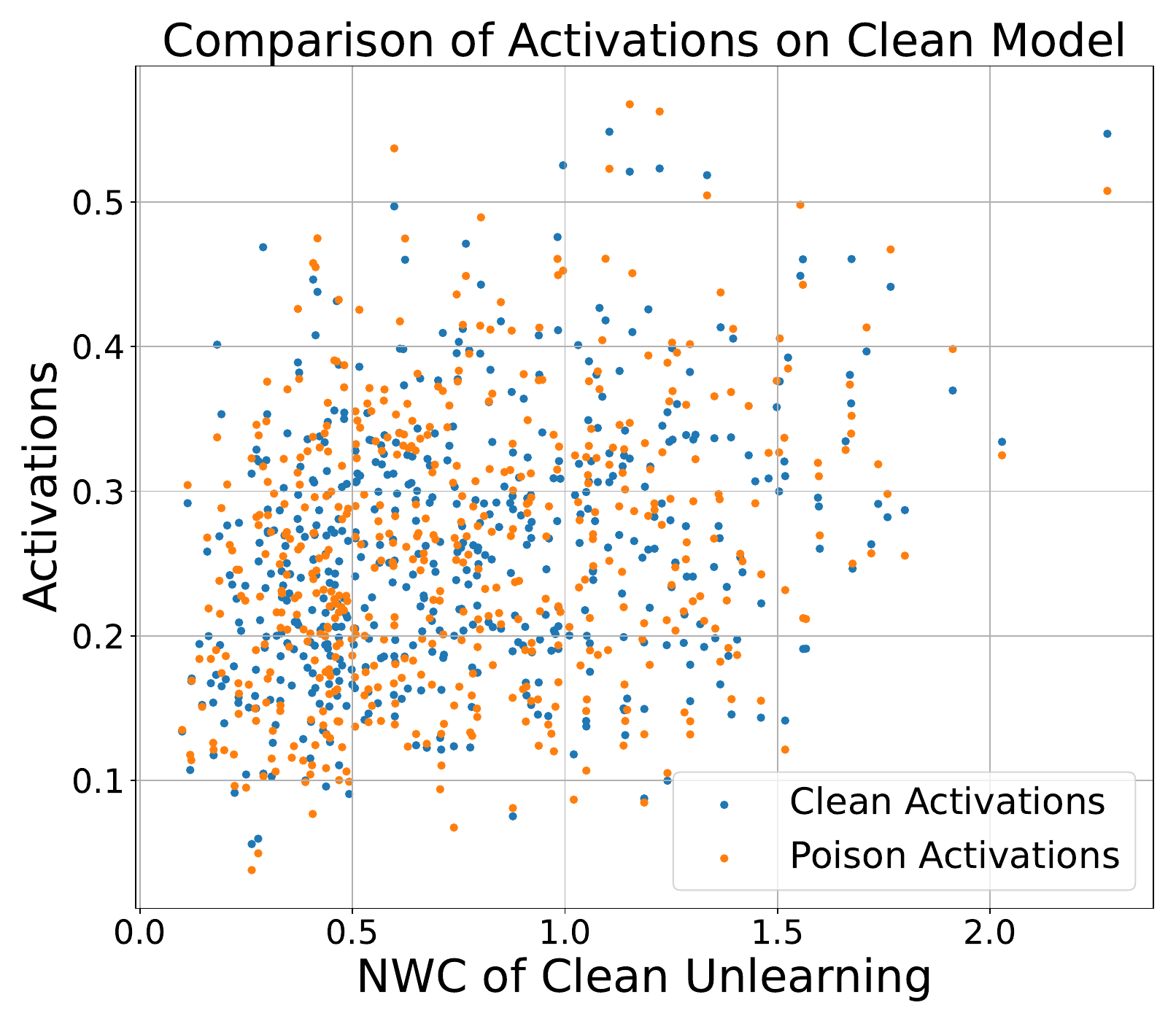}
        \caption{Original clean model}
        \vspace{-1mm}
    \end{subfigure}
    \begin{subfigure}{0.33\linewidth}
        \includegraphics[width=\linewidth]{ 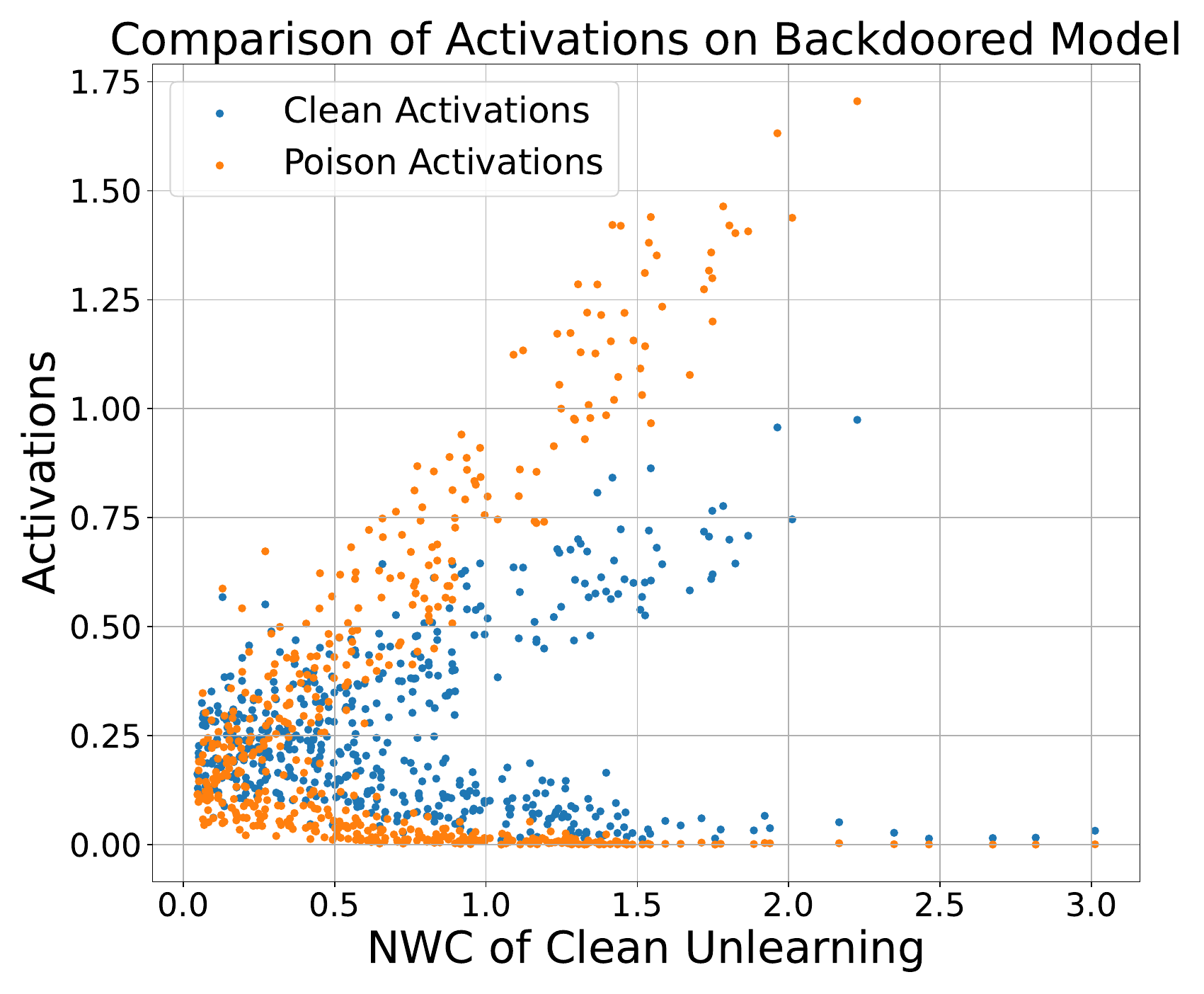}
        \caption{Original backdoored model}
        \vspace{-1mm}
    \end{subfigure}
    \begin{subfigure}{0.32\linewidth}
        \includegraphics[width=\linewidth]{ 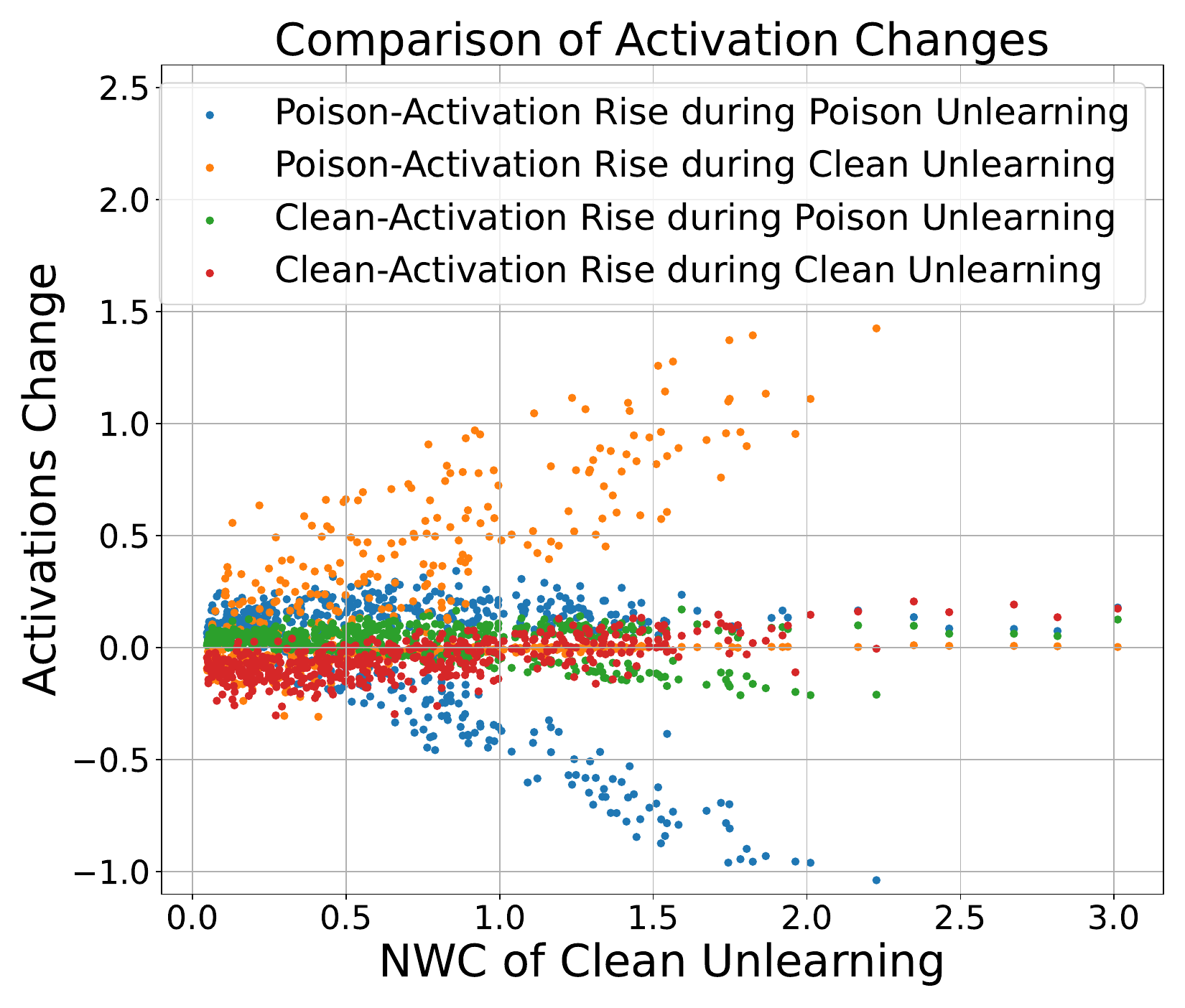}
        \caption{Backdoored-model Unlearning}
        \vspace{-1mm}
    \end{subfigure}

    \caption{Illustration of clean and poison activations of each neuron. (a) and (b) represent the activations on the original clean and backdoored model, respectively. (c) shows the activation changes during the clean and poison unlearning on backdoored model. Activations are captured from the last convolutional layer with an additional \textit{Relu} activation function on PreAct-ResNet18~\cite{he2016identity}.}
    \label{fig:act}
    \vspace{-5mm}
\end{figure}

\subsection{Further Investigations \& Insights} 
\label{subsec:further_insight}

Here, we offer insights from the perspective of neuron activations, trying to answer two important questions: \textbf{[Q1]} What causes the clean unlearning \nameMetrics to exhibit a positive correlation with those in poison unlearning, and \textbf{[Q2]} What motivates the neurons more active in the backdoored model.


\textbf{Neuron Activations \& Activation Rise.} The neuron activation is determined by computing the average value of all inputs to the specific neuron, \eg, $h^{(l)k} \approx \sigma( \boldsymbol{\theta}^{(l)k} \boldsymbol{h}^{(l-1)})$ for simplicity, where $\sigma(\cdot)$ is the activation function.  
In line with the terminology used in FP~\cite{liu2018fine}, \textit{clean activation} denotes the scenario where all the input samples are clean while \textit{poison activation} refers to the presence of poisoned inputs. 
To better observe the changes in activation during unlearning, we calculate the activation rise from the original model to the unlearned model, \ie, $\Delta h^{(l)k} = h^{(l)k}_{ul} - h^{(l)k}_{bd}$. A positive value indicates an increase in activation, while a negative value signifies a decrease.

\textbf{Relationship between \nameMetric and Activation Change.} Considering that a backdoored model has learned two tasks from the clean and poisoned data~\cite{li2021anti}, the main influence of \nameMetric on a neuron can be roughly attributed to its activation change on both clean and poisoned inputs. For neuron $k$ in layer $l$, we can formulate it as $\mathrm{\nameMetric}^{(l)k} \propto |\Delta h^{(l)k}_c| + |\Delta h^{(l)k}_p|$, where $\Delta h^{(l)k}_c$ and $\Delta h^{(l)k}_p$ represent the activation rise on clean and poisoned inputs, respectively.

Figure~\ref{fig:act} illustrates the clean and poison activations in (a) the original clean model, (b) the original backdoored model, and (c) the backdoored-model unlearning. 
We now try to answer the above two questions from these observations.
\textbf{[A1]} We can observe that poison activations are the main factors affected during both clean unlearning (increase) and poison unlearning (decrease), while clean activations are only slightly influenced (see Figure~\ref{fig:act} (c)), \ie, $\mathrm{\nameMetric}^{(l)k}_\uparrow \rightarrow |\Delta h^{(l)k}_c|_{\approx} + |\Delta h^{(l)k}_p|_\uparrow$. Also, the growing \nameMetric during clean unlearning can indicate larger poison and clean activations (where $h^{(l)k}_p > h^{(l)k}_c$) to some extent (see Figure~\ref{fig:act} (b)). Thus, we deduce that the co-function of clean and poison activations dominates the performance on both tasks, while the higher values of poison activation in the backdoored model make it an easier target for modification. In this case, the neurons with higher poison activations tend to decrease their values during poison unlearning, thereby reducing the attack success rate.
Conversely, during clean unlearning, these neurons increase poison activations, which suppresses the function of clean activations and reduces clean accuracy.
\textbf{[A2]} Similarly, the significantly lower values of mixed clean and poison activations (maximum: 0.5676) on the clean model (see Figure~\ref{fig:act} (a)) indicate that it is less active compared to the backdoored model (maximum: 1.7053), where a similar pattern can also be seen on the bottom left of Figure~\ref{fig:act} (b).

\subsection{Two-Stage Backdoor Defense Framework} 
\label{subsec:framework}
Based on the above observations, we now propose a defense framework incorporating \textit{Neuron Weight Change-based Backdoor Reinitialization} (including \textit{Clean Unlearning}, \textit{Neuron Weight Change Calculation} and \textit{Zero Reinitialization}), and \textit{Activeness-aware Fine-tuning}. The detailed defense process is illustrated in Figure~\ref{fig:framework} and Algorithm~\ref{alg:overall} (found in Appendix~\ref{sec:overall_algo}). 

\textbf{Stage 1) Neuron Weight Change-based Backdoor Reinitialization.}
We aim to mitigate the backdoor effect with acceptable clean-accuracy sacrificed in this stage.
\textit{[a. Clean Unlearning.]} To identify the backdoor-related neurons, we first conduct a full clean unlearning using the available clean data $\mathcal{D}_{c}$ on the backdoored model. 
\textit{[b. Neuron Weight Change Calculation.]}
Then, we record the subweight changes and calculate the \nameMetric for each neuron as described in Section~\ref{subsec:facts}. The resulting sorted order of neurons reflects the backdoor strength. 
\textit{[c. Zero Reinitialization.]}
After that, we can now eliminate the backdoor effect through zero reinitialization. 
Based on the \nameMetric neuron order, we identify the top-$n\%$ neurons as strongly backdoor-related. As suggested in Section~\ref{subsec:further_insight}, high-\nameMetric neurons may also contribute to clean accuracy to some extent. Therefore, we further choose to reinitialize the subweights of the most-changing $m\%$ among the selected neurons to zero in the backdoored model, while leaving the others unchanged.
The reinitialized model parameter is denoted as $\hat{\boldsymbol{\theta}}$.

\textbf{Stage 2) Activeness-Aware Fine-tuning.}
To further repair the reinitialized subweights and avoid recovering the backdoor effect again, we conduct Activeness-aware fine-tuning on the reinitialized model ($\hat{\boldsymbol{\theta}}$) using the clean dataset, $\mathcal{D}_{c}$. This involves incorporating gradient-norm regulation into the original loss function, such as the cross-entropy loss $\mathcal{L}_{ce}$, to penalize high gradient values. This regulation serves to suppress neuron activity during fine-tuning. The final loss function is: 
\begin{equation}
    \label{equ:loss_ori}
    \mathcal{L}_{ft}(\hat{\boldsymbol{\theta}})=\mathcal{L}_{ce}(\hat{\boldsymbol{\theta}})+\lambda \cdot\|\nabla_{\hat{\boldsymbol{\theta}}} \mathcal{L}_{ce}(\hat{\boldsymbol{\theta}})\|_2,
\end{equation}
where $\|\nabla_{\hat{\boldsymbol{\theta}}} \mathcal{L}_{ce}(\hat{\boldsymbol{\theta}})\|_2$ represents the $L_2$ norm of gradients, and $\lambda$ is the \emph{penalty coefficient} controlling its impact.   
Hence, the objective of fine-tuning is to minimize the loss function $\mathcal{L}_{ft}(\hat{\boldsymbol{\theta}})$ using the available clean data $\mathcal{D}_{c}$: 
\begin{equation}
    \label{equ:ft}
    \begin{aligned}
    &\underset{\hat{\boldsymbol{\theta}}}{\min }
    \mathbb{E}_{(\boldsymbol{x}_{c}, y_{c}) \in \mathcal{D}_{c}}[  \mathcal{L}_{ft}(f(\boldsymbol{x}_{c}; \hat{\boldsymbol{\theta}}), y_{c})].
    \end{aligned}
\end{equation}
During practical optimization for computational efficiency, we adopt the approximation scheme in \cite{zhao2022penalizing}, which can be formulated as: 
\begin{equation}
    \label{equ:loss_appro}   
    \begin{aligned}
    \nabla_{\hat{\boldsymbol{\theta}}} \mathcal{L}_{ft}(\hat{\boldsymbol{\theta}}) 
    & \approx (1-\alpha) \nabla_{\hat{\boldsymbol{\theta}}} \mathcal{L}_{ce}(\hat{\boldsymbol{\theta}}) + \alpha \nabla_{\hat{\boldsymbol{\theta}}} \mathcal{L}_{ce} (\hat{\boldsymbol{\theta}}+r \frac{\nabla_{\hat{\boldsymbol{\theta}}} \mathcal{L}_{ce}(\hat{\boldsymbol{\theta}})}{ \|\nabla_{\hat{\boldsymbol{\theta}}} \mathcal{L}_{ce}(\hat{\boldsymbol{\theta}}) \|_2} ).
    \end{aligned}
\end{equation}
Here, $r$ is used for appropriating the Hessian multiplication operation, and $\alpha=\frac{\lambda}{r}$ is the \emph{balance coefficient}. 
Due to the space limit, the detailed derivation and algorithm are provided in Appendix~\ref{sec:ft_deri}. 
After the activeness-aware fine-tuning stage, we can obtain a repaired clean model, which demonstrates outstanding performance in the experiments.

\section{Experiments}
\label{sec:experiments}
\subsection{Experimental Setup}
\textbf{Attack Setup.}
We consider 8 SOTA backdoor attacks in the main experiment. There are BadNets~\cite{gu2019badnets}, Blended~\cite{chen2017targeted}, Input-aware~\cite{nguyen2020input}, LF~\cite{zeng2021rethinking}, SIG~\cite{barni2019new}, SSBA~\cite{li2021invisible}, Trojan~\cite{liu2018trojaning} and WaNet~\cite{nguyen2021wanet}. For a fair comparison, we follow the default attack configuration as in BackdoorBench~\cite{wu2022backdoorbench}, including the trigger pattern, trigger size, the target label (\textit{i.e.}, the $0^{th}$ label), \etc. We choose 10\% poisoning ratio as the default setting. To fully evaluate the effectiveness of our proposed framework, all the attacks are implemented on three benchmark datasets, \textit{i.e.}, CIFAR-10~\cite{krizhevsky2009learning}, Tiny ImageNet~\cite{le2015tiny}, and GTSRB~\cite{stallkamp2011german}, over two popular DNNs, \textit{i.e.}, PreAct-ResNet18~\cite{he2016identity} and VGG19-BN~\cite{simonyan2014very}. In particular, SIG is only applied to CIFAR-10 since it cannot reach a 10\% poisoning ratio on Tiny ImageNet and GTSRB.
Besides, we also evaluate the defense methods under 5\% and 1\% poisoning ratios to verify the robustness of our proposed framework. Due to the space limit, we only exhibit parts of the main results in this section. More implementation details can be found in Appendix~\ref{sec:implem_details}.

\textbf{Defense Setup.}
We compare the proposed framework with 8 SOTA backdoor defense methods: Fine-tuning (FT), Fine-pruning (FP)~\cite{liu2018fine}, NAD~\cite{li2021neural}, NC~\cite{wang2019neural}, ANP~\cite{wu2021adversarial}, 
CLP~\cite{zheng2022data}, i-BAU~\cite{zeng2021adversarial}, and RNP~\cite{li2023reconstructive}. We use the recommended configurations in BackdoorBench~\cite{wu2022backdoorbench}. 
Since the defense settings for training-stage defenses are different~\cite{li2021anti, huang2022backdoor}, we only compare the post-training defenses, where 5\% clean data can be accessed following the previous settings~\cite{wu2021adversarial}. 
For \nameFramework, we set the learning rates for clean unlearning to $10^{-4}$ and fine-tuning to $10^{-2}$.
The default neuron ratio $n$ and weight ratio $m$ are set to 0.15 and 0.7, respectively. We follow the suggested settings of hyper-parameters $r=0.05$ and $\alpha=0.7$ for fine-tuning~\cite{zhao2022penalizing}.

\textbf{Evaluation Metrics.} 
We use three metrics to evaluate the performance of each defense method: Accuracy on clean data (\textbf{ACC}), Attack Success Rate (\textbf{ASR}), and Defense Effectiveness Rating (\textbf{DER})~\cite{zhu2023enhancing}.
Specifically, ACC measures the proportion of clean data correctly predicted; ASR measures the proportion of poisoned data misclassified to the target label; DER $\in [0,1]$ evaluates the cost of ACC for reducing ASR, which is defined as: $\mathrm{DER}=[\max (0, \Delta \mathrm{ASR})-\max (0, \Delta \mathrm{ACC})+1] / 2,$
where $\Delta \mathrm{ASR}$ and $\Delta \mathrm{ACC}$ are the drop in ASR and ACC after applying defense on the backdoored model, respectively.
Larger ACC, DER, and smaller ASR are desired for a successful defense. 
Note that in the following result tables, ``-'' indicates that the value is inapplicable. The \textbf{boldface} values indicate the best performance and the {\ul underline} values denote the second-best result.

\begin{table*}[]
\caption{Comparison with the SOTA defenses on \textbf{CIFAR-10} dataset with PreAct-ResNet18 (\%).}
\label{tab:cifar10_preact_10}
\vspace{-2mm}
\centering
\resizebox{1\linewidth}{!}{
\begin{tabular}{c|ccc|ccc|ccc|ccc|ccc}
\hline
\multirow{2}{*}{\begin{tabular}[c]{@{}c@{}}Backdoor\\ Attacks\end{tabular}} & \multicolumn{3}{c|}{No Defense}                    & \multicolumn{3}{c|}{FT}                            & \multicolumn{3}{c|}{FP~\cite{liu2018fine}}                            & \multicolumn{3}{c|}{NAD~\cite{li2021neural}}                           & \multicolumn{3}{c}{NC~\cite{wang2019neural}}                                            \\ \cline{2-16} 
                                                                            & ACC $\uparrow$ & ASR $\downarrow$ & DER $\uparrow$ & ACC $\uparrow$ & ASR $\downarrow$ & DER $\uparrow$ & ACC $\uparrow$ & ASR $\downarrow$ & DER $\uparrow$ & ACC $\uparrow$ & ASR $\downarrow$ & DER $\uparrow$ & ACC $\uparrow$      & ASR $\downarrow$      & DER $\uparrow$      \\ \hline
BadNets~\cite{gu2019badnets}                                                                     & 91.32          & 95.03            & -              & 89.96          & 1.48             & {\ul 96.10}    & \textbf{91.31} & 57.13            & 68.95          & 89.87          & 2.14             & 95.72          & 89.05               & {\ul 1.27}            & 95.75               \\
Blended~\cite{chen2017targeted}                                                                     & 93.47          & 99.92            & -              & 92.78          & 96.11            & 51.56          & {\ul 93.17}    & 99.26            & 50.18          & 92.17          & 97.69            & 50.47          & \textbf{93.47}      & 99.92                 & 50.00               \\
Input-aware~\cite{nguyen2020input}                                                                 & 90.67          & 98.26            & -              & {\ul 93.12}    & 1.72             & 98.27          & 91.74          & \textbf{0.04}    & \textbf{99.11} & \textbf{93.18} & 1.68             & 98.29          & 92.61               & {\ul 0.76}            & {\ul 98.75}         \\
LF~\cite{zeng2021rethinking}                                                                          & 93.19          & 99.28            & -              & 92.37          & 78.44            & 60.01          & \textbf{92.90} & 98.97            & 50.01          & 92.37          & 47.83            & 75.31          & 91.62               & \textbf{1.41}         & \textbf{98.15}      \\
SIG~\cite{barni2019new}                                                                         & 84.48          & 98.27            & -              & \textbf{90.80} & {\ul 2.37}       & {\ul 97.95}    & 89.10          & 26.20            & 86.03          & 90.02          & 10.66            & 93.81          & 84.48               & 98.27                 & 50.00               \\
SSBA~\cite{li2021invisible}                                                                        & 92.88          & 97.86            & -              & 92.14          & 74.79            & 61.16          & {\ul 92.54}    & 83.50            & 57.01          & 91.91          & 77.40            & 59.74          & 90.99               & \textbf{0.58}         & \textbf{97.69}      \\
Trojan~\cite{liu2018trojaning}                                                                      & 93.42          & 100.00           & -              & 92.42          & 5.99             & 96.51          & 92.46          & 71.17            & 63.94          & 91.88          & 3.73             & \textbf{97.36} & 91.76               & 8.22                  & 95.06               \\
WaNet~\cite{nguyen2021wanet}                                                                       & 91.25          & 89.73            & -              & \textbf{93.48} & 17.10            & 86.32          & 91.46          & 1.09             & {\ul 94.32}    & 93.17          & 22.98            & 83.38          & 91.80               & 7.53                  & 91.10               \\ \hline
Average                                                                     & 91.34          & 97.29            & -              & \textbf{92.13} & 34.75            & 80.98          & {\ul 91.84}    & 54.67            & 71.19          & 91.82          & 33.01            & 81.76          & 90.72               & 27.24                 & 84.56               \\ \hline
\multirow{2}{*}{\begin{tabular}[c]{@{}c@{}}Backdoor\\ Attacks\end{tabular}} & \multicolumn{3}{c|}{ANP~\cite{wu2021adversarial}}                           & \multicolumn{3}{c|}{CLP~\cite{zheng2022data}}                           & \multicolumn{3}{c|}{i-BAU~\cite{zeng2021adversarial}}                         & \multicolumn{3}{c|}{RNP~\cite{li2023reconstructive}}                           & \multicolumn{3}{c}{\nameFramework \textbf{\textcolor{red}{(Ours)}}} \\ \cline{2-16} 
                                                                            & ACC $\uparrow$ & ASR $\downarrow$ & DER $\uparrow$ & ACC $\uparrow$ & ASR $\downarrow$ & DER $\uparrow$ & ACC $\uparrow$ & ASR $\downarrow$ & DER $\uparrow$ & ACC $\uparrow$ & ASR $\downarrow$ & DER $\uparrow$ & ACC $\uparrow$      & ASR $\downarrow$      & DER $\uparrow$      \\ \hline
BadNets~\cite{gu2019badnets}                                                                     & {\ul 90.94}    & 5.91             & 94.37          & 90.06          & 77.50            & 58.14          & 89.15          & \textbf{1.21}    & 95.83          & 89.81          & 24.97            & 84.28          & 90.72               & 1.31                  & \textbf{96.53}      \\
Blended~\cite{chen2017targeted}                                                                     & 93.00          & 84.90            & 57.28          & 91.32          & 99.74            & 49.01          & 87.00          & {\ul 50.53}      & {\ul 71.46}    & 88.76          & 79.74            & 57.73          & 91.61               & \textbf{2.61}         & \textbf{97.73}      \\
Input-aware~\cite{nguyen2020input}                                                                 & 91.04          & 1.32             & 98.47          & 90.30          & 2.17             & 97.86          & 89.17          & 27.08            & 84.84          & 90.52          & 1.84             & 98.13          & 93.06               & 1.94                  & 98.16               \\
LF~\cite{zeng2021rethinking}                                                                          & 92.83          & 54.99            & 71.96          & {\ul 92.84}    & 99.18            & 49.88          & 84.36          & 44.96            & 72.75          & 88.43          & 7.02             & 93.75          & 91.20               & {\ul 2.64}            & {\ul 97.32}         \\
SIG~\cite{barni2019new}                                                                         & 83.36          & 36.43            & 80.36          & 83.80          & 98.91            & 49.66          & 85.67          & 3.68             & 97.29          & 84.48          & 98.27            & 50.00          & {\ul 90.41}         & \textbf{1.27}         & \textbf{98.50}      \\
SSBA~\cite{li2021invisible}                                                                        & \textbf{92.67} & 60.16            & 68.74          & 91.38          & 68.13            & 64.11          & 87.67          & 3.97             & 94.34          & 88.60          & 17.89            & 87.84          & 91.57               & {\ul 1.66}            & {\ul 97.44}         \\
Trojan~\cite{liu2018trojaning}                                                                      & {\ul 92.97}    & 46.27            & 76.64          & \textbf{92.98} & 100.00           & 49.78          & 90.37          & \textbf{2.91}    & {\ul 97.02}    & 90.89          & {\ul 3.59}       & 96.94          & 91.76               & 5.06                  & 96.64               \\
WaNet~\cite{nguyen2021wanet}                                                                       & 91.32          & 2.22             & 93.76          & 81.91          & 78.42            & 50.99          & 89.49          & 5.21             & 91.38          & 90.43          & {\ul 0.96}       & 93.98          & {\ul 93.26}         & \textbf{0.88}         & \textbf{94.43}      \\ \hline
Average                                                                     & 91.02          & 36.53            & 80.20          & 89.32          & 78.01            & 58.68          & 87.86          & {\ul 17.44}      & {\ul 88.11}    & 88.99          & 29.28            & 82.83          & 91.70               & \textbf{2.18}         & \textbf{97.09}      \\ \hline
\end{tabular}}
\vspace{-3mm}
\end{table*}

\begin{table*}[]
\caption{Comparison with the SOTA defenses on \textbf{Tiny ImageNet} dataset with PreAct-ResNet18 (\%).}
\label{tab:tiny_preact_10}
\vspace{-2mm}
\centering
\resizebox{1\linewidth}{!}{
\begin{tabular}{c|ccc|ccc|ccc|ccc|ccc}
\hline
\multirow{2}{*}{\begin{tabular}[c]{@{}c@{}}Backdoor\\ Attacks\end{tabular}} & \multicolumn{3}{c|}{No Defense}                    & \multicolumn{3}{c|}{FT}                            & \multicolumn{3}{c|}{FP~\cite{liu2018fine}}                            & \multicolumn{3}{c|}{NAD~\cite{li2021neural}}                           & \multicolumn{3}{c}{NC~\cite{wang2019neural}}                                            \\ \cline{2-16} 
                                                                            & ACC $\uparrow$ & ASR $\downarrow$ & DER $\uparrow$ & ACC $\uparrow$ & ASR $\downarrow$ & DER $\uparrow$ & ACC $\uparrow$ & ASR $\downarrow$ & DER $\uparrow$ & ACC $\uparrow$ & ASR $\downarrow$ & DER $\uparrow$ & ACC $\uparrow$      & ASR $\downarrow$      & DER $\uparrow$      \\ \hline
BadNets~\cite{gu2019badnets}                                                                     & 56.23          & 100.00           & -              & {\ul 55.18}    & {\ul 0.09}       & \textbf{99.43} & 51.73          & 99.99            & 47.76          & 46.37          & 0.27             & 94.93          & 48.26               & 0.10                  & 95.96               \\
Blended~\cite{chen2017targeted}                                                                     & 56.03          & 99.71            & -              & {\ul 55.04}    & 97.73            & 50.49          & 51.89          & 95.94            & 49.81          & 46.89          & 95.00            & 47.79          & 52.55               & 93.21                 & 51.51               \\
Input-aware~\cite{nguyen2020input}                                                                 & 57.45          & 98.85            & -              & {\ul 57.45}    & 1.65             & {\ul 98.60}    & 55.28          & 62.92            & 66.88          & 47.91          & 1.86             & 93.73          & 56.20               & {\ul 0.09}            & \textbf{98.76}      \\
LF~\cite{zeng2021rethinking}                                                                          & 55.97          & 98.57            & -              & {\ul 54.80}    & 94.87            & 51.26          & 51.44          & 95.25            & 49.40          & 45.45          & 50.49            & 68.78          & 52.99               & 85.56                 & 55.02               \\
SSBA~\cite{li2021invisible}                                                                        & 55.22          & 97.71            & -              & {\ul 54.80}    & 91.57            & 52.86          & 50.47          & 88.87            & 52.04          & 45.32          & 57.32            & 65.25          & 52.47               & 53.47                 & 70.75               \\
Trojan~\cite{liu2018trojaning}                                                                      & 55.89          & 99.98            & -              & {\ul 55.42}    & 0.50             & \textbf{99.50} & 50.22          & 8.82             & 92.74          & 48.48          & 0.83             & 95.87          & 52.69               & {\ul 0.15}            & 98.31               \\
WaNet~\cite{nguyen2021wanet}                                                                       & 56.78          & 99.49            & -              & \textbf{56.74} & {\ul 0.19}       & \textbf{99.63} & 53.84          & 3.94             & 96.30          & 46.98          & 0.43             & 94.63          & 52.33               & 0.23                  & 97.40               \\ \hline
Average                                                                     & 56.22          & 99.19            & -              & {\ul 55.63}    & 40.94            & 78.83          & 52.12          & 65.11            & 64.99          & 46.77          & 29.46            & 80.14          & 52.50               & 33.26                 & 81.10               \\ \hline
\multirow{2}{*}{\begin{tabular}[c]{@{}c@{}}Backdoor\\ Attacks\end{tabular}} & \multicolumn{3}{c|}{ANP~\cite{wu2021adversarial}}                           & \multicolumn{3}{c|}{CLP~\cite{zheng2022data}}                           & \multicolumn{3}{c|}{i-BAU~\cite{zeng2021adversarial}}                         & \multicolumn{3}{c|}{RNP~\cite{li2023reconstructive}}                           & \multicolumn{3}{c}{\nameFramework \textbf{\textcolor{red}{(Ours)}}} \\ \cline{2-16} 
                                                                            & ACC $\uparrow$ & ASR $\downarrow$ & DER $\uparrow$ & ACC $\uparrow$ & ASR $\downarrow$ & DER $\uparrow$ & ACC $\uparrow$ & ASR $\downarrow$ & DER $\uparrow$ & ACC $\uparrow$ & ASR $\downarrow$ & DER $\uparrow$ & ACC $\uparrow$      & ASR $\downarrow$      & DER $\uparrow$      \\ \hline
BadNets~\cite{gu2019badnets}                                                                     & 50.55          & 7.74             & 93.29          & \textbf{55.94} & 100.00           & 49.86          & 51.48          & 97.36            & 48.95          & 21.91          & \textbf{0.00}    & 82.84          & 53.72               & 0.34                  & {\ul 98.58}         \\
Blended~\cite{chen2017targeted}                                                                     & 54.99          & 84.61            & 57.03          & \textbf{55.70} & 99.68            & 49.85          & 53.03          & 91.90            & 52.40          & 34.60          & \textbf{0.11}    & {\ul 89.08}    & 53.62               & {\ul 2.30}            & \textbf{97.50}      \\
Input-aware~\cite{nguyen2020input}                                                                 & 53.17          & 0.17             & 97.20          & \textbf{57.75} & 99.58            & 50.00          & 52.48          & 72.98            & 60.45          & 15.57          & \textbf{0.00}    & 78.49          & 55.38               & 0.10                  & 98.34               \\
LF~\cite{zeng2021rethinking}                                                                          & 54.66          & 95.39            & 50.94          & \textbf{55.61} & 98.49            & 49.86          & 51.13          & 85.32            & 54.21          & 49.18          & \textbf{0.00}    & {\ul 95.89}    & 52.47               & {\ul 1.90}            & \textbf{96.59}      \\
SSBA~\cite{li2021invisible}                                                                        & 52.83          & 91.44            & 51.94          & \textbf{55.17} & 97.65            & 50.01          & 49.86          & 81.90            & 55.22          & 37.64          & \textbf{0.00}    & {\ul 90.06}    & 52.93               & {\ul 1.38}            & \textbf{97.02}      \\
Trojan~\cite{liu2018trojaning}                                                                      & 50.37          & 1.40             & 96.53          & \textbf{55.86} & 8.39             & 95.78          & 52.65          & 98.49            & 49.12          & 46.27          & \textbf{0.00}    & 95.18          & 53.66               & 0.31                  & {\ul 98.72}         \\
WaNet~\cite{nguyen2021wanet}                                                                       & 53.87          & 0.75             & 97.91          & 56.21          & 98.50            & 50.21          & 53.71          & 75.23            & 60.60          & 20.50          & \textbf{0.00}    & 81.60          & 55.01               & 0.71                  & {\ul 98.50}         \\ \hline
Average                                                                     & 52.92          & 40.21            & 77.83          & \textbf{56.03} & 86.04            & 56.51          & 52.05          & 86.17            & 54.42          & 32.24          & \textbf{0.02}    & {\ul 87.59}    & 53.83               & {\ul 1.01}            & \textbf{97.89}      \\ \hline
\end{tabular}}
\vspace{-3mm}
\end{table*}

\subsection{Main Results}
We validate the effectiveness of our proposed framework on 8 SOTA backdoor attacks and compare it with 8 defenses. In this section, we present the main results on CIFAR-10 and Tiny ImageNet with a 10\% poisoning ratio on PreAct-ResNet18 for illustration, which is shown in Table~\ref{tab:cifar10_preact_10} and Table~\ref{tab:tiny_preact_10}. More results on GTSRB and VGG19-BN can be found in Appendix~\ref{sec:perform_gtsrb} and \ref{sec:perform_vgg}, respectively.

\textbf{Results on CIFAR-10.}
Table~\ref{tab:cifar10_preact_10} shows the results on CIFAR-10. Results show that our \nameFramework outperforms all the other SOTA defenses on the average of ASR (2.18\%) and DER (97.09\%), as well as a promising ACC (91.70\%) higher than the original ``No Defense'' models (91.34\%), which indicates its effectiveness in removing the backdoor effect with the least cost. Though most defenses fail in strong attacks Blended, LF, or SSBA, \eg, FT, FP, NAD, ANP, CLP, i-BAU, and RNP, our proposed \nameFramework success with the best ASR and DER on Blended and second best ASR and DER on LF and SSBA. FT and NAD perform similarly on each attack with a promising ACC, but they also fail on WaNet with a high ASR except for the mentioned strong attacks. FP and ANP perform well in ACC among all the defenses, while the defense performances on ASR and DER are unstable, which may be due to the unsuccessful backdoor-related neuron locating. CLP fails on most of the attacks with high ASR and low DER, which may be due to the structure constraint of computing channel Lipschitz only on the convolutional-batch normalization layer combination. i-BAU performs the second best on average ASR and DER, with failures on three attacks.
\nameFramework outperforms RNP on almost all performances, which indicates that using clean unlearning with \nameMetric is more effective than the unlearn-recovery process in RNP.

\textbf{Results on Tiny ImageNet.}
Table~\ref{tab:tiny_preact_10} presents the results on Tiny ImageNet with PreAct-ResNet18. We observe that most defenses also fail on Blended, SSBA, and LF with high ASR. Similar performances of other attacks are shown on most of the defenses compared to CIFAR-10, where FT and FP also perform well in ACC and CLP fails on most attacks. 
Although i-BAU can successfully defend against most of the attacks in CIFAR-10, it fails on all attacks here, indicating that its adversarial training fails with large classification categories. For RNP, though it performs well in ASR on almost all attacks, the ACC is sacrificed too much to be unacceptable, indicating an unbalanced defense performance. In comparison, our \nameFramework can achieve SOTA on the average of DER, and perform second best on the average of ASR, which validates its superior defense performance.

\vspace{-2mm}
\subsection{Ablation Studies}
\label{subsec:ablation_study}

\begin{wrapfigure}{R}{0.3\textwidth}
    \centering
    \includegraphics[width=1\linewidth]{ 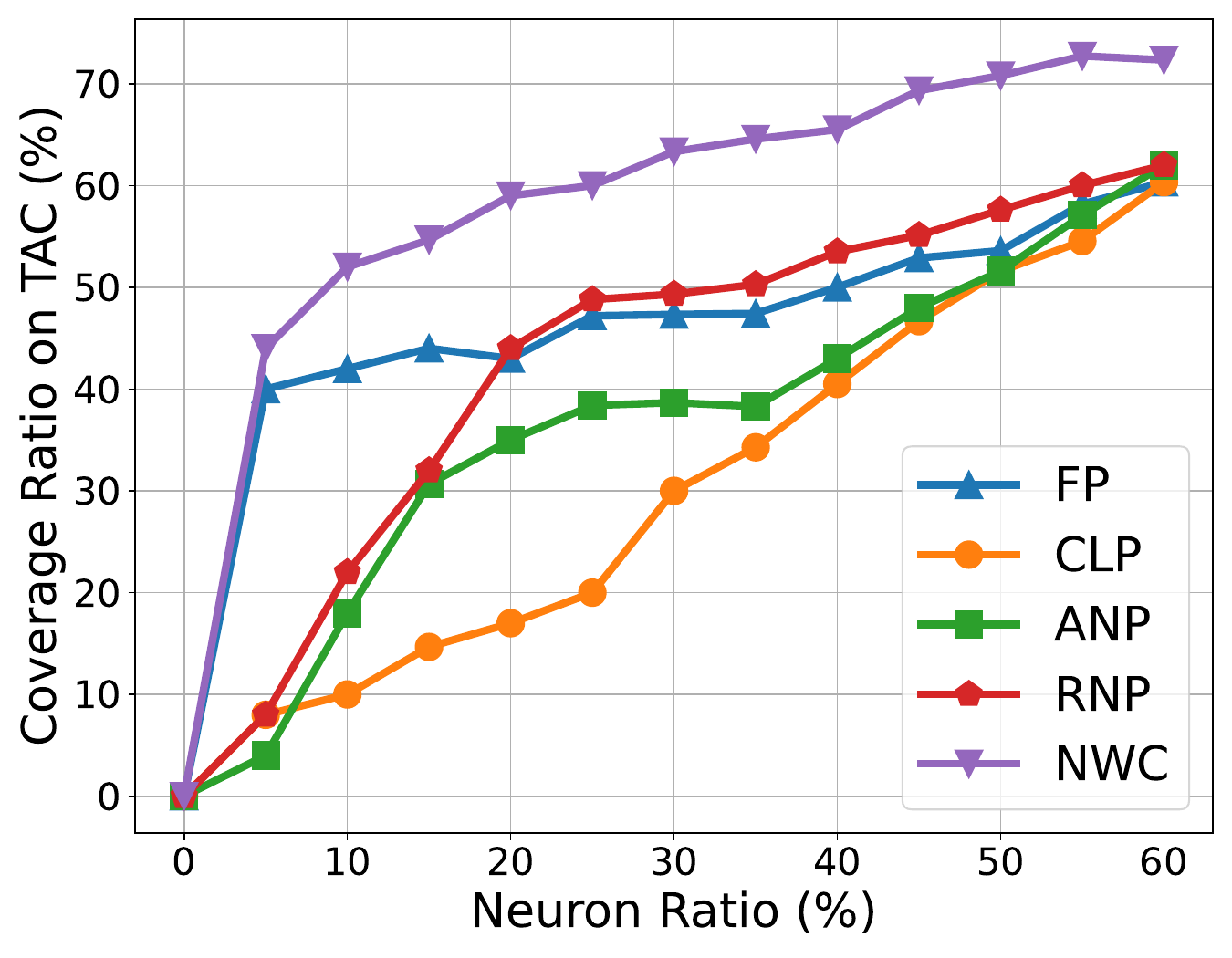}
    \vspace{-6mm}
    \caption{Comparison of neuron coverage ratio on TAC under different neuron ratios.}
    \label{fig:Fig_neuron_coverage}
\vspace{-5mm}
\end{wrapfigure}

\textbf{Effectiveness of \nameMetric order for Backdoor Strength.}
To verify the effectiveness of employing clean-unlearning \nameMetric order in gauging the backdoor strength, we borrow the \textit{Trigger-activated Change} (TAC)~\cite{zheng2022data} order as the ground truth and compare the neuron coverage ratio on TAC under different proportions, \ie, measuring the overlap of the selected neurons on both metrics. Specifically, TAC measures the change in neuron activation before and after the input image is attached with a trigger, where the larger value indicates a stronger backdoor effect. We select the following SOTA metrics for comparison: 1) the average neuron activations in FP~\cite{liu2018fine}; 2) the channel Lipschitz in CLP~\cite{zheng2022data}; 2) the perturb-recovery learned mask in ANP~\cite{wu2021adversarial}; 3) the unlearn-recovery learned mask in RNP~\cite{li2023reconstructive}. 
The result is illustrated in Figure~\ref{fig:Fig_neuron_coverage}, where the x-axis represents the (reinitializing/pruning) neuron ratio and the y-axis represents the neuron coverage ratio on TAC. The higher values on the y-axis indicate a better matching of the current metric and the TAC metric, \ie, more backdoor-related neurons are chosen. We can observe that \nameMetric is always the best under different neuron ratios compared to others.
It validates the effectiveness of using \nameMetric order as the metric for neuron reinitialization/pruning. 
Moreover, we also test the performance of using \nameMetric in FP to show its superiority in assisting other defenses. The results can be found in Appendix~\ref{sec:perform_other_defense}.

\begin{wraptable}{r}{0.5\textwidth}
\caption{Comparison of different reinitialization schemes on CIFAR-10 with PreAct-ResNet18 (\%).}
\label{tab:ablation_reinit}
\vspace{-2mm}
\centering
\resizebox{\linewidth}{!}{
\begin{tabular}{c|cc|cc|cc}
\hline
\multirow{2}{*}{\begin{tabular}[c]{@{}c@{}}Backdoor\\ Attacks\end{tabular}} & \multicolumn{2}{c|}{$\textbf{V}_1$}           & \multicolumn{2}{c|}{$\textbf{V}_2$}           & \multicolumn{2}{c}{$\textbf{V}_3$ (Ours)}            \\ \cline{2-7} 
                                                                            & ACC $\uparrow$ & ASR $\downarrow$ & ACC $\uparrow$ & ASR $\downarrow$ & ACC $\uparrow$ & ASR $\downarrow$ \\ \hline
BadNets~\cite{gu2019badnets}                                                                     & 69.53          & 0.00             & 68.20          & 0.00             & \textbf{80.94}          & 0.00             \\
Blended~\cite{chen2017targeted}                                                                     & 76.13          & 0.00             & 64.20          & 0.05             &\textbf{82.72}          & 0.00             \\
LF~\cite{zeng2021rethinking}                                                                          & 76.41          & 0.00             & 78.64          & 0.02             &\textbf{82.42}          & 0.00             \\
SSBA~\cite{li2021invisible}                                                                        & 74.55          & 0.00             & 75.43          & 0.02             & \textbf{83.10}          & 0.00             \\ \hline
\end{tabular}}
\vspace{-5mm}
\end{wraptable}

\textbf{Effectiveness of Zero Reinitialization on Subweight.}
We compare different combinations of reinitializing neurons and weights to validate the effectiveness of zero reinitialization on the selected subweights. Specifically, three versions are designed for comparison: (1) $\textbf{V}_1$: all weights on the selected top-\nameMetric neurons are reinitialized to zero. (2) $\textbf{V}_2$: $m\%$ (70\% is used as default) of the top weights on each selected top-\nameMetric neuron are reinitialized to zero. (3) $\textbf{V}_3$ (Ours): our final version, where $m\%$ of the top weights among all selected top-\nameMetric neurons are reinitialized to zero. As the ACC and ASR drop monotonously with more neurons selected to be reinitialized, we record the results when ASR first drops to almost zero (\ie, lower equal than 0.05\%). Table~\ref{tab:ablation_reinit} shows the performances on four attacks. It validates that reinitializing the top weights among the selected neurons may help alleviate the hurt of clean accuracy. The failure of $\textbf{V}_2$ may be blamed on the different backdoor strength of each neuron, \ie, some strong backdoor-related neurons fail to be removed thoroughly and thus more neurons are reinitialized to reach zero ASR.

\begin{wraptable}{r}{0.5\textwidth}
\caption{Comparison of different fine-tuning schemes on CIFAR-10 with PreAct-ResNet18 (\%).}
\label{tab:ablation_ft}
\vspace{-2mm}
\centering
\resizebox{\linewidth}{!}{
\begin{tabular}{c|cc|cc|cc}
\hline
\multirow{2}{*}{\begin{tabular}[c]{@{}c@{}}Backdoor\\ Attacks\end{tabular}} & \multicolumn{2}{c|}{No-FT}           & \multicolumn{2}{c|}{Vanilla-FT}           & \multicolumn{2}{c}{Aa-FT (Ours)}            \\ \cline{2-7} 
                                                                            & ACC $\uparrow$ & ASR $\downarrow$ & ACC $\uparrow$ & ASR $\downarrow$ & ACC $\uparrow$ & ASR $\downarrow$ \\ \hline
BadNets~\cite{gu2019badnets}                                                                     & 80.94          & 0.00             & 90.66          & 1.69             & 90.72          & 1.37             \\
Blended~\cite{chen2017targeted}                                                                     & 82.72          & 0.00             & 91.40         & 6.39            &91.61          & 2.61             \\
LF~\cite{zeng2021rethinking}                                                                          & 82.42          & 0.00             & 91.51          & 5.57             &91.20          & 2.64             \\
SSBA~\cite{li2021invisible}                                                                        & 83.10          & 0.00             & 91.12          & 7.6             & 91.57          & 1.66             \\ \hline
\end{tabular}}
\vspace{-5mm}
\end{wraptable}
\textbf{Effectiveness of Gradient-norm Regulation in Fine-tuning.}
To validate the essential role of gradient-norm regulation in the fine-tuning stage, we compare the performance on ``no fine-tuning'' (No-FT), ``vanilla fine-tuning'' (Vanilla-FT), and our ``activeness-aware fine-tuning'' (Aa-FT). For a fair comparison, we follow the default settings and choose to fine-tune the reinitialized model after the second stage. The results are shown in Table~\ref{tab:ablation_ft}. It shows that fine-tuning is effective in improving clean accuracy, which is sacrificed to erase the backdoored effect in reinitialization, though it is prone to bring back some extent of ASR. Furthermore, compared to vanilla fine-tuning, our implemented activeness-aware fine-tuning can suppress the rise of ASR as well as improve ACC to the same level.

\vspace{-2mm}

\subsection{Further Analysis}
\label{subsec:further_analysis}
\begin{figure}[]
    \centering
        \includegraphics[width=0.24\linewidth]{ 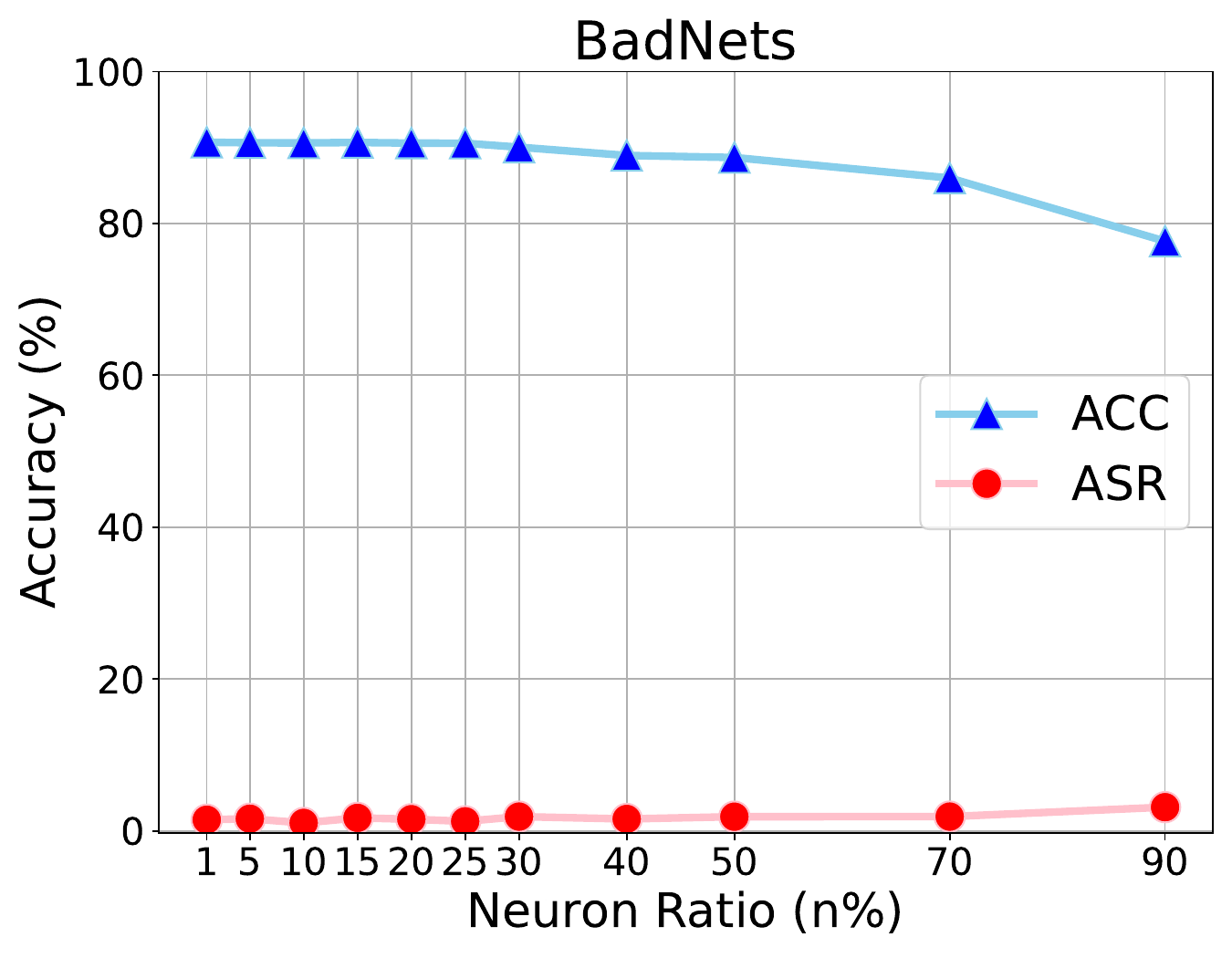}
        \includegraphics[width=0.24\linewidth]{ 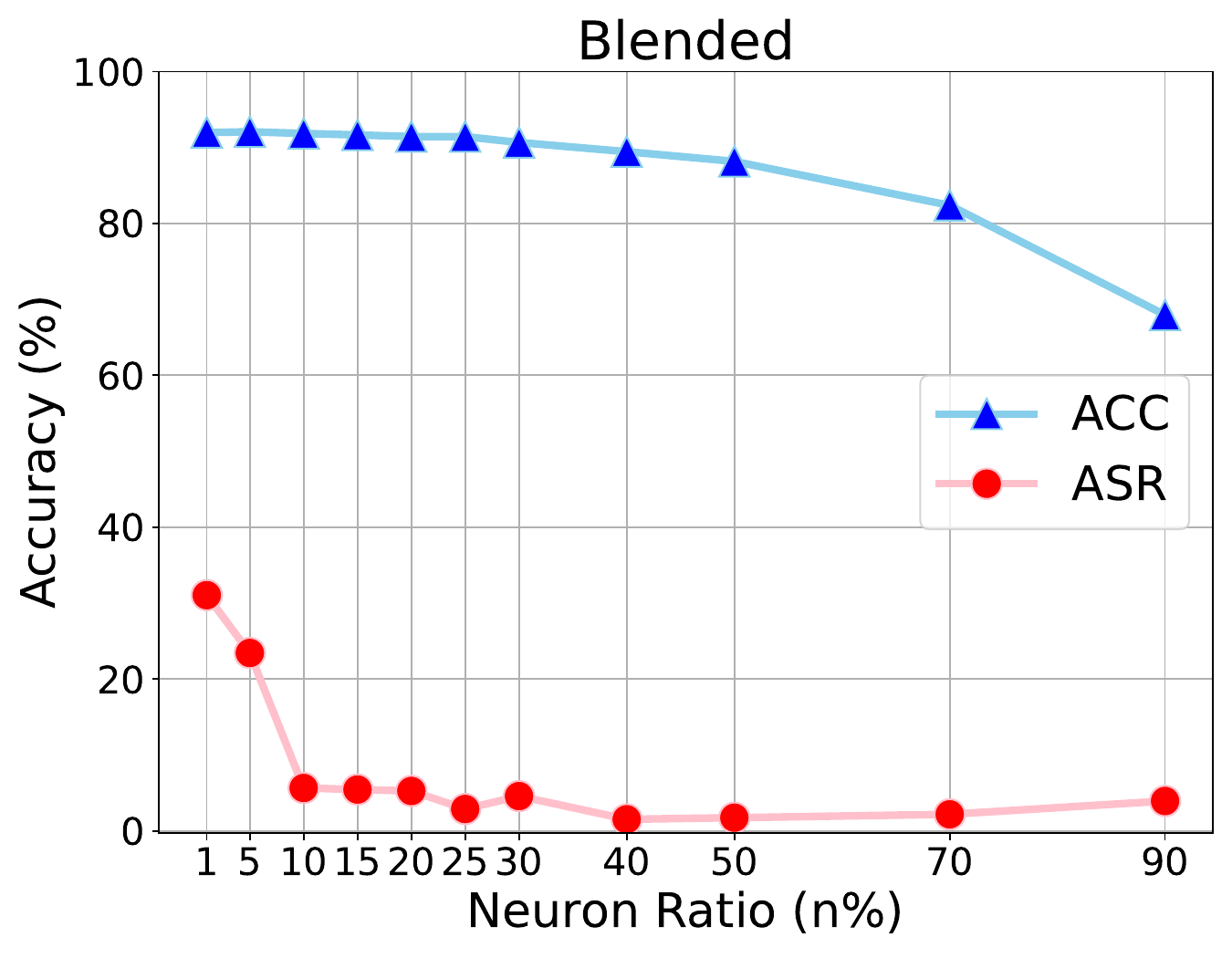}
        \includegraphics[width=0.24\linewidth]{ 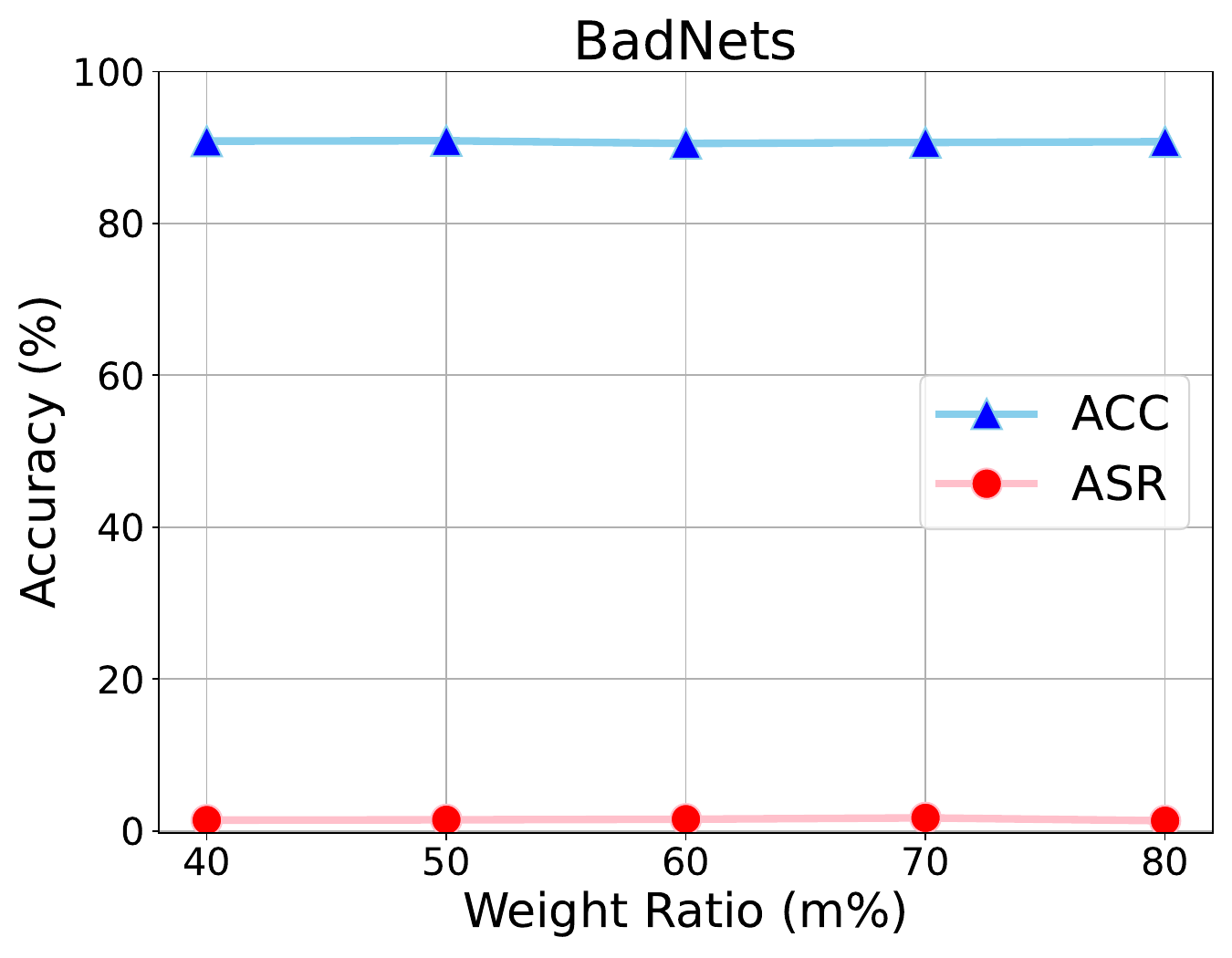}
        \includegraphics[width=0.24\linewidth]{ 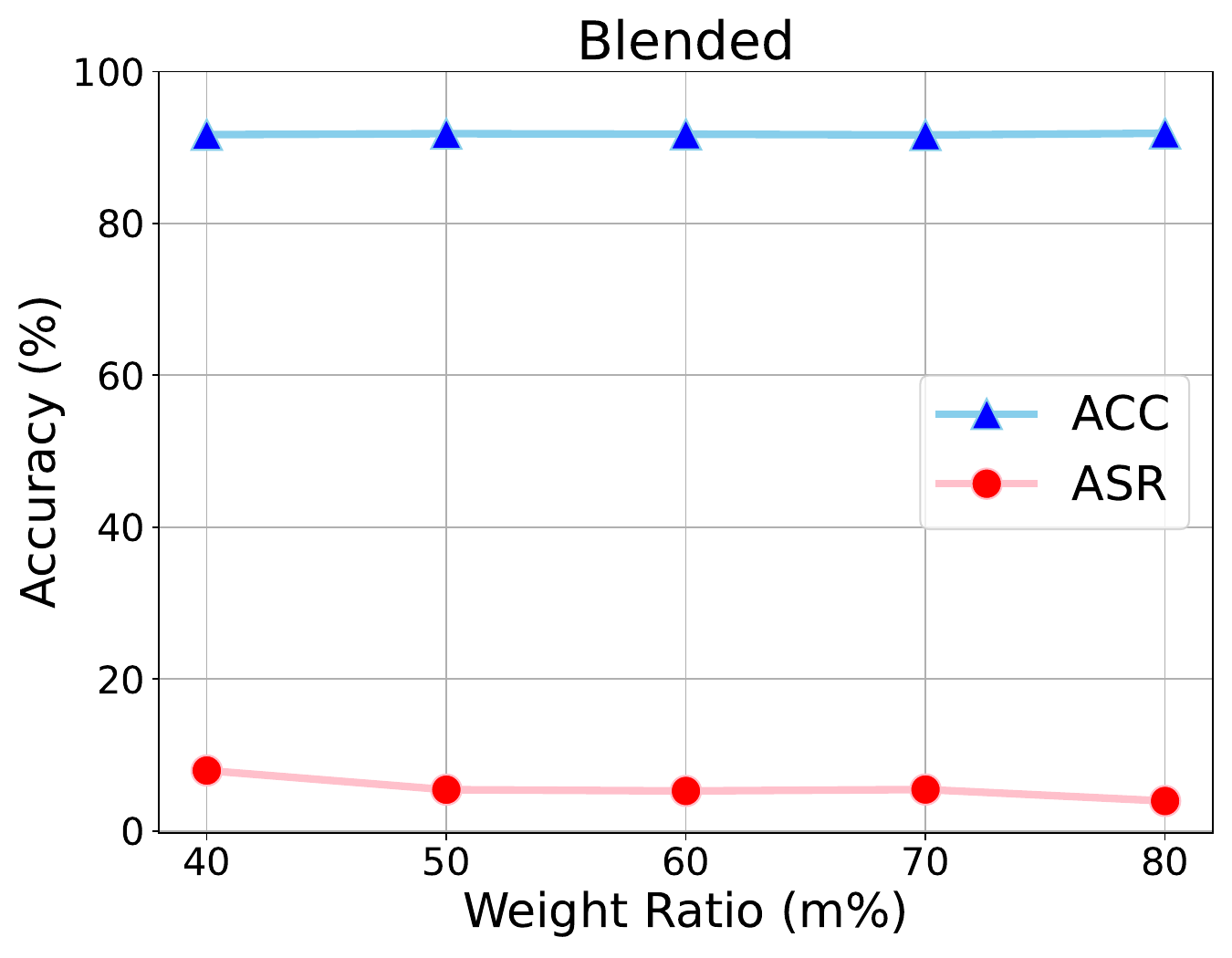}
    \vspace{-2mm}
    \caption{Performance with different neuron ratios (two subfigures on the left) and weight ratios (two subfigures on the right) under the attacks of BadNets and Blended.}
    \label{fig:analysis_ratio}
\vspace{-5mm}
\end{figure}
To test the sensitivity of our method towards hyper-parameter tuning, we record the performance of \nameFramework under a wide range of neuron ratio ($n\%$) and weight ratio ($m\%$) tuning. The experiments are conducted on CIFAR-10 with a 10\% poisoning ratio on PreAct-ResNet18. The neuron ratios are tuned from 1\% to 90\%, and weight ratios are tuned from 40\% to 80\%. 
The results of BadNets and Blended attacks are depicted in Figure~\ref{fig:analysis_ratio}. It shows that \nameFramework is insensitive to both neuron ratio and weight ratio, where the ASR is kept at a very low level, while ACC is maintained at a top level under a wide range of tuning. The high consistency in the performances may come from the promising recovery ability of the fine-tuning stage. More results on other attacks are shown in the Appendix~\ref{sec:perform_ratio}.


\textbf{More Experiments and Analysis.} Due to the space limit, we postpone the detailed discussion of the poisoning ratio, the clean data ratio, and the fine-tuning learning rate to Appendix~\ref{sec:perform_poison_ratio}, Appendix~\ref{sec:perform_clean_ratio} and Appendix~\ref{sec:perform_ftlr}, respectively. We also evaluate defense on the clean model in Appendix~\ref{sec:clean_model}.


\section{Conclusion}
\label{sec:conclusion}
In this work, we propose an effective two-stage backdoor defense method, \nameFramework, to eliminate the backdoor effect in DNNs. Our research reveals two important observations regarding the backdoored models to support our method. First, there is a positive correlation between weight changes during poison and clean unlearning in backdoored models.  
This finding enables us to identify and eliminate backdoor-related neurons through clean unlearning and zero reinitialization. 
Second, neurons in backdoored models are more active compared to those in clean models
, which suggests regulating the gradient norm during fine-tuning. 
Furthermore, we also provide insights into these two observations from the perspective of neuron activations, which may be a valuable contribution to the field of backdoor defense. Extensive experiments demonstrate the superiority of our method over recent defenses. 
One current challenge as well as promising future work involves defending against backdoor attacks without any accessible clean data. The data generation techniques and data-free techniques may be the potential solutions.

{
\small
\bibliographystyle{unsrtnat}
\bibliography{egbib}

\begin{thebibliography}{53}
\providecommand{\natexlab}[1]{#1}
\providecommand{\url}[1]{\texttt{#1}}
\expandafter\ifx\csname urlstyle\endcsname\relax
  \providecommand{\doi}[1]{doi: #1}\else
  \providecommand{\doi}{doi: \begingroup \urlstyle{rm}\Url}\fi

\bibitem[Taigman et~al.(2014)Taigman, Yang, Ranzato, and Wolf]{taigman2014deepface}
Yaniv Taigman, Ming Yang, Marc'Aurelio Ranzato, and Lior Wolf.
\newblock Deepface: Closing the gap to human-level performance in face verification.
\newblock In \emph{CVPR}, 2014.

\bibitem[Parmar and Mehta(2014)]{parmar2014face}
Divyarajsinh~N Parmar and Brijesh~B Mehta.
\newblock Face recognition methods \ applications.
\newblock \emph{arXiv preprint arXiv:1403.0485}, 2014.

\bibitem[Ibrahim and Zin(2011)]{ibrahim2011study}
Ratnawati Ibrahim and Zalhan~Mohd Zin.
\newblock Study of automated face recognition system for office door access control application.
\newblock In \emph{ICCSN}, 2011.

\bibitem[Chen et~al.(2021)Chen, Zhang, Liu, Feng, Dong, Luo, and Wan]{chen2021uscl}
Yixiong Chen, Chunhui Zhang, Li~Liu, Cheng Feng, Changfeng Dong, Yongfang Luo, and Xiang Wan.
\newblock Uscl: pretraining deep ultrasound image diagnosis model through video contrastive representation learning.
\newblock In \emph{MICCAI}, 2021.

\bibitem[Chen et~al.(2023)Chen, Liu, Li, Jiang, Ding, and Zhou]{chen2023metalr}
Yixiong Chen, Li~Liu, Jingxian Li, Hua Jiang, Chris Ding, and Zongwei Zhou.
\newblock Metalr: Meta-tuning of learning rates for transfer learning in medical imaging.
\newblock In \emph{MICCAI}, 2023.

\bibitem[Yurtsever et~al.(2020)Yurtsever, Lambert, Carballo, and Takeda]{yurtsever2020survey}
Ekim Yurtsever, Jacob Lambert, Alexander Carballo, and Kazuya Takeda.
\newblock A survey of autonomous driving: Common practices and emerging technologies.
\newblock \emph{IEEE access}, 2020.

\bibitem[Caesar et~al.(2020)Caesar, Bankiti, Lang, Vora, Liong, Xu, Krishnan, Pan, Baldan, and Beijbom]{caesar2020nuscenes}
Holger Caesar, Varun Bankiti, Alex~H Lang, Sourabh Vora, Venice~Erin Liong, Qiang Xu, Anush Krishnan, Yu~Pan, Giancarlo Baldan, and Oscar Beijbom.
\newblock nuscenes: A multimodal dataset for autonomous driving.
\newblock In \emph{CVPR}, 2020.

\bibitem[Gu et~al.(2019)Gu, Liu, Dolan-Gavitt, and Garg]{gu2019badnets}
Tianyu Gu, Kang Liu, Brendan Dolan-Gavitt, and Siddharth Garg.
\newblock Badnets: Evaluating backdooring attacks on deep neural networks.
\newblock \emph{IEEE Access}, 2019.

\bibitem[Li et~al.(2021{\natexlab{a}})Li, Li, Wu, Li, He, and Lyu]{li2021invisible}
Yuezun Li, Yiming Li, Baoyuan Wu, Longkang Li, Ran He, and Siwei Lyu.
\newblock Invisible backdoor attack with sample-specific triggers.
\newblock In \emph{ICCV}, 2021{\natexlab{a}}.

\bibitem[Wu et~al.(2023)Wu, Liu, Zhu, Liu, He, and Lyu]{wu2023adversarial}
Baoyuan Wu, Li~Liu, Zihao Zhu, Qingshan Liu, Zhaofeng He, and Siwei Lyu.
\newblock Adversarial machine learning: A systematic survey of backdoor attack, weight attack and adversarial example.
\newblock \emph{arXiv preprint arXiv:2302.09457}, 2023.

\bibitem[Zheng et~al.(2022{\natexlab{a}})Zheng, Tang, Li, and Liu]{zheng2022pre}
Runkai Zheng, Rongjun Tang, Jianze Li, and Li~Liu.
\newblock Pre-activation distributions expose backdoor neurons.
\newblock In \emph{NeurIPS}, 2022{\natexlab{a}}.

\bibitem[He et~al.(2016)He, Zhang, Ren, and Sun]{he2016identity}
Kaiming He, Xiangyu Zhang, Shaoqing Ren, and Jian Sun.
\newblock Identity mappings in deep residual networks.
\newblock In \emph{ECCV}, 2016.

\bibitem[Krizhevsky et~al.(2009)Krizhevsky, Hinton, et~al.]{krizhevsky2009learning}
Alex Krizhevsky, Geoffrey Hinton, et~al.
\newblock Learning multiple layers of features from tiny images.
\newblock 2009.

\bibitem[Li et~al.(2021{\natexlab{b}})Li, Lyu, Koren, Lyu, Li, and Ma]{li2021anti}
Yige Li, Xixiang Lyu, Nodens Koren, Lingjuan Lyu, Bo~Li, and Xingjun Ma.
\newblock Anti-backdoor learning: Training clean models on poisoned data.
\newblock In \emph{NeurIPS}, 2021{\natexlab{b}}.

\bibitem[Chen et~al.(2018)Chen, Carvalho, Baracaldo, Ludwig, Edwards, Lee, Molloy, and Srivastava]{chen2018detecting}
Bryant Chen, Wilka Carvalho, Nathalie Baracaldo, Heiko Ludwig, Benjamin Edwards, Taesung Lee, Ian Molloy, and Biplav Srivastava.
\newblock Detecting backdoor attacks on deep neural networks by activation clustering.
\newblock \emph{arXiv preprint arXiv:1811.03728}, 2018.

\bibitem[Tran et~al.(2018)Tran, Li, and Madry]{tran2018spectral}
Brandon Tran, Jerry Li, and Aleksander Madry.
\newblock Spectral signatures in backdoor attacks.
\newblock In \emph{NeurIPS}, 2018.

\bibitem[Qi et~al.(2023)Qi, Xie, Wang, Wu, Mahloujifar, and Mittal]{qi2023towards}
Xiangyu Qi, Tinghao Xie, Jiachen~T Wang, Tong Wu, Saeed Mahloujifar, and Prateek Mittal.
\newblock Towards a proactive $\{$ML$\}$ approach for detecting backdoor poison samples.
\newblock In \emph{USENIX Security}, 2023.

\bibitem[Liu et~al.(2023)Liu, Sangiovanni-Vincentelli, and Yue]{liu2023beating}
Min Liu, Alberto Sangiovanni-Vincentelli, and Xiangyu Yue.
\newblock Beating backdoor attack at its own game.
\newblock In \emph{ICCV}, 2023.

\bibitem[Chen et~al.(2022{\natexlab{a}})Chen, Wu, and Wang]{chen2022effective}
Weixin Chen, Baoyuan Wu, and Haoqian Wang.
\newblock Effective backdoor defense by exploiting sensitivity of poisoned samples.
\newblock In \emph{NeurIPS}, 2022{\natexlab{a}}.

\bibitem[Wang et~al.(2019)Wang, Yao, Shan, Li, Viswanath, Zheng, and Zhao]{wang2019neural}
Bolun Wang, Yuanshun Yao, Shawn Shan, Huiying Li, Bimal Viswanath, Haitao Zheng, and Ben~Y Zhao.
\newblock Neural cleanse: Identifying and mitigating backdoor attacks in neural networks.
\newblock In \emph{SP}, 2019.

\bibitem[Zeng et~al.(2021{\natexlab{a}})Zeng, Chen, Park, Mao, Jin, and Jia]{zeng2021adversarial}
Yi~Zeng, Si~Chen, Won Park, Zhuoqing Mao, Ming Jin, and Ruoxi Jia.
\newblock Adversarial unlearning of backdoors via implicit hypergradient.
\newblock In \emph{International Conference on Learning Representations}, 2021{\natexlab{a}}.

\bibitem[Li et~al.(2023)Li, Lyu, Ma, Koren, Lyu, Li, and Jiang]{li2023reconstructive}
Yige Li, Xixiang Lyu, Xingjun Ma, Nodens Koren, Lingjuan Lyu, Bo~Li, and Yu-Gang Jiang.
\newblock Reconstructive neuron pruning for backdoor defense.
\newblock \emph{arXiv preprint arXiv:2305.14876}, 2023.

\bibitem[Nguyen and Tran(2020)]{nguyen2020input}
Tuan~Anh Nguyen and Anh Tran.
\newblock Input-aware dynamic backdoor attack.
\newblock In \emph{NeurIPS}, 2020.

\bibitem[Nguyen and Tran(2021)]{nguyen2021wanet}
Anh Nguyen and Anh Tran.
\newblock Wanet--imperceptible warping-based backdoor attack.
\newblock \emph{arXiv preprint arXiv:2102.10369}, 2021.

\bibitem[Chen et~al.(2017)Chen, Liu, Li, Lu, and Song]{chen2017targeted}
Xinyun Chen, Chang Liu, Bo~Li, Kimberly Lu, and Dawn Song.
\newblock Targeted backdoor attacks on deep learning systems using data poisoning.
\newblock \emph{arXiv preprint arXiv:1712.05526}, 2017.

\bibitem[Barni et~al.(2019)Barni, Kallas, and Tondi]{barni2019new}
Mauro Barni, Kassem Kallas, and Benedetta Tondi.
\newblock A new backdoor attack in cnns by training set corruption without label poisoning.
\newblock In \emph{ICIP}, 2019.

\bibitem[Shafahi et~al.(2018)Shafahi, Huang, Najibi, Suciu, Studer, Dumitras, and Goldstein]{shafahi2018poison}
Ali Shafahi, W~Ronny Huang, Mahyar Najibi, Octavian Suciu, Christoph Studer, Tudor Dumitras, and Tom Goldstein.
\newblock Poison frogs! targeted clean-label poisoning attacks on neural networks.
\newblock In \emph{NeurIPS}, 2018.

\bibitem[Zhao et~al.(2020)Zhao, Ma, Zheng, Bailey, Chen, and Jiang]{zhao2020clean}
Shihao Zhao, Xingjun Ma, Xiang Zheng, James Bailey, Jingjing Chen, and Yu-Gang Jiang.
\newblock Clean-label backdoor attacks on video recognition models.
\newblock In \emph{CVPR}, 2020.

\bibitem[Doan et~al.(2021{\natexlab{a}})Doan, Lao, Zhao, and Li]{doan2021lira}
Khoa Doan, Yingjie Lao, Weijie Zhao, and Ping Li.
\newblock Lira: Learnable, imperceptible and robust backdoor attacks.
\newblock In \emph{ICCV}, 2021{\natexlab{a}}.

\bibitem[Bagdasaryan and Shmatikov(2021)]{bagdasaryan2021blind}
Eugene Bagdasaryan and Vitaly Shmatikov.
\newblock Blind backdoors in deep learning models.
\newblock In \emph{USENIX Security}, 2021.

\bibitem[Doan et~al.(2021{\natexlab{b}})Doan, Lao, and Li]{doan2021backdoor}
Khoa Doan, Yingjie Lao, and Ping Li.
\newblock Backdoor attack with imperceptible input and latent modification.
\newblock In \emph{NeurIPS}, 2021{\natexlab{b}}.

\bibitem[Huang et~al.(2022)Huang, Li, Wu, Qin, and Ren]{huang2022backdoor}
Kunzhe Huang, Yiming Li, Baoyuan Wu, Zhan Qin, and Kui Ren.
\newblock Backdoor defense via decoupling the training process.
\newblock In \emph{ICLR}, 2022.

\bibitem[Guo et~al.(2021)Guo, Li, and Liu]{guo2021aeva}
Junfeng Guo, Ang Li, and Cong Liu.
\newblock Aeva: Black-box backdoor detection using adversarial extreme value analysis.
\newblock \emph{arXiv preprint arXiv:2110.14880}, 2021.

\bibitem[Jiang et~al.(2022)Jiang, Wen, Zhan, Wang, Song, and Bian]{jiang2022critical}
Wei Jiang, Xiangyu Wen, Jinyu Zhan, Xupeng Wang, Ziwei Song, and Chen Bian.
\newblock Critical path-based backdoor detection for deep neural networks.
\newblock \emph{TNNLS}, 2022.

\bibitem[Kolouri et~al.(2020)Kolouri, Saha, Pirsiavash, and Hoffmann]{kolouri2020universal}
Soheil Kolouri, Aniruddha Saha, Hamed Pirsiavash, and Heiko Hoffmann.
\newblock Universal litmus patterns: Revealing backdoor attacks in cnns.
\newblock In \emph{CVPR}, 2020.

\bibitem[Xu et~al.(2021)Xu, Wang, Li, Borisov, Gunter, and Li]{xu2021detecting}
Xiaojun Xu, Qi~Wang, Huichen Li, Nikita Borisov, Carl~A Gunter, and Bo~Li.
\newblock Detecting ai trojans using meta neural analysis.
\newblock In \emph{SP}, 2021.

\bibitem[Liu et~al.(2018{\natexlab{a}})Liu, Dolan-Gavitt, and Garg]{liu2018fine}
Kang Liu, Brendan Dolan-Gavitt, and Siddharth Garg.
\newblock Fine-pruning: Defending against backdooring attacks on deep neural networks.
\newblock In \emph{RAID}, 2018{\natexlab{a}}.

\bibitem[Zheng et~al.(2022{\natexlab{b}})Zheng, Tang, Li, and Liu]{zheng2022data}
Runkai Zheng, Rongjun Tang, Jianze Li, and Li~Liu.
\newblock Data-free backdoor removal based on channel lipschitzness.
\newblock In \emph{ECCV}, 2022{\natexlab{b}}.

\bibitem[Chen et~al.(2022{\natexlab{b}})Chen, Zhang, Zhang, Chang, Liu, and Wang]{chen2022quarantine}
Tianlong Chen, Zhenyu Zhang, Yihua Zhang, Shiyu Chang, Sijia Liu, and Zhangyang Wang.
\newblock Quarantine: Sparsity can uncover the trojan attack trigger for free.
\newblock In \emph{CVPR}, 2022{\natexlab{b}}.

\bibitem[Guan et~al.(2022)Guan, Tu, He, and Tao]{guan2022few}
Jiyang Guan, Zhuozhuo Tu, Ran He, and Dacheng Tao.
\newblock Few-shot backdoor defense using shapley estimation.
\newblock In \emph{CVPR}, 2022.

\bibitem[Wu and Wang(2021)]{wu2021adversarial}
Dongxian Wu and Yisen Wang.
\newblock Adversarial neuron pruning purifies backdoored deep models.
\newblock In \emph{NeurIPS}, 2021.

\bibitem[Chai and Chen(2022)]{chai2022one}
Shuwen Chai and Jinghui Chen.
\newblock One-shot neural backdoor erasing via adversarial weight masking.
\newblock In \emph{NeurIPS}, 2022.

\bibitem[Li et~al.(2021{\natexlab{c}})Li, Lyu, Koren, Lyu, Li, and Ma]{li2021neural}
Yige Li, Xixiang Lyu, Nodens Koren, Lingjuan Lyu, Bo~Li, and Xingjun Ma.
\newblock Neural attention distillation: Erasing backdoor triggers from deep neural networks.
\newblock \emph{arXiv preprint arXiv:2101.05930}, 2021{\natexlab{c}}.

\bibitem[Wei et~al.(2024)Wei, Zhang, Zha, and Wu]{wei2024shared}
Shaokui Wei, Mingda Zhang, Hongyuan Zha, and Baoyuan Wu.
\newblock Shared adversarial unlearning: Backdoor mitigation by unlearning shared adversarial examples.
\newblock \emph{Advances in Neural Information Processing Systems}, 36, 2024.

\bibitem[Zhao et~al.(2022)Zhao, Zhang, and Hu]{zhao2022penalizing}
Yang Zhao, Hao Zhang, and Xiuyuan Hu.
\newblock Penalizing gradient norm for efficiently improving generalization in deep learning.
\newblock In \emph{International Conference on Machine Learning}, pages 26982--26992. PMLR, 2022.

\bibitem[Zeng et~al.(2021{\natexlab{b}})Zeng, Park, Mao, and Jia]{zeng2021rethinking}
Yi~Zeng, Won Park, Z~Morley Mao, and Ruoxi Jia.
\newblock Rethinking the backdoor attacks' triggers: A frequency perspective.
\newblock In \emph{ICCV}, 2021{\natexlab{b}}.

\bibitem[Liu et~al.(2018{\natexlab{b}})Liu, Ma, Aafer, Lee, Zhai, Wang, and Zhang]{liu2018trojaning}
Yingqi Liu, Shiqing Ma, Yousra Aafer, Wen-Chuan Lee, Juan Zhai, Weihang Wang, and Xiangyu Zhang.
\newblock Trojaning attack on neural networks.
\newblock In \emph{NDSS Symposium}, 2018{\natexlab{b}}.

\bibitem[Wu et~al.(2022)Wu, Chen, Zhang, Zhu, Wei, Yuan, and Shen]{wu2022backdoorbench}
Baoyuan Wu, Hongrui Chen, Mingda Zhang, Zihao Zhu, Shaokui Wei, Danni Yuan, and Chao Shen.
\newblock Backdoorbench: A comprehensive benchmark of backdoor learning.
\newblock In \emph{NeurIPS Datasets and Benchmarks Track}, 2022.

\bibitem[Le and Yang(2015)]{le2015tiny}
Ya~Le and Xuan Yang.
\newblock Tiny imagenet visual recognition challenge.
\newblock \emph{CS 231N}, 2015.

\bibitem[Stallkamp et~al.(2011)Stallkamp, Schlipsing, Salmen, and Igel]{stallkamp2011german}
Johannes Stallkamp, Marc Schlipsing, Jan Salmen, and Christian Igel.
\newblock The german traffic sign recognition benchmark: a multi-class classification competition.
\newblock In \emph{IJCNN}, 2011.

\bibitem[Simonyan and Zisserman(2014)]{simonyan2014very}
Karen Simonyan and Andrew Zisserman.
\newblock Very deep convolutional networks for large-scale image recognition.
\newblock \emph{arXiv preprint arXiv:1409.1556}, 2014.

\bibitem[Zhu et~al.(2023)Zhu, Wei, Shen, Fan, and Wu]{zhu2023enhancing}
Mingli Zhu, Shaokui Wei, Li~Shen, Yanbo Fan, and Baoyuan Wu.
\newblock Enhancing fine-tuning based backdoor defense with sharpness-aware minimization.
\newblock In \emph{ICCV}, 2023.

\bibitem[Wei et~al.(2023)Wei, Zhang, Zha, and Wu]{wei2023shared}
Shaokui Wei, Mingda Zhang, Hongyuan Zha, and Baoyuan Wu.
\newblock Shared adversarial unlearning: Backdoor mitigation by unlearning shared adversarial examples.
\newblock In \emph{NeurIPS}, 2023.

\end{thebibliography}
}


\appendix
\clearpage
\setcounter{section}{0}

\section*{Appendix Outline}
This appendix is organized as follows:
\begin{itemize}
    \item In Section~\ref{sec:overall_algo}, we detail the algorithm of \nameFramework.
    \item In Section~\ref{sec:ft_deri}, we detail the approximation process of the fine-tuning optimizations.
    \item In Section~\ref{sec:implem_details}, we introduce the implementation details, including the details of datasets, models, attacks, and defenses with our proposed method. 
    \item In Section~\ref{sec:perform_gtsrb}, we compare the defense results on the GTSRB dataset.
    \item In Section~\ref{sec:perform_vgg}, we compare the defense results on VGG19-BN structure.
    \item In Section~\ref{sec:perform_other_defense}, we show the effectiveness of using \nameMetric in other defense.
    \item In Section~\ref{sec:perform_ratio}, we show the comprehensive results with different neuron ratios and weight ratios.
    \item In Section~\ref{sec:perform_poison_ratio}, we discuss different poisoning ratios on the defense performance.
    \item In Section~\ref{sec:perform_clean_ratio}, we discuss different clean data ratios on the defense performance.
    \item In Section~\ref{sec:perform_ftlr}, we discuss different fine-tuning learning rates on the defense performance.
    \item In Section~\ref{sec:clean_model}, we evaluate the influence of using our method on the clean model.
\end{itemize}

\section{Detailed Algorithm of \nameFramework}
\label{sec:overall_algo}
To clearly illustrate our proposed method, we provide the detailed algorithm of the entire process of \nameFramework, which is shown in Algorithm~\ref{alg:overall}.

\section{Details of the Approximated Fine-Tuning Optimizations}
\label{sec:ft_deri}
As illustrated in Section~\ref{subsec:framework}, our proposed Activaness-aware fine-tuning involves calculating an additional gradient-norm regulation in the loss function, making it computationally inefficient. Therefore, we adopt the approximation scheme from~\cite{zhao2022penalizing} during practical optimization. 
Here, we provide the details of the approximated fine-tuning optimization. The approximation deviation is shown in the following. 
Specifically, for each step of the gradient calculation, we formulate it as:
\begin{equation}
    \label{equ:loss_appro}   
    \begin{aligned}
    \nabla_{\hat{\boldsymbol{\theta}}} \mathcal{L}_{ft}(\hat{\boldsymbol{\theta}}) 
    & =\nabla_{\hat{\boldsymbol{\theta}}} \mathcal{L}_{ce}(\hat{\boldsymbol{\theta}})+\lambda \cdot \nabla_{\hat{\boldsymbol{\theta}}}^2 \mathcal{L}_{ce}(\hat{\boldsymbol{\theta}}) \frac{\nabla_{\hat{\boldsymbol{\theta}}} \mathcal{L}_{ce}(\hat{\boldsymbol{\theta}})}{ \|\nabla_{\hat{\boldsymbol{\theta}}} \mathcal{L}_{ce}(\hat{\boldsymbol{\theta}}) \|_2} \\
    & 
     \approx \nabla_{\hat{\boldsymbol{\theta}}} \mathcal{L}_{ce}(\hat{\boldsymbol{\theta}}) 
     +\frac{\lambda}{r} \cdot (\nabla_{\hat{\boldsymbol{\theta}}} \mathcal{L}_{ce} (\hat{\boldsymbol{\theta}}+r \frac{\nabla_{\hat{\boldsymbol{\theta}}} \mathcal{L}_{ce}(\hat{\boldsymbol{\theta}})}{ \|\nabla_{\hat{\boldsymbol{\theta}}} \mathcal{L}_{ce}(\hat{\boldsymbol{\theta}}) \|_2} )-\nabla_{\hat{\boldsymbol{\theta}}} \mathcal{L}_{ce}(\hat{\boldsymbol{\theta}}) ) \\
    & = (1-\alpha) \nabla_{\hat{\boldsymbol{\theta}}} \mathcal{L}_{ce}(\hat{\boldsymbol{\theta}}) + \alpha \nabla_{\hat{\boldsymbol{\theta}}} \mathcal{L}_{ce} (\hat{\boldsymbol{\theta}}+r \frac{\nabla_{\hat{\boldsymbol{\theta}}} \mathcal{L}_{ce}(\hat{\boldsymbol{\theta}})}{ \|\nabla_{\hat{\boldsymbol{\theta}}} \mathcal{L}_{ce}(\hat{\boldsymbol{\theta}}) \|_2} ).
    \end{aligned}
\end{equation}
To avoid the Hessian computation, the second term is further approximated through an additional parameter update: \begin{equation}
    \label{equ:loss_appro_second}  
    \nabla_{\hat{\boldsymbol{\theta}}} \mathcal{L}_{ce} (\hat{\boldsymbol{\theta}}+r \frac{\nabla_{\hat{\boldsymbol{\theta}}} \mathcal{L}_{ce}(\hat{\boldsymbol{\theta}})}{ \|\nabla_{\hat{\boldsymbol{\theta}}} \mathcal{L}_{ce}(\hat{\boldsymbol{\theta}}) \|_2} ) \approx \nabla_{\hat{\boldsymbol{\theta}}} \mathcal{L}_{ce}(\hat{\boldsymbol{\theta}})|_{\hat{\boldsymbol{\theta}}=\hat{\boldsymbol{\theta}}+r \frac{\nabla_{\hat{\boldsymbol{\theta}}} \mathcal{L}_{ce}(\hat{\boldsymbol{\theta}})}{ \|\nabla_{\hat{\boldsymbol{\theta}}} \mathcal{L}_{ce}(\hat{\boldsymbol{\theta}}) \|_2}}.
\end{equation}

Based on the approximation, the practical fine-tuning process is illustrated in Algorithm~\ref{alg:approx_finetune}.

\section{More Implementation Details}
\label{sec:implem_details}
We further illustrate the implementations here, covering the details of datasets, models, attacks, and defenses.

\textbf{Dataset Details.} 
The experiments are conducted on CIFAR-10~\cite{krizhevsky2009learning}, Tiny ImageNet~\cite{le2015tiny}, and GTSRB~\cite{stallkamp2011german}. 
\begin{itemize}
    \item \textit{CIFAR-10.} It contains 60,000 32$\times$32 colored images with 10 classes. Each class owns 6,000 images, consisting of 5,000 for training and 1,000 for testing.
    \item \textit{Tiny ImageNet.} It is a subset of the full ImageNet, consisting of 100,000 training data and 10,000 testing data. There are 200 classes in total and 500 images per class for training. All images are 64$\times$64 with color.
    \item \textit{GTSRB.} GTSRB (German Traffic Sign Recognition Benchmark) contains 39,209 images for training and 12,630 images for testing with 43 classes. All images are 32$\times$32 colored images.
\end{itemize}

\textbf{Models.}
We choose PreAct-ResNet18~\cite{he2016identity} and VGG19-BN~\cite{simonyan2014very} as the target models to conduct attacks and defenses following the default configurations in BackdoorBench~\cite{wu2022backdoorbench}. Both of them contain convolutional layers and batch normalization layers, which can be implemented with all kinds of defense methods, \textit{e.g.}, FP~\cite{liu2018fine} for the last convolutional layer and ANP~\cite{wu2021adversarial} for the batch normalization layers. 
The extensive experiments on ablation study and further analysis are conducted on PreAct-ResNet18 by default.

\begin{figure}[]
    \centering
        \begin{subfigure}{0.11\linewidth}
            \includegraphics[width=\linewidth]{ 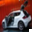}
            \caption*{BadNets}
        \end{subfigure}
        \begin{subfigure}{0.11\linewidth}
            \includegraphics[width=\linewidth]{ 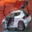}
            \caption*{Blended}
        \end{subfigure}
        \begin{subfigure}{0.11\linewidth}
            \includegraphics[width=\linewidth]{ 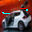}
            \caption*{Input-aware}
        \end{subfigure}
        \begin{subfigure}{0.11\linewidth}
            \includegraphics[width=\linewidth]{ 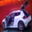}
            \caption*{LF}
        \end{subfigure}
        \begin{subfigure}{0.11\linewidth}
            \includegraphics[width=\linewidth]{ 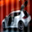}
            \caption*{SIG}
        \end{subfigure}
        \begin{subfigure}{0.11\linewidth}
            \includegraphics[width=\linewidth]{ 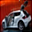}
            \caption*{SSBA}
        \end{subfigure}
        \begin{subfigure}{0.11\linewidth}
            \includegraphics[width=\linewidth]{ 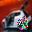}
            \caption*{Trojan}
        \end{subfigure}
        \begin{subfigure}{0.11\linewidth}
            \includegraphics[width=\linewidth]{ 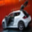}
            \caption*{WaNet}
        \end{subfigure}
    \caption{Examples of 8 backdoor-attack triggers on the same image of CIFAR-10.}
    \label{fig:attack_example}
\end{figure}

\textbf{Attack Details.}
We conduct 8 SOTA attacks for comprehensive testing, consisting of BadNets~\cite{gu2019badnets}, Blended~\cite{chen2017targeted}, Input-aware~\cite{nguyen2020input}, LF~\cite{zeng2021rethinking}, SIG~\cite{barni2019new}, SSBA~\cite{li2021invisible}, Trojan~\cite{liu2018trojaning} and WaNet~\cite{nguyen2021wanet}. All the attacks follow BackdoorBench's default configurations. 
Figure~\ref{fig:attack_example} shows all 8 attack triggers with the same example of CIFAR-10.
Specifically, for BadNets, a 3$\times$3 white square is patched at the bottom-right corner of the images for CIFAR-10 and GTSRB, and a 6$\times$6 white square is for Tiny ImageNet. For Blended, a Hello-Ketty image is blended in the images with a 0.2 transparent ratio.
We choose the 10\% poisoning ratio and 0$^{th}$ label as the default setting to conduct attacks and test all defenses following the previous works~\cite{zhu2023enhancing,wei2023shared}. 5\% and 1\% poisoning ratios are conducted only for testing our proposed method.

\textbf{Defense Details.}
We conduct 8 SOTA defenses for a comprehensive comparison, containing Fine-tuning (FT), Fine-pruning (FP)~\cite{liu2018fine}, NAD~\cite{li2021neural}, NC~\cite{wang2019neural}, ANP~\cite{wu2021adversarial}, 
CLP~\cite{zheng2022data}, i-BAU~\cite{zeng2021adversarial}, and RNP~\cite{li2023reconstructive}.
The defenses also follow the BackdoorBench's default configurations. Note that we compare only the post-training defenses with 5\% benign data provided. The learning rate for all methods is set to $10^{-2}$, the batch size is set to 256. For RNP with no implementation on BackdoorBench, we adapt it into the benchmark framework from the officially released code\footnote{\href{https://github.com/bboylyg/RNP}{https://github.com/bboylyg/RNP}}. Particularly, the clean data ratio is set to 0.5\% since we found that the RNP failed to defend well under the 5\% setting. 
For our proposed method, we set the default learning rates for unlearning as $10^{-4}$ and fine-tuning as $10^{-2}$. The unlearning is stopped when clean accuracy drops to or below 10\%. The default fine-tuning epoch is set to 20. The default neuron ratio $n$ and weight ratio $m$ are set to 0.15 and 0.7, respectively. We follow the suggested settings of hyper-parameters $r=0.05$ and $\alpha=0.7$ for fine-tuning~\cite{zhao2022penalizing}. Other settings are set following the default BackdoorBench configuration.

All experiments are conducted on a server with GPU RTX 3090 and CPU AMD EPYC 7543 32-Core Processor. These experiments were successfully executed using less than 24G of memory on a single GPU card.

\begin{algorithm}[t]
    \caption{ Two-Stage Backdoor Defense}
    \label{alg:overall}
    \raggedright
    {\bf Input}: Small clean set $\mathcal{D}_{c}$, backdoored model with parameter $\boldsymbol{\theta}_{bd}$, max iteration number $T$, neuron ratio $n\%$ and weight ratio $m\%$ for reinitialization\\
    {\bf Output}: Clean model with parameter $\boldsymbol{\theta}^*$\\
    \begin{algorithmic} [1]
    \STATE 
    /* \textbf{Neuron Weight Change-based Backdoor Reinitialization} */
    \\// \textit{a. Clean Unlearning}
    \WHILE{Clean accuracy on $ \boldsymbol{\theta}_{ul} > 10\%$} 
        \STATE Sample a mini-batch $\mathcal{B}_{c}$ from $\mathcal{D}_{c}$
        \STATE $\boldsymbol{\theta}_{ul} \leftarrow \max_{\boldsymbol{\theta}_{bd}}\mathcal{L}\left(f\left(\mathcal{B}_{c};\boldsymbol{\theta}_{bd}\right)\right)$
    \ENDWHILE
    \\// \textit{b. Neuron Weight Change Calculation.}
    \STATE Record subweight changes and calculate \nameMetric by Equation~(\ref{equ:ucn}) \wrt $\boldsymbol{\theta}_{ul}$ and $\boldsymbol{\theta}_{bd}$
    \\// \textit{c. Zero Reinitialization.}
    \STATE Obtain and sort the weight changes from the top-$n\%$ neurons \wrt \nameMetric
    \STATE $\hat{\boldsymbol{\theta}} \leftarrow$ reinitialize $m\%$ of the most-changing weights into zero on $\boldsymbol{\theta}_{bd}$
    \\\STATE /* \textbf{Activeness-aware Fine-tuning} */
    \FOR{$t=1\ \TO\ T$}
        \STATE Sample a mini-batch $\mathcal{B}_{c}$ from $\mathcal{D}_{c}$
        \STATE Update $\hat{\boldsymbol{\theta}}$ by Equation~(\ref{equ:ft}) with the approximated Equation~(\ref{equ:loss_appro})
    \ENDFOR
    \STATE $\boldsymbol{\theta}^* \leftarrow$ fine-tuned $\hat{\boldsymbol{\theta}}$
    \end{algorithmic}
\end{algorithm}

\begin{algorithm}[t]
    \caption{The approximated Optimization of Activeness-aware Fine-tuning}
    \label{alg:approx_finetune}
    \raggedright
    {\bf Input}: Small clean set $\mathcal{D}_{c}$, the reinitialized model with parameter $\hat{\boldsymbol{\theta}}$, max iteration number $T$, approximation scalar $r$, balance coefficient $\alpha$\\
    {\bf Output}: Clean model with parameter $\boldsymbol{\theta}^*$\\
    \begin{algorithmic} [1]
    \STATE 
    /* \textbf{Activeness-aware Fine-tuning} */
    \FOR{$t=1\ \TO\ T$}
        \STATE Sample a mini-batch $\mathcal{B}_{c}$ from $\mathcal{D}_{c}$
        \STATE Calculate the gradient $g_1=\nabla_{\hat{\boldsymbol{\theta}}} \mathcal{L}_{ce}(\hat{\boldsymbol{\theta}})$ with $\mathcal{B}_{c}$
        \STATE Temporally update the current parameter $\hat{\boldsymbol{\theta}}' = \hat{\boldsymbol{\theta}} + r \frac{\nabla_{\hat{\boldsymbol{\theta}}} \mathcal{L}_{ce}(\hat{\boldsymbol{\theta}})}{ \|\nabla_{\hat{\boldsymbol{\theta}}} \mathcal{L}_{ce}(\hat{\boldsymbol{\theta}}) \|_2}$
        \STATE Calculate the gradient $g_2=\nabla_{\hat{\boldsymbol{\theta}}'} \mathcal{L}_{ce}(\hat{\boldsymbol{\theta}}')$ with $\mathcal{B}_{c}$
        \STATE Calculate the final gradient $g=(1-\alpha)g_1 + \alpha g_2$
        \STATE Update the parameter $\hat{\boldsymbol{\theta}}$ based on $g$
    \ENDFOR
    \STATE $\boldsymbol{\theta}^* \leftarrow$ fine-tuned $\hat{\boldsymbol{\theta}}$
    \end{algorithmic}
\end{algorithm}

\begin{table*}[]
\caption{Comparison with the SOTA defenses on GTSRB dataset with PreAct-ResNet18 (\%).}
\label{tab:gtsrb_preact_10}
\centering
\resizebox{\linewidth}{!}{
\begin{tabular}{c|ccc|ccc|ccc|ccc|ccc}
\hline
\multirow{2}{*}{\begin{tabular}[c]{@{}c@{}}Backdoor\\ Attacks\end{tabular}} & \multicolumn{3}{c|}{No Defense}                    & \multicolumn{3}{c|}{FT}                            & \multicolumn{3}{c|}{FP~\cite{liu2018fine}}                            & \multicolumn{3}{c|}{NAD~\cite{li2021neural}}                           & \multicolumn{3}{c}{NC~\cite{wang2019neural}}                                            \\ \cline{2-16} 
                                                                            & ACC $\uparrow$ & ASR $\downarrow$ & DER $\uparrow$ & ACC $\uparrow$ & ASR $\downarrow$ & DER $\uparrow$ & ACC $\uparrow$ & ASR $\downarrow$ & DER $\uparrow$ & ACC $\uparrow$ & ASR $\downarrow$ & DER $\uparrow$ & ACC $\uparrow$      & ASR $\downarrow$      & DER $\uparrow$      \\ \hline
BadNets~\cite{gu2019badnets}                                                                     & 97.24          & 59.25            & -              & \textbf{98.73} & 5.09             & 77.08          & 98.21          & 0.09             & 79.58          & {\ul 98.69}    & 0.63             & 79.31          & 97.48               & {\ul 0.01}            & {\ul 79.62}         \\
Blended~\cite{chen2017targeted}                                                                     & 98.58          & 99.99            & -              & 98.57          & 100.00           & 50.00          & 98.38          & 100.00           & 49.90          & 98.61          & 100.00           & 50.00          & 97.76               & {\ul 8.03}            & \textbf{95.57}      \\
Input-aware~\cite{nguyen2020input}                                                                 & 97.26          & 92.74            & -              & 98.40          & 29.81            & 81.46          & 98.08          & 2.32             & 95.21          & 98.27          & 40.65            & 76.04          & 98.55               & {\ul 0.01}            & {\ul 96.36}         \\
LF~\cite{zeng2021rethinking}                                                                          & 97.93          & 99.57            & -              & {\ul 98.01}    & 79.98            & 59.80          & 97.59          & 99.70            & 49.83          & \textbf{98.14} & 51.83            & 73.87          & 97.97               & {\ul 1.34}            & {\ul 99.11}         \\
SSBA~\cite{li2021invisible}                                                                        & 97.98          & 99.56            & -              & 97.92          & 99.10            & 50.20          & 97.75          & 99.46            & 49.94          & {\ul 97.95}    & 99.39            & 50.07          & 97.72               & {\ul 0.29}            & \textbf{99.50}      \\
Trojan~\cite{liu2018trojaning}                                                                      & 98.57          & 100.00           & -              & {\ul 98.54}    & 0.02             & \textbf{99.97} & 98.31          & 71.30            & 64.22          & 98.20          & 0.11             & {\ul 99.76}    & 87.53               & 0.83                  & 94.07               \\
WaNet~\cite{nguyen2021wanet}                                                                       & 97.74          & 94.25            & -              & 98.54          & 0.29             & 96.98          & 97.62          & 88.07            & 53.03          & {\ul 98.61}    & 0.56             & 96.84          & 98.25               & \textbf{0.00}         & \textbf{97.12}      \\ \hline
Average                                                                     & 97.90          & 92.19            & -              & \textbf{98.39} & 44.90            & 73.64          & 97.99          & 65.85            & 63.10          & {\ul 98.35}    & 41.88            & 75.13          & 96.47               & {\ul 1.50}            & {\ul 94.48}         \\ \hline
\multirow{2}{*}{\begin{tabular}[c]{@{}c@{}}Backdoor\\ Attacks\end{tabular}} & \multicolumn{3}{c|}{ANP~\cite{wu2021adversarial}}                           & \multicolumn{3}{c|}{CLP~\cite{zheng2022data}}                           & \multicolumn{3}{c|}{i-BAU~\cite{zeng2021adversarial}}                         & \multicolumn{3}{c|}{RNP~\cite{li2023reconstructive}}                           & \multicolumn{3}{c}{\nameFramework \textbf{\textcolor{red}{(Ours)}}} \\ \cline{2-16} 
                                                                            & ACC $\uparrow$ & ASR $\downarrow$ & DER $\uparrow$ & ACC $\uparrow$ & ASR $\downarrow$ & DER $\uparrow$ & ACC $\uparrow$ & ASR $\downarrow$ & DER $\uparrow$ & ACC $\uparrow$ & ASR $\downarrow$ & DER $\uparrow$ & ACC $\uparrow$      & ASR $\downarrow$      & DER $\uparrow$      \\ \hline
BadNets~\cite{gu2019badnets}                                                                     & 96.89          & 0.06             & 79.42          & 97.67          & 66.85            & 50.00          & 96.47          & 0.02             & 79.23          & 97.69          & 0.67             & 79.29          & 98.38               & \textbf{0.00}         & \textbf{79.63}      \\
Blended~\cite{chen2017targeted}                                                                     & {\ul 98.75}    & 99.82            & 50.09          & 98.48          & 100.00           & 49.95          & 92.35          & 86.35            & 53.71          & \textbf{98.84} & 99.64            & 50.18          & 95.57               & \textbf{5.94}         & {\ul 95.52}         \\
Input-aware~\cite{nguyen2020input}                                                                 & {\ul 99.14}    & \textbf{0.00}    & \textbf{96.37} & 98.64          & 96.01            & 50.00          & 97.09          & 0.52             & 96.03          & 97.17          & \textbf{0.00}    & 96.32          & \textbf{99.23}      & 0.29                  & 96.22               \\
LF~\cite{zeng2021rethinking}                                                                          & 97.80          & 81.38            & 59.03          & 97.70          & 99.50            & 49.92          & 95.78          & 16.15            & 90.64          & 97.91          & 99.06            & 50.25          & 96.83               & \textbf{0.03}         & \textbf{99.22}      \\
SSBA~\cite{li2021invisible}                                                                        & 97.86          & 98.93            & 50.26          & \textbf{98.08} & 99.24            & 50.16          & 96.14          & 1.88             & 97.92          & 97.78          & 99.43            & 49.96          & 96.06               & \textbf{0.10}         & {\ul 98.77}         \\
Trojan~\cite{liu2018trojaning}                                                                      & 98.08          & \textbf{0.00}    & 99.75          & 98.17          & 95.20            & 52.20          & 95.98          & {\ul 0.02}       & 98.70          & \textbf{98.56} & 100.00           & 50.00          & 97.16               & 0.15                  & 99.22               \\
WaNet~\cite{nguyen2021wanet}                                                                       & 97.08          & \textbf{0.00}    & 96.80          & 7.16           & 100.00           & 4.71           & 96.72          & \textbf{0.00}    & 96.62          & 96.77          & \textbf{0.00}    & 96.64          & \textbf{98.80}      & {\ul 0.03}            & {\ul 97.11}         \\ \hline
Average                                                                     & 97.94          & 40.03            & 75.96          & 85.13          & 93.83            & 43.85          & 95.79          & 14.99            & 87.55          & 97.81          & 56.97            & 67.52          & 97.43               & \textbf{0.93}         & \textbf{95.10}      \\ \hline
\end{tabular}}
\end{table*}

\begin{table*}[]
\caption{Comparison with the SOTA defenses on CIFAR-10 dataset with VGG19-BN(\%).}
\label{tab:cifar10_vgg_10}
\centering
\resizebox{\linewidth}{!}{
\begin{tabular}{c|ccc|ccc|ccc|ccc|ccc}
\hline
\multirow{2}{*}{\begin{tabular}[c]{@{}c@{}}Backdoor\\ Attacks\end{tabular}} & \multicolumn{3}{c|}{No Defense}                    & \multicolumn{3}{c|}{FT}                            & \multicolumn{3}{c|}{FP~\cite{liu2018fine}}                            & \multicolumn{3}{c|}{NAD~\cite{li2021neural}}                           & \multicolumn{3}{c}{NC~\cite{wang2019neural}}                                           \\ \cline{2-16} 
                                                                            & ACC $\uparrow$ & ASR $\downarrow$ & DER $\uparrow$ & ACC $\uparrow$ & ASR $\downarrow$ & DER $\uparrow$ & ACC $\uparrow$ & ASR $\downarrow$ & DER $\uparrow$ & ACC $\uparrow$ & ASR $\downarrow$ & DER $\uparrow$ & ACC $\uparrow$      & ASR $\downarrow$      & DER $\uparrow$     \\ \hline
BadNets~\cite{gu2019badnets}                                                                     & 90.42          & 94.43            & -              & 88.19          & 27.59            & 82.31          & 88.96          & 10.23            & 91.37          & 86.48          & 5.47             & 92.51          & {\ul 89.21}         & 11.31                 & 90.96              \\
Blended~\cite{chen2017targeted}                                                                     & 91.91          & 99.50            & -              & 90.08          & 86.82            & 55.42          & 89.95          & 87.46            & 55.04          & 88.60          & 83.86            & 56.17          & 90.07               & 83.33                 & 57.16              \\
Input-aware~\cite{nguyen2020input}                                                                 & 88.66          & 94.58            & -              & \textbf{91.56} & 13.08            & 90.75          & 91.35          & 6.08             & 94.25          & 91.00          & 14.11            & 90.23          & 89.70               & 97.02                 & 50.00              \\
LF~\cite{zeng2021rethinking}                                                                          & 83.28          & 13.83            & -              & 87.67          & 1.82             & 56.01          & {\ul 88.31}    & 1.23             & 56.30          & 83.72          & 1.14             & {\ul 56.34}    & 86.64               & 1.36                  & 56.24              \\
SIG~\cite{barni2019new}                                                                         & 83.48          & 98.87            & -              & 88.01          & 4.28             & 97.29          & \textbf{88.34} & 15.26            & 91.81          & 86.07          & 7.39             & 95.74          & 83.48               & 98.87                 & 50.00              \\
SSBA~\cite{li2021invisible}                                                                        & 90.85          & 95.11            & -              & 89.26          & 70.22            & 61.65          & 89.28          & 65.80            & 63.87          & 88.33          & 56.64            & 67.97          & \textbf{90.85}      & 95.11                 & 50.00              \\
Trojan~\cite{liu2018trojaning}                                                                      & 91.57          & 100.00           & -              & 89.60          & 7.17             & 95.43          & 89.74          & 50.90            & 73.64          & 87.29          & {\ul 2.30}       & {\ul 96.71}    & 89.28               & 7.76                  & 94.98              \\
WaNet~\cite{nguyen2021wanet}                                                                       & 84.58          & 96.49            & -              & \textbf{91.35} & 5.72             & 95.39          & 91.12          & 4.74             & 95.88          & 90.73          & 10.33            & 93.08          & 91.20               & 6.88                  & 94.81              \\ \hline
Average                                                                     & 88.09          & 86.60            & -              & 89.47          & 27.09            & 79.28          & 89.63          & 30.21            & 77.77          & 87.78          & 22.65            & 81.10          & 88.80               & 50.20                 & 68.02              \\ \hline
\multirow{2}{*}{\begin{tabular}[c]{@{}c@{}}Backdoor\\ Attacks\end{tabular}} & \multicolumn{3}{c|}{ANP~\cite{wu2021adversarial}}                           & \multicolumn{3}{c|}{CLP~\cite{zheng2022data}}                           & \multicolumn{3}{c|}{i-BAU~\cite{zeng2021adversarial}}                         & \multicolumn{3}{c|}{RNP~\cite{li2023reconstructive}}                           & \multicolumn{3}{c}{\nameFramework \textbf{\textcolor{red}{(Ours)}}} \\ \cline{2-16} 
                                                                            & ACC $\uparrow$ & ASR $\downarrow$ & DER $\uparrow$ & ACC $\uparrow$ & ASR $\downarrow$ & DER $\uparrow$ & ACC $\uparrow$ & ASR $\downarrow$ & DER $\uparrow$ & ACC $\uparrow$ & ASR $\downarrow$ & DER $\uparrow$ & ACC $\uparrow$      & ASR $\downarrow$      & DER $\uparrow$     \\ \hline
BadNets~\cite{gu2019badnets}                                                                     & 88.39          & {\ul 0.48}       & \textbf{95.96} & \textbf{89.38} & 6.61             & 93.39          & 86.01          & 2.28             & 93.87          & 79.90          & \textbf{0.12}    & 91.90          & 88.76               & 3.58                  & {\ul 94.60}        \\
Blended~\cite{chen2017targeted}                                                                     & 89.19          & \textbf{4.77}    & \textbf{96.01} & \textbf{90.66} & 98.59            & 49.83          & 87.58          & 69.90            & 62.64          & 26.92          & 43.20            & 45.66          & {\ul 90.51}         & {\ul 6.68}            & {\ul 95.71}        \\
Input-aware~\cite{nguyen2020input}                                                                 & 86.56          & \textbf{0.71}    & \textbf{95.88} & 87.54          & {\ul 2.66}       & {\ul 95.40}    & 88.29          & 69.56            & 62.33          & 60.76          & 87.33            & 39.67          & {\ul 91.52}         & 13.29                 & 90.64              \\
LF~\cite{zeng2021rethinking}                                                                          & 84.33          & {\ul 0.08}       & \textbf{56.88} & 81.99          & 14.61            & 49.36          & 87.68          & 1.47             & 56.18          & 79.89          & \textbf{0.00}    & 55.22          & \textbf{88.71}      & 1.77                  & 56.03              \\
SIG~\cite{barni2019new}                                                                         & 82.69          & \textbf{0.00}    & \textbf{99.04} & 82.12          & 98.68            & 49.41          & 83.41          & 5.37             & 96.72          & 35.36          & {\ul 0.01}       & 75.37          & {\ul 88.14}         & 2.58                  & {\ul 98.14}        \\
SSBA~\cite{li2021invisible}                                                                        & {\ul 89.81}    & {\ul 1.34}       & \textbf{96.36} & 85.82          & 98.56            & 47.49          & 87.56          & 22.26            & 84.78          & 69.02          & \textbf{0.03}    & 86.62          & 88.97               & 3.41                  & {\ul 94.91}        \\
Trojan~\cite{liu2018trojaning}                                                                      & 89.39          & \textbf{0.00}    & \textbf{98.91} & \textbf{90.56} & 99.70            & 49.65          & 88.63          & 8.23             & 94.41          & 53.40          & 57.30            & 52.27          & {\ul 90.22}         & 6.63                  & 96.01              \\
WaNet~\cite{nguyen2021wanet}                                                                       & 78.04          & \textbf{0.03}    & 94.96          & 88.46          & 1.75             & {\ul 97.37}    & 89.76          & {\ul 1.61}       & \textbf{97.44} & 87.90          & 81.03            & 57.73          & {\ul 91.28}         & 2.50                  & 97.00              \\ \hline
Average                                                                     & 86.05          & \textbf{0.93}    & \textbf{91.75} & 87.07          & 52.64            & 66.49          & 87.37          & 22.58            & 81.05          & 61.64          & 33.63            & 63.05          & \textbf{89.76}      & {\ul 5.06}            & {\ul 90.38}        \\ \hline
\end{tabular}}
\end{table*}

\section{Evaluations on GTSRB dataset}
\label{sec:perform_gtsrb}
We validate the effectiveness of our proposed method in the dataset GTSRB other than CIFAR-10 and Tiny ImageNet. Table~\ref{tab:gtsrb_preact_10} shows the corresponding performance on PreAct-ResNet18 with 10\% poisoning ratio and 5\% clean data ratio. We can observe that \nameFramework performs consistently with the lowest average ASR and the largest average DER as in CIFAR-10. Most of the defenses also fail in the strong attacks Blended, LF, and Trojan, while NC performs the second best with comparable average ASR and DER. Compared to the other two datasets, \nameFramework performs much superior with most ASRs lower than 0.5\% except for Blended attack. This suggests that \nameFramework might perform better when the model is learned with the input images with similar characteristics.

\section{Performance on VGG19-BN model structure}
\label{sec:perform_vgg}
Except for PreAct-ResNet18, we also test another model, VGG19-BN. The performance on CIFAR-10 with 10\% poisoning ratio and 5\% clean data ratio is illustrated in Table~\ref{tab:cifar10_vgg_10}. We follow similar settings in PreAct-ResNet18 on CIFAR-10, conducting 8 attacks and comparing our method with 8 defenses. The performance is also evaluated by ACC, ASR, and DER. 
As shown in the table, \nameFramework owns SOTA performance on VGG19-BN with the best average ACC and second-best average ASR and DER. Most of the other performances are in a similar pattern as on PreAct-ResNet18. Although ANP performs almost the best in ASR and DER for all attacks, it damages the corresponding ACC as well, especially for the WaNet attacked model, where ACC decreases from 84.58\% to 78.04\%.
On the contrary, our methods succeed in most attacks with high ACC and comparable ASR.

\begin{figure}[]
    \centering
        \includegraphics[width=0.32\linewidth]{ 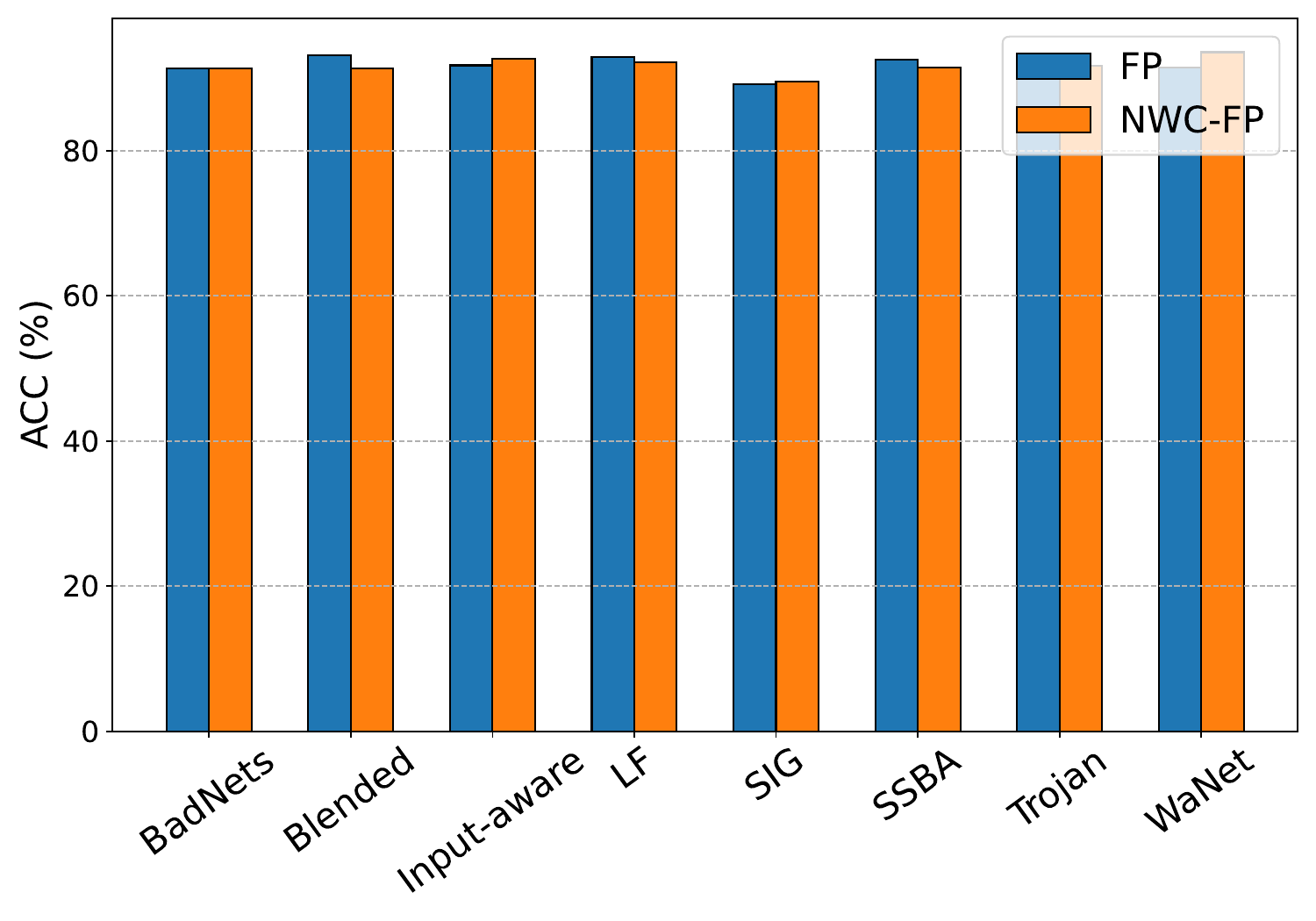}
        \includegraphics[width=0.32\linewidth]{ 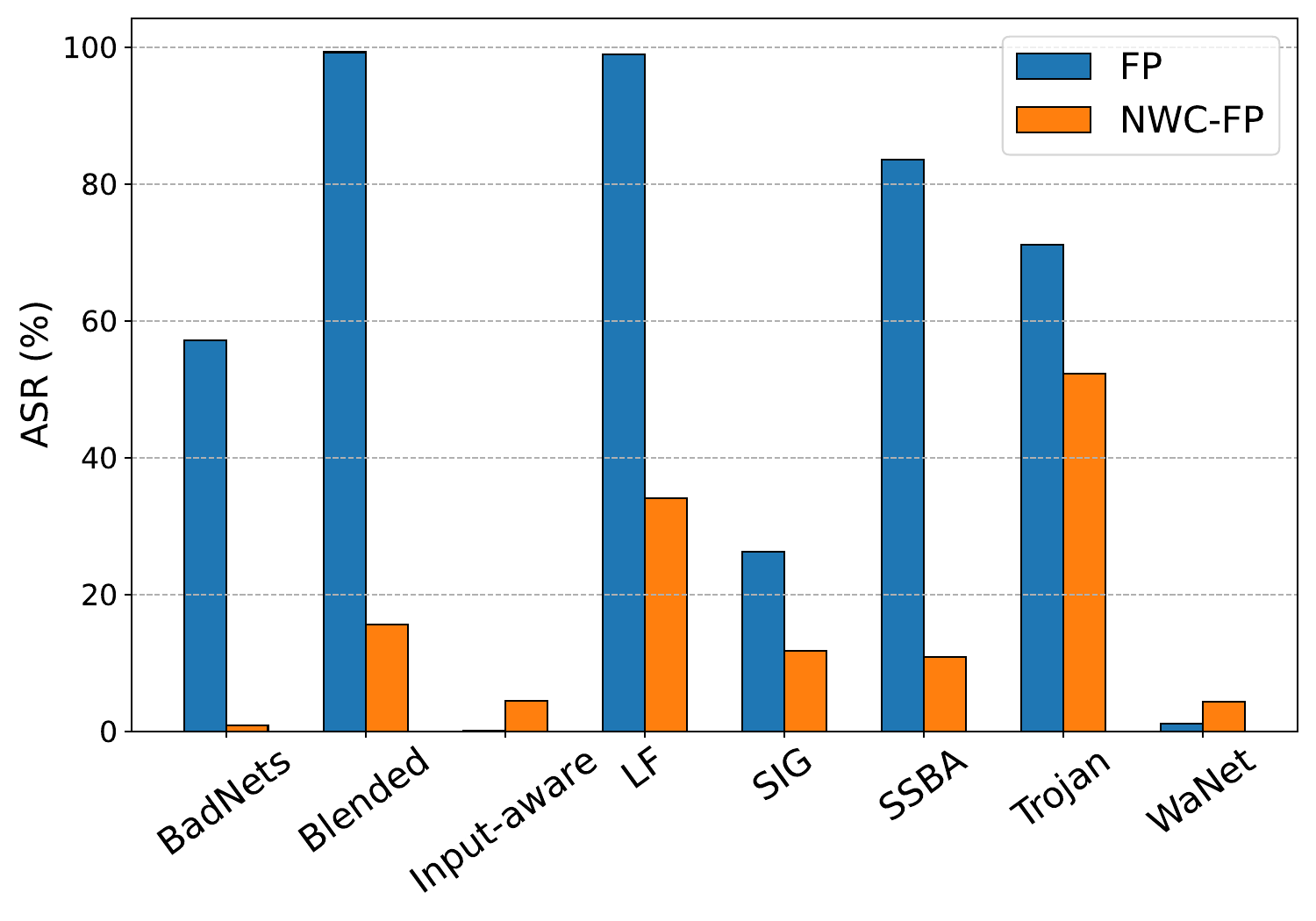}
        \includegraphics[width=0.32\linewidth]{ 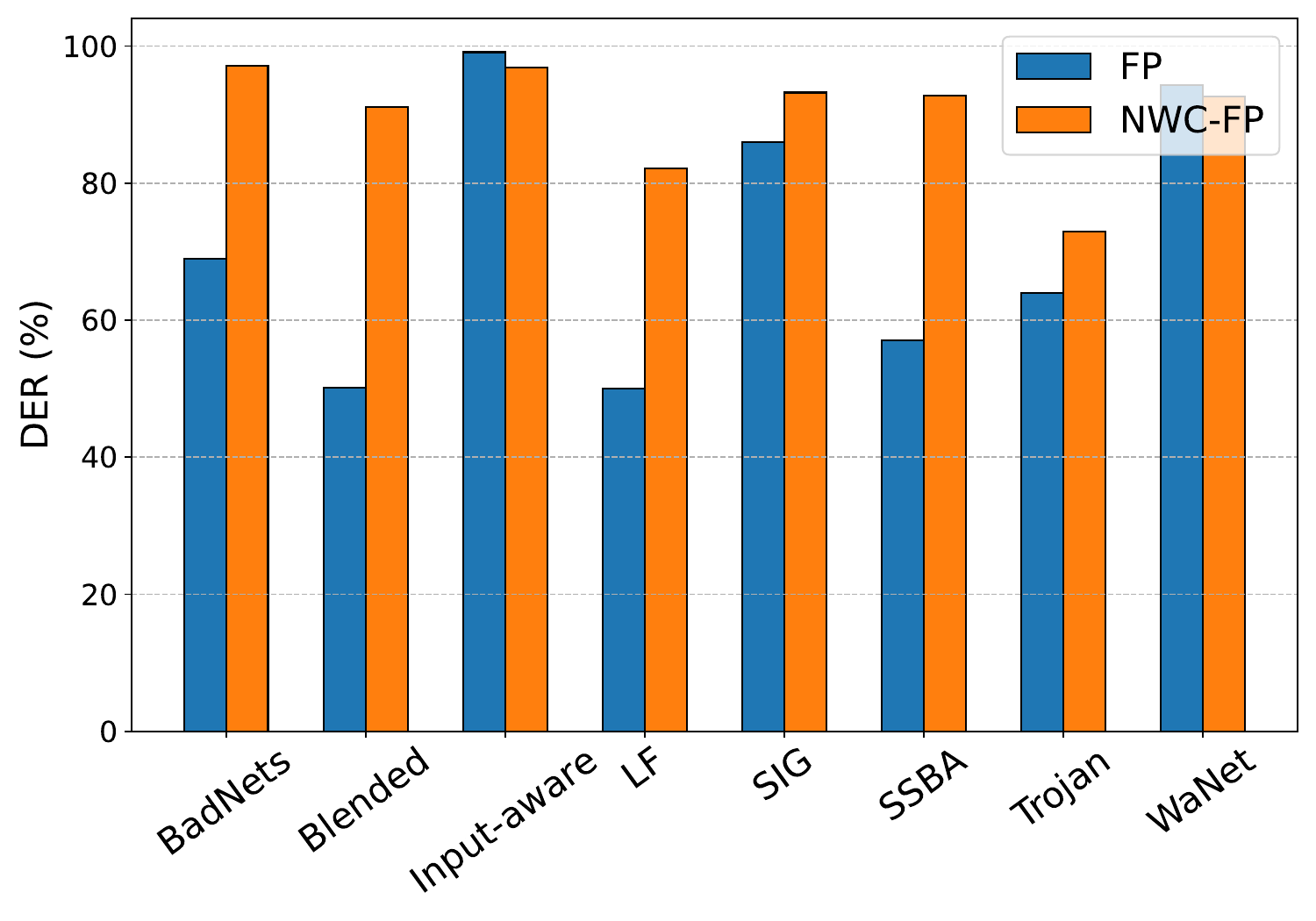}
    \vspace{-2mm}
    \caption{Performance comparisons between the original FP and the FP with \nameMetric (\nameMetric-FP) under 8 attacks. Left: ACC; Middle: ASR; Right: DER.}
    \label{fig:app_nwc_fp}
\end{figure}
\section{Effectiveness of using \nameMetric in Other Defense}
\label{sec:perform_other_defense}
As illustrated in Section~\ref{subsec:ablation_study}, it is effective to utilize \nameMetric order in gauging the backdoor strength. To further test its ability to improve other defenses, we substitute the average neuron activation in FP to \nameMetric (denoted as \nameMetric-FP) and conduct the experiments on the default settings in PreAct-ResNet18 on CIFAR-10. Different from zero reinitialization, the pruned neurons will not be updated on the following fine-tuning. The performances of 8 attacks are shown in Figure~\ref{fig:app_nwc_fp}. We can investigate that, by using \nameMetric in FP, performances on ASR and DER are improved on most of the attacks, with few effects on ACC. It exhibits the potential of using our \nameMetric in more defense methods.

\begin{figure}[]
    \centering
        \includegraphics[width=0.24\linewidth]{ figure/fig_ratio/Fig_neuron_ratio_BadNets.pdf}
        \includegraphics[width=0.24\linewidth]{ figure/fig_ratio/Fig_neuron_ratio_Blended.pdf}
        \includegraphics[width=0.24\linewidth]{ 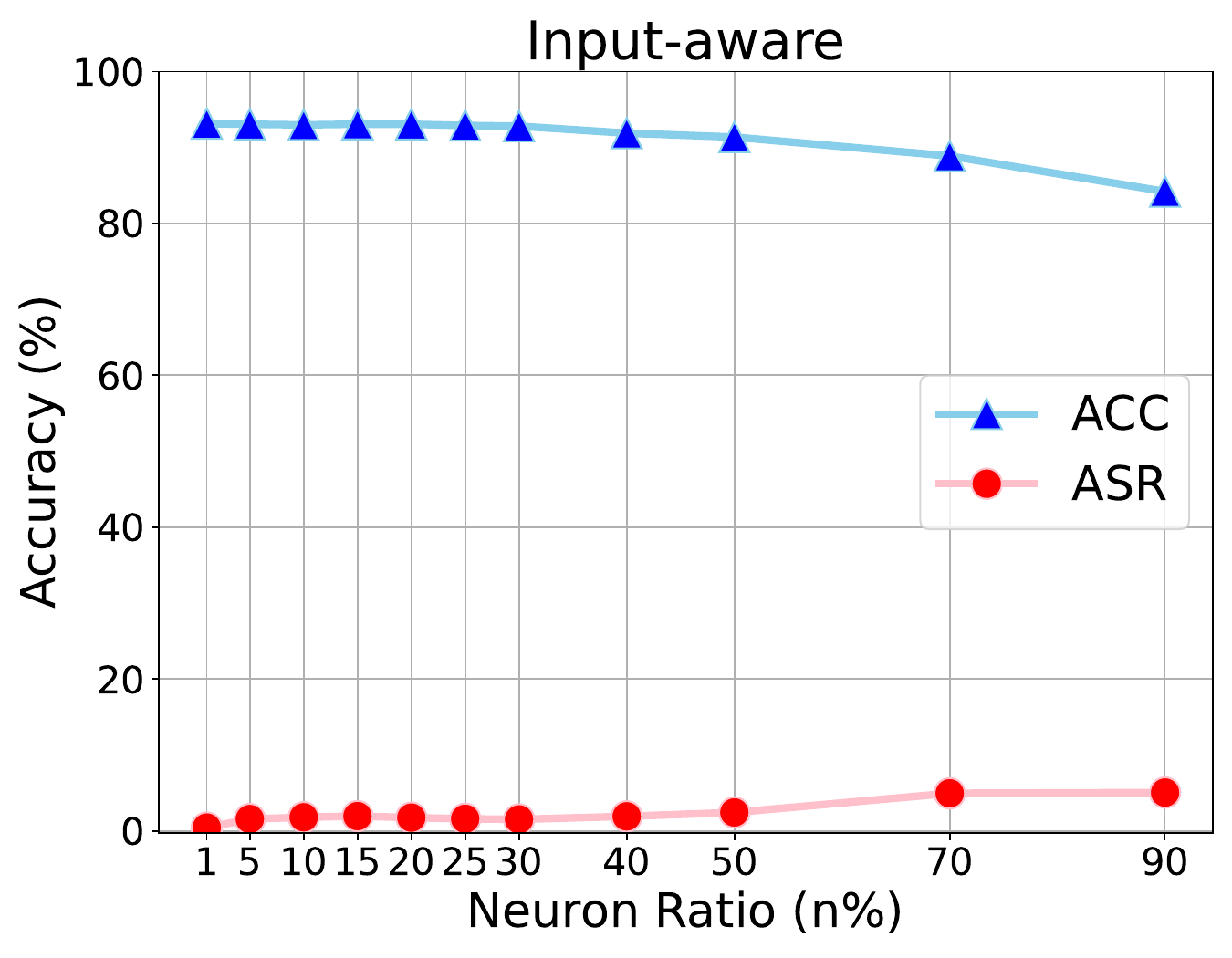}
        \includegraphics[width=0.24\linewidth]{ 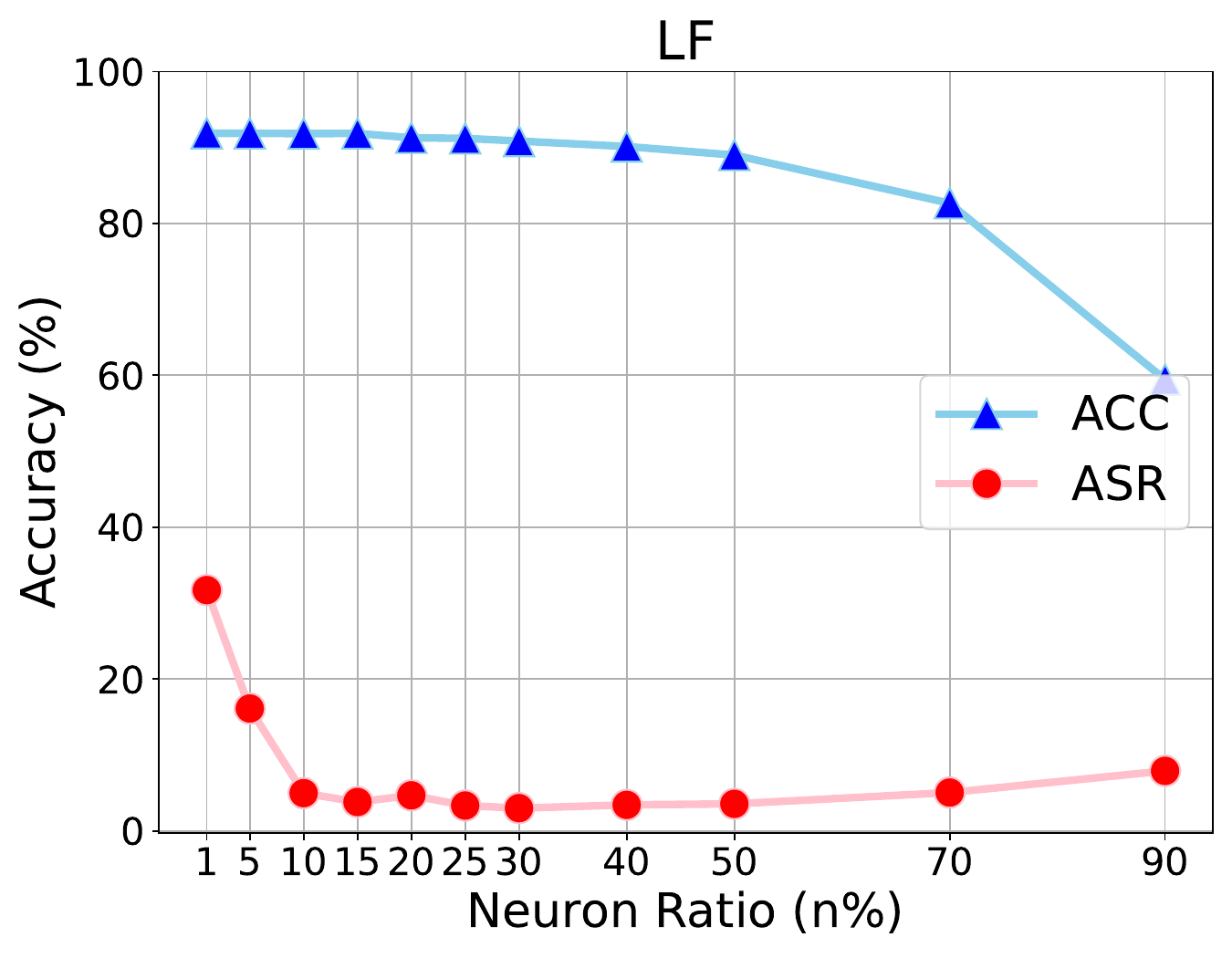}
        \includegraphics[width=0.24\linewidth]{ 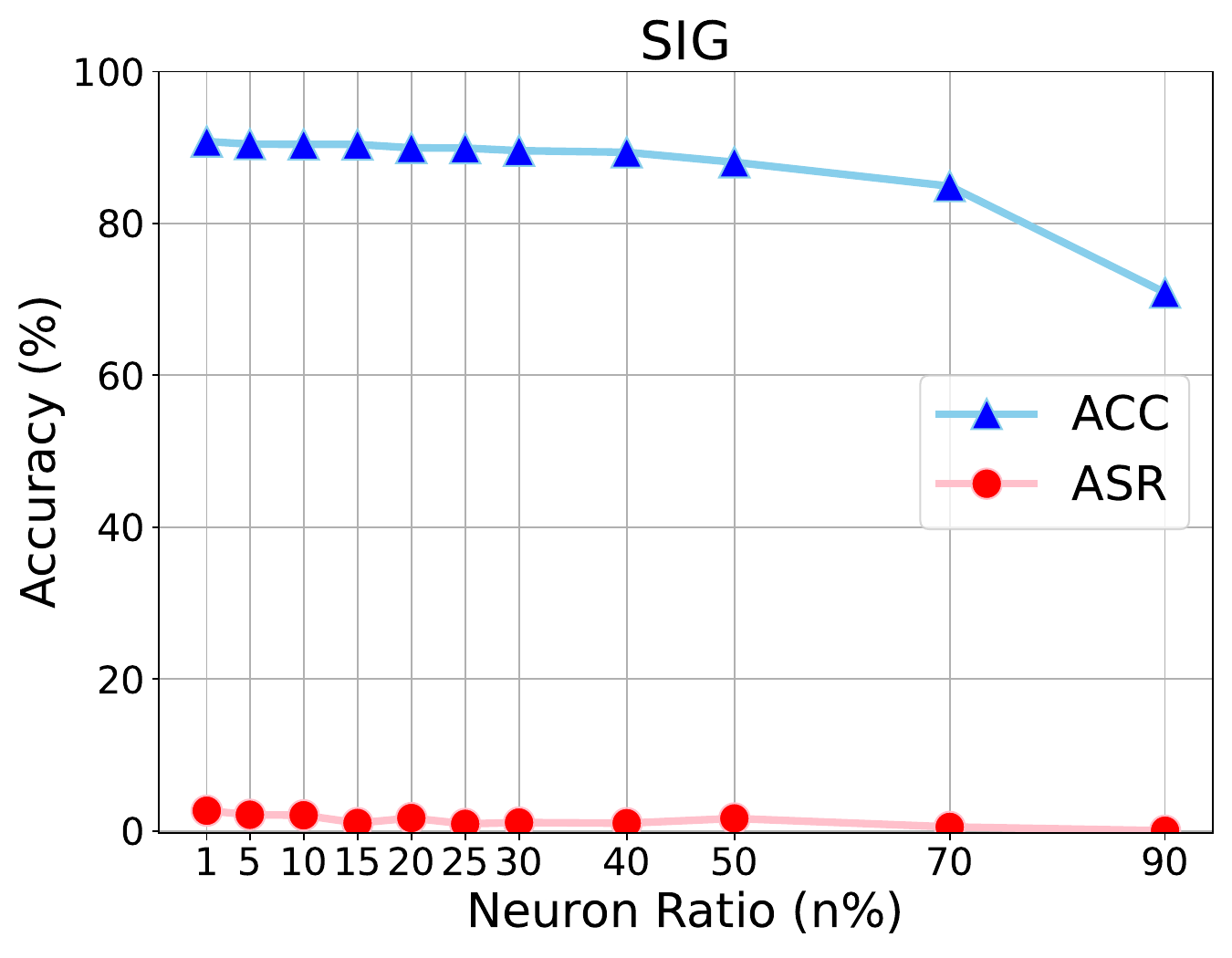}
        \includegraphics[width=0.24\linewidth]{ 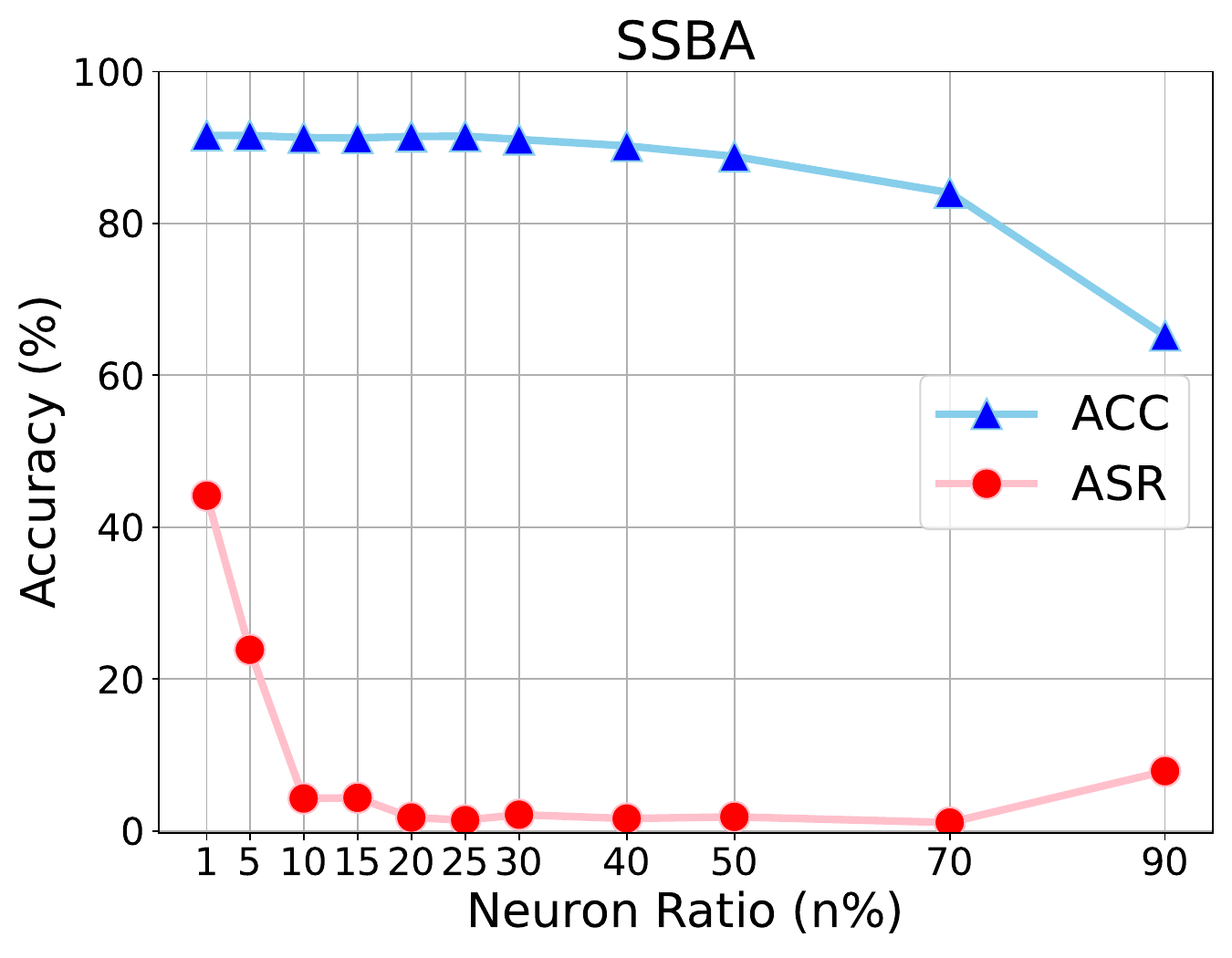}
        \includegraphics[width=0.24\linewidth]{ 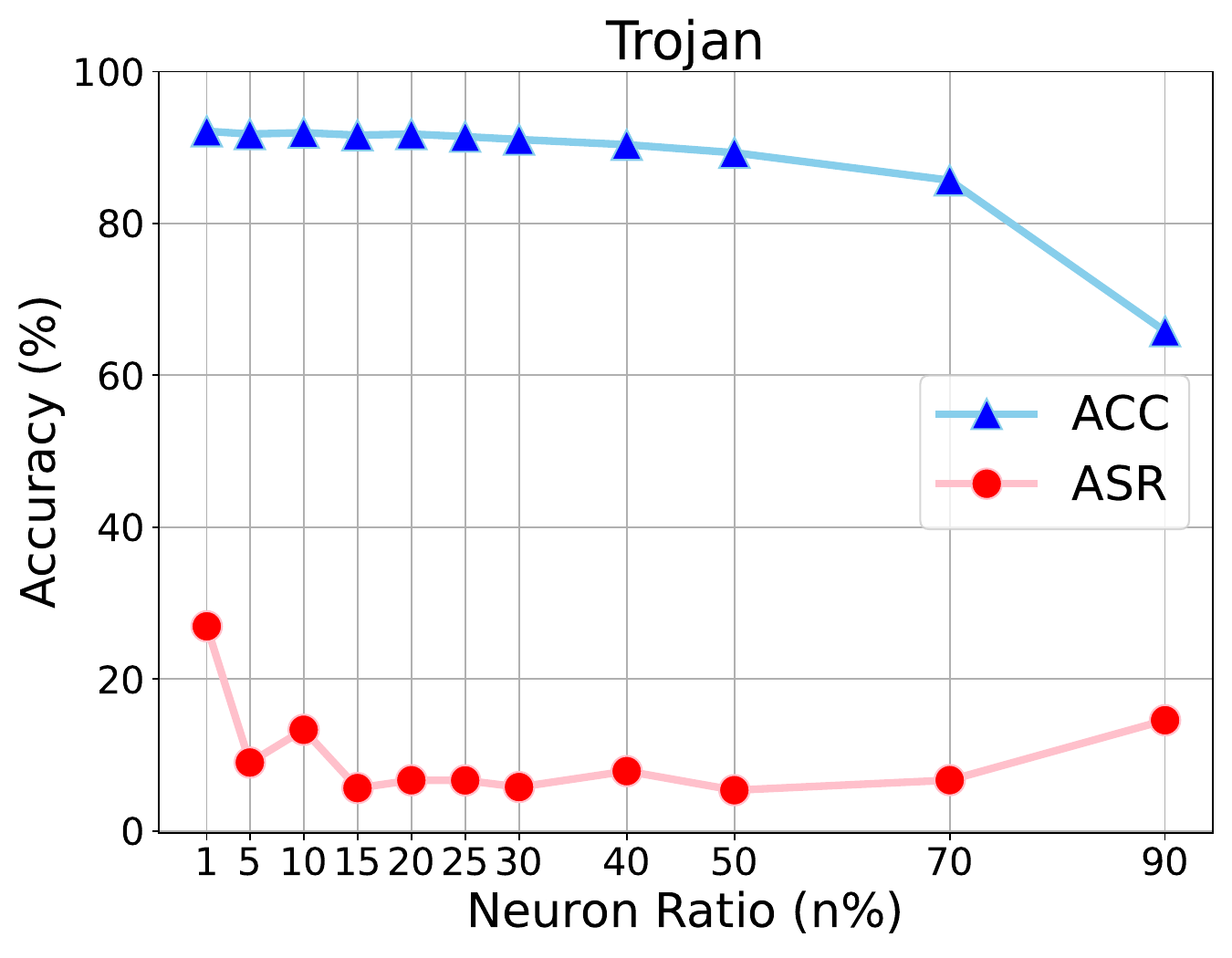}
        \includegraphics[width=0.24\linewidth]{ 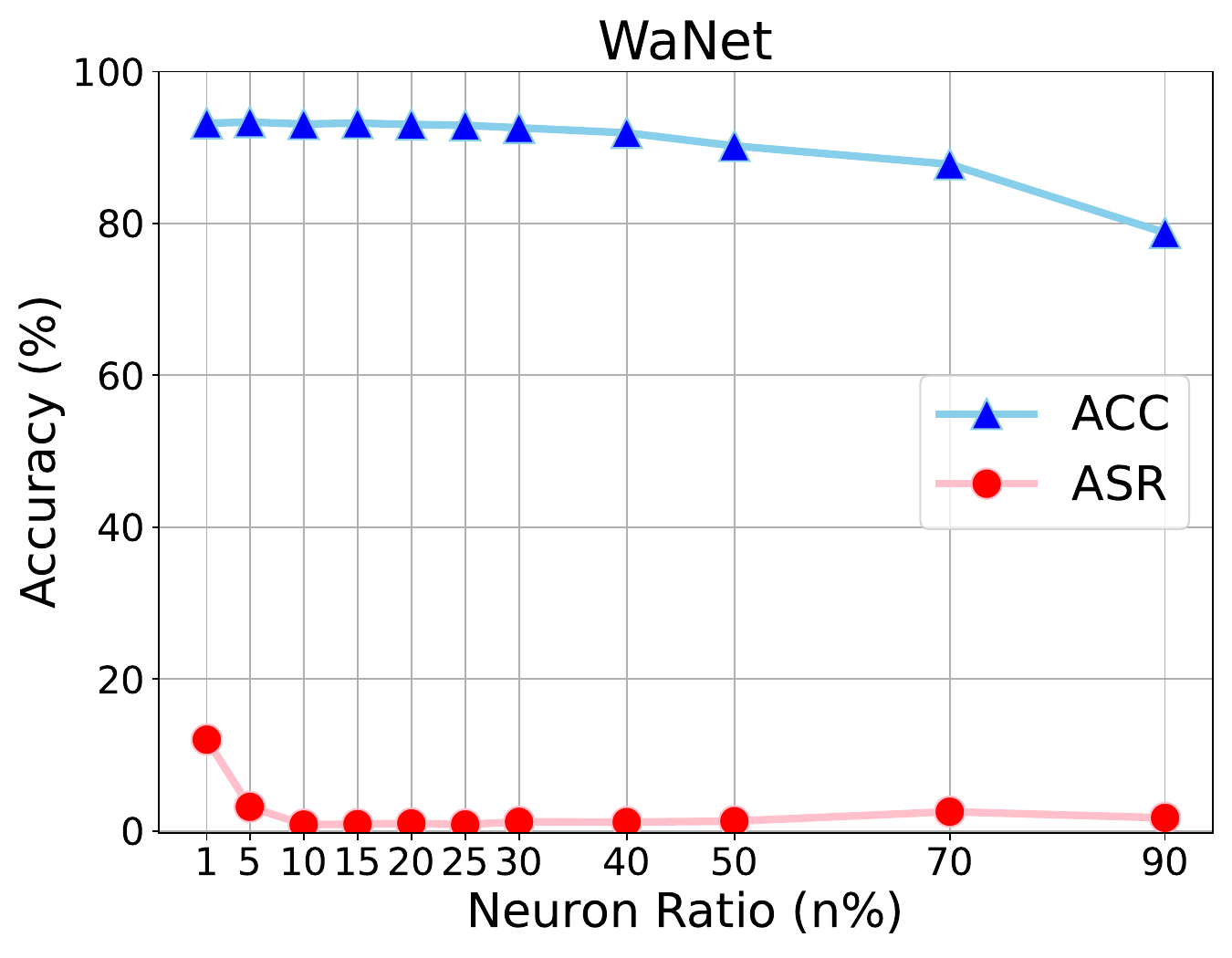}
    \caption{Performance with different neuron ratios under 8 attacks.}
    \label{fig:app_neuron_ratio}
\end{figure}
\begin{figure}[]
    \centering
        \includegraphics[width=0.24\linewidth]{ figure/fig_ratio/Fig_weight_ratio_BadNets.pdf}
        \includegraphics[width=0.24\linewidth]{ figure/fig_ratio/Fig_weight_ratio_Blended.pdf}
        \includegraphics[width=0.24\linewidth]{ 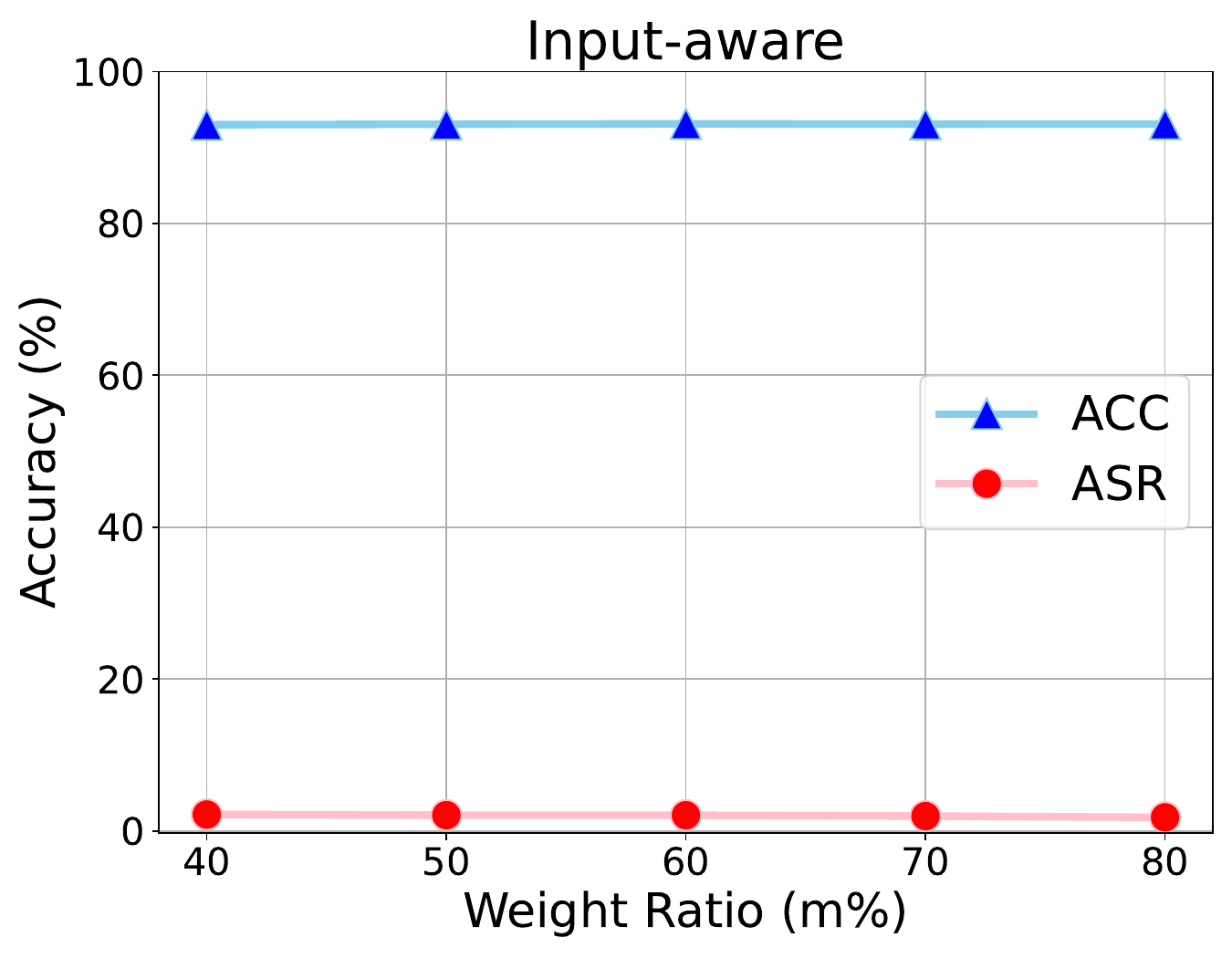}
        \includegraphics[width=0.24\linewidth]{ 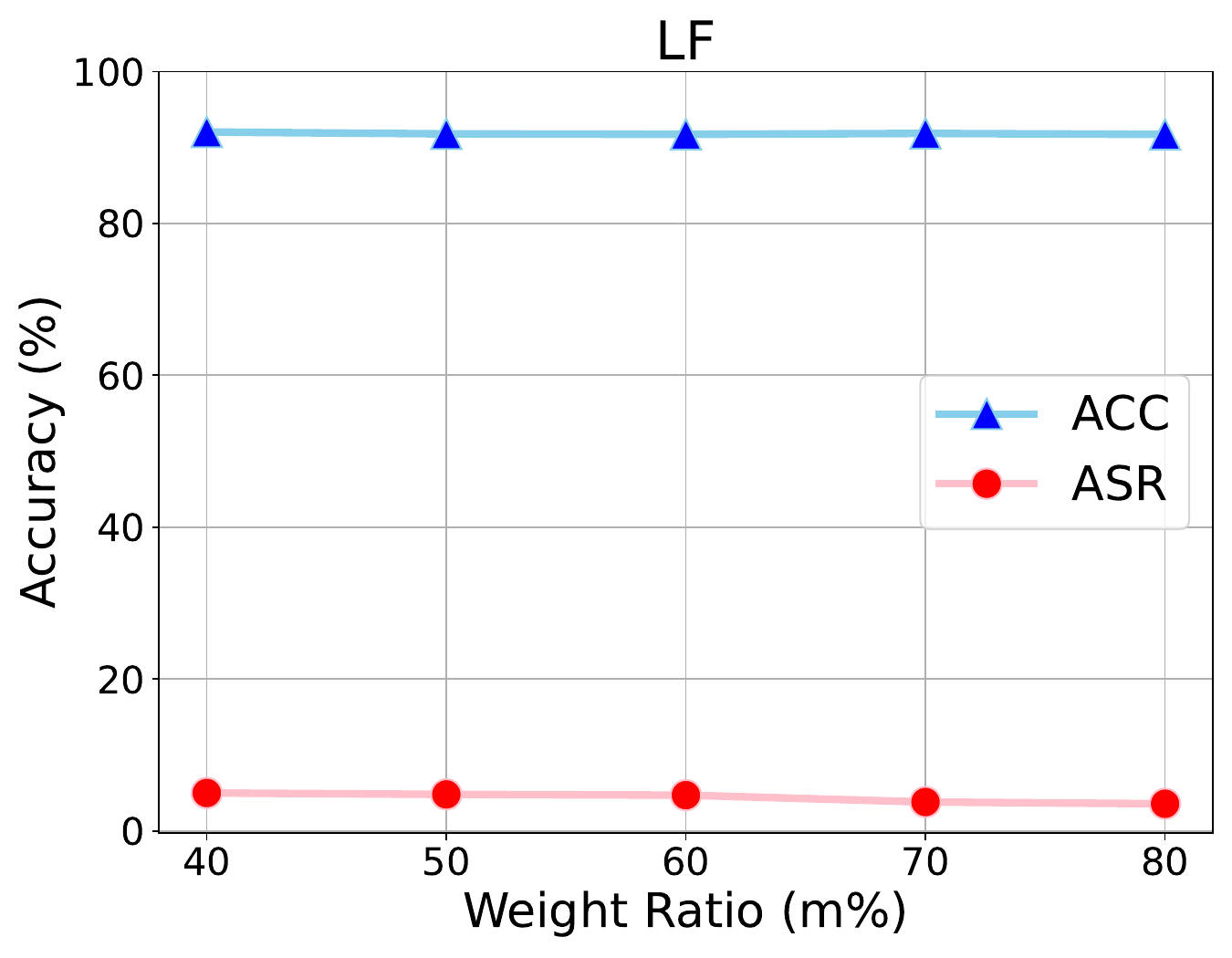}
        \includegraphics[width=0.24\linewidth]{ 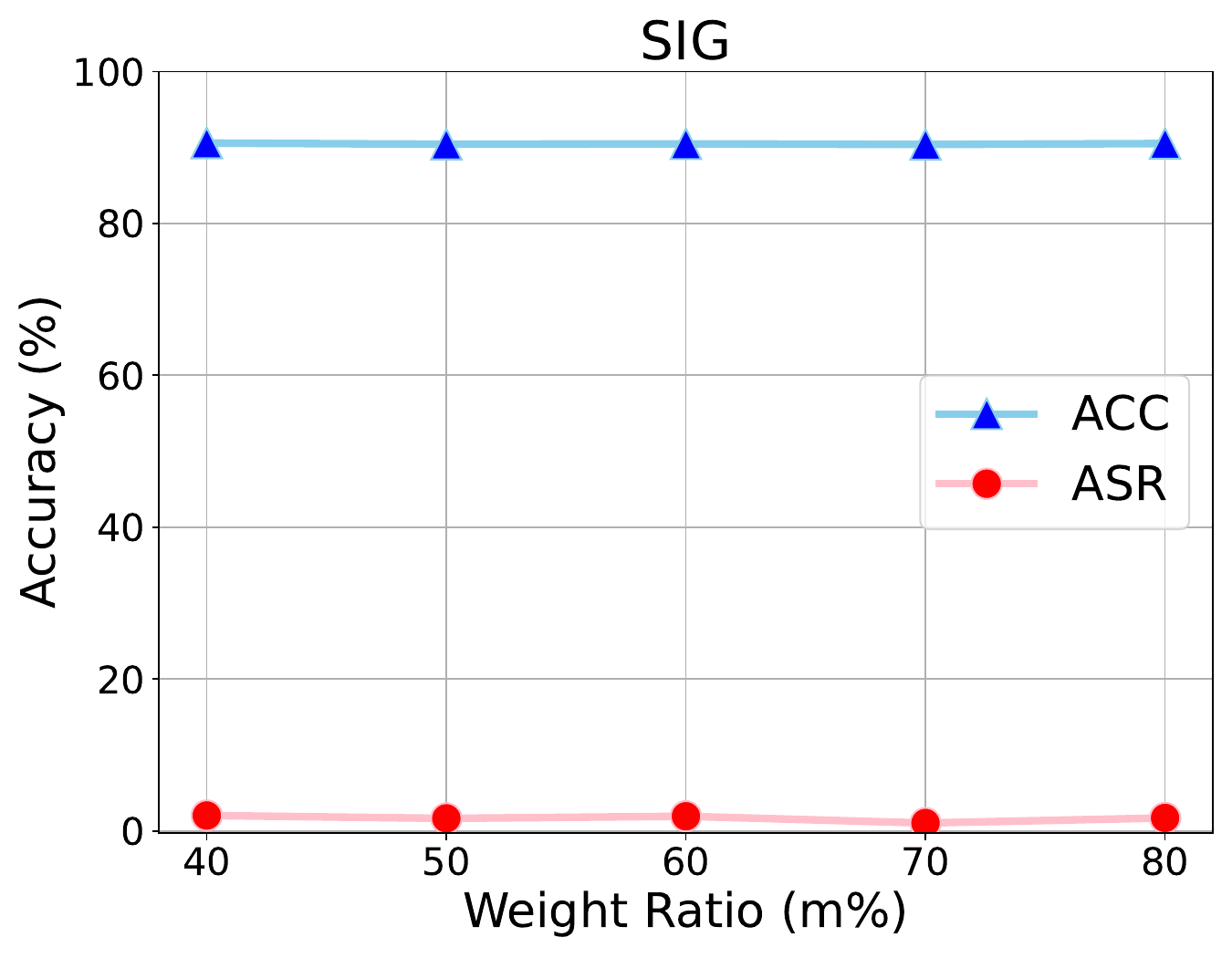}
        \includegraphics[width=0.24\linewidth]{ 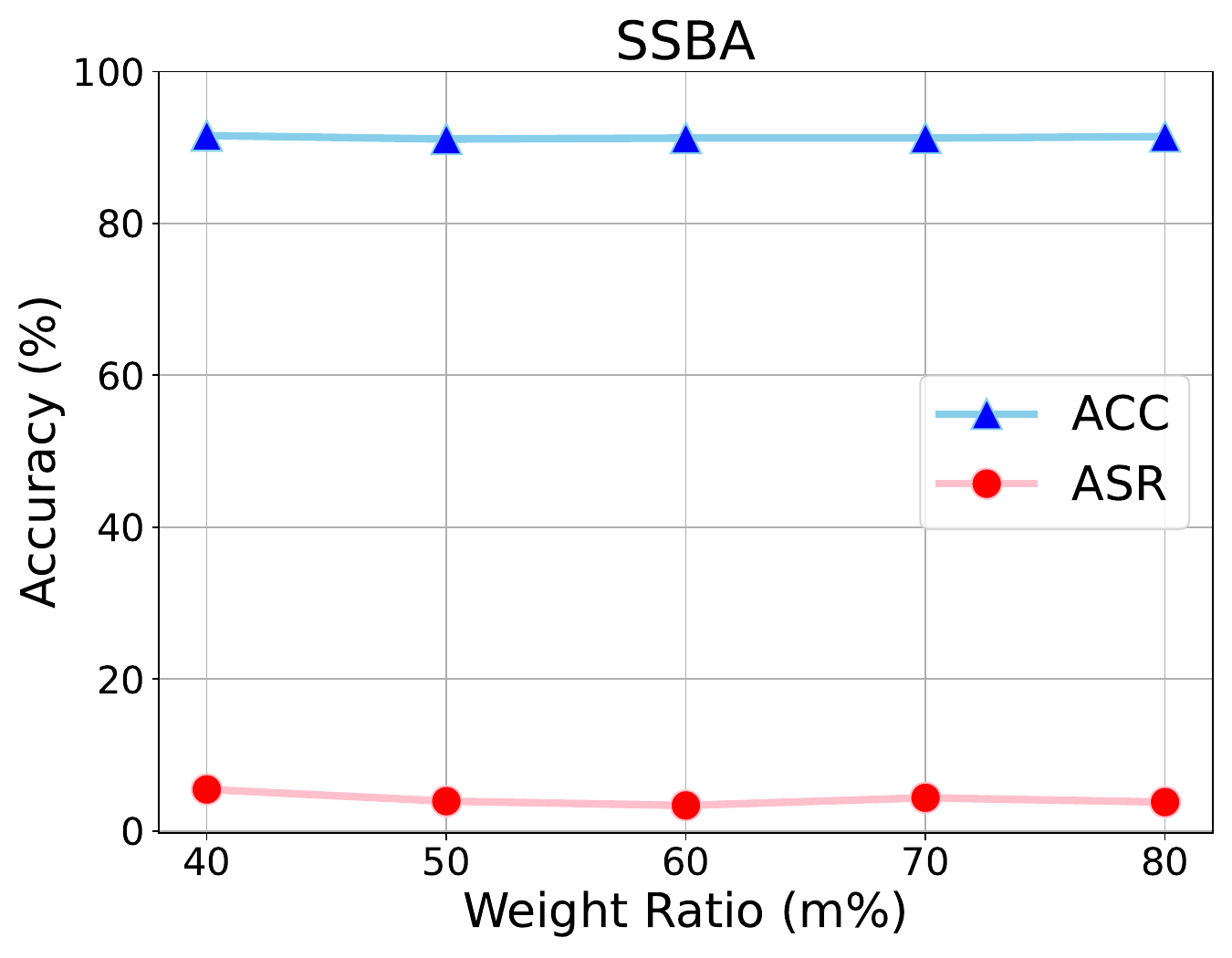}
        \includegraphics[width=0.24\linewidth]{ 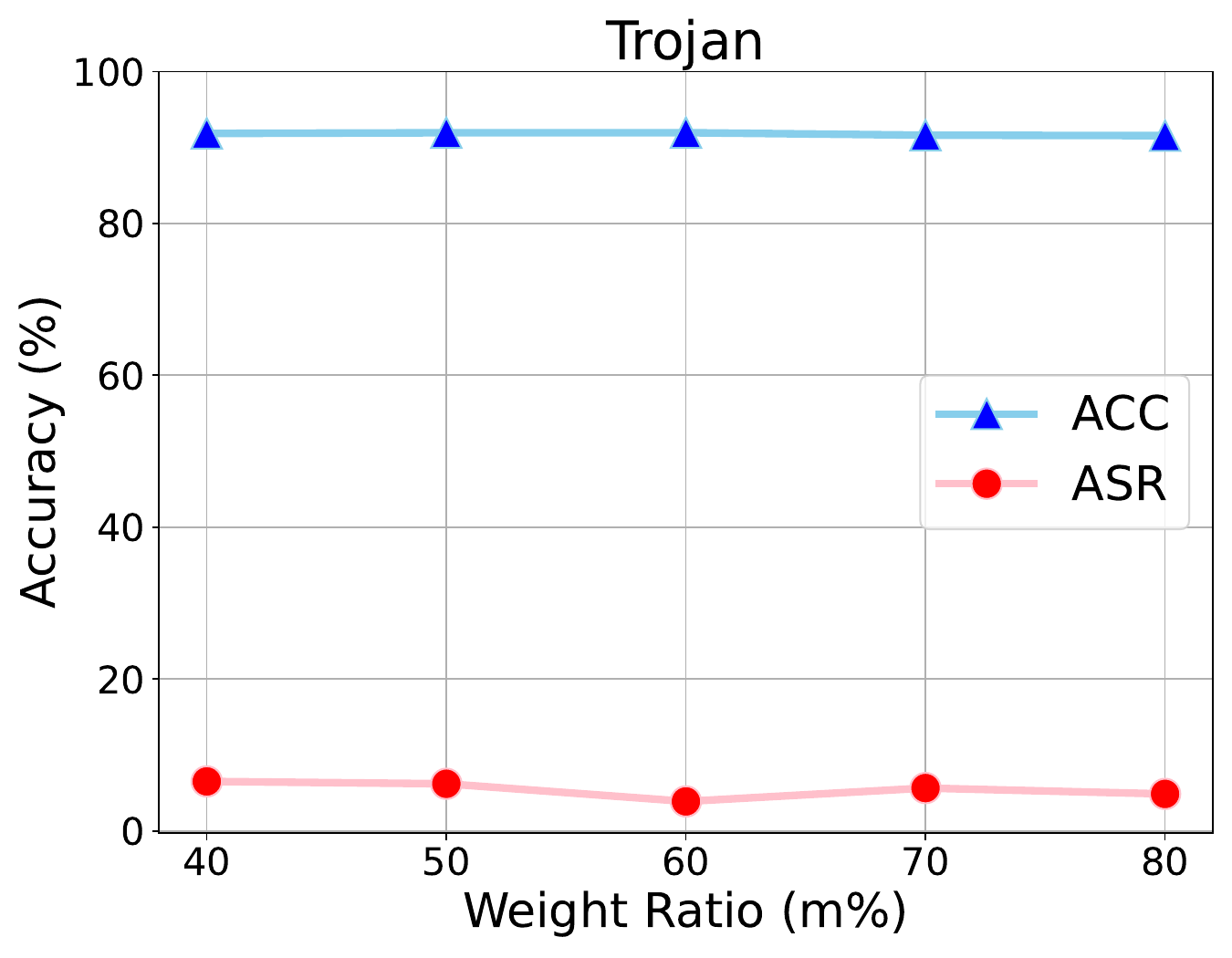}
        \includegraphics[width=0.24\linewidth]{ 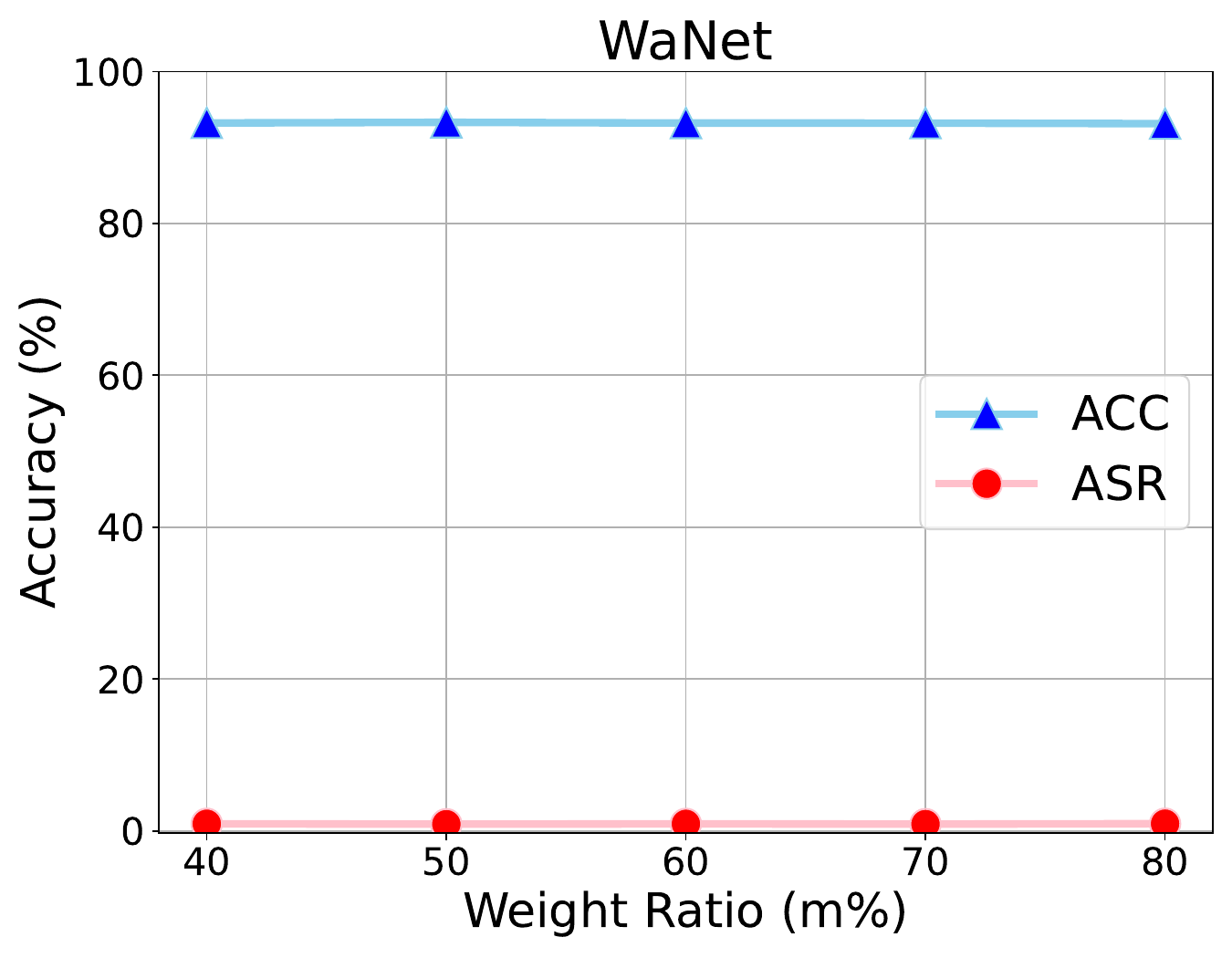}
    \caption{Performance with different weight ratios under 8 attacks.}
    \label{fig:app_weight_ratio}
\end{figure}
\section{Performance with Different Neuron Ratio and Weight Ratio}
\label{sec:perform_ratio}
Section~\ref{subsec:further_analysis} shows some of the performances with different neuron ratios and weight ratios and verifies the robustness of the hyper-parameter tuning in our method. Here, we exhibit the full results under all 8 attacks in PreAct-ResNet18 on CIFAR-10 with 10\% poisoning ratio. Figure~\ref{fig:app_neuron_ratio} and Figure~\ref{fig:app_weight_ratio} exhibit the tuning of neuron ratio and weight ratio, respectively. All performances are as good as expected.

\begin{wrapfigure}{R}{0.4\textwidth}
    \centering
    \includegraphics[width=1\linewidth]{ 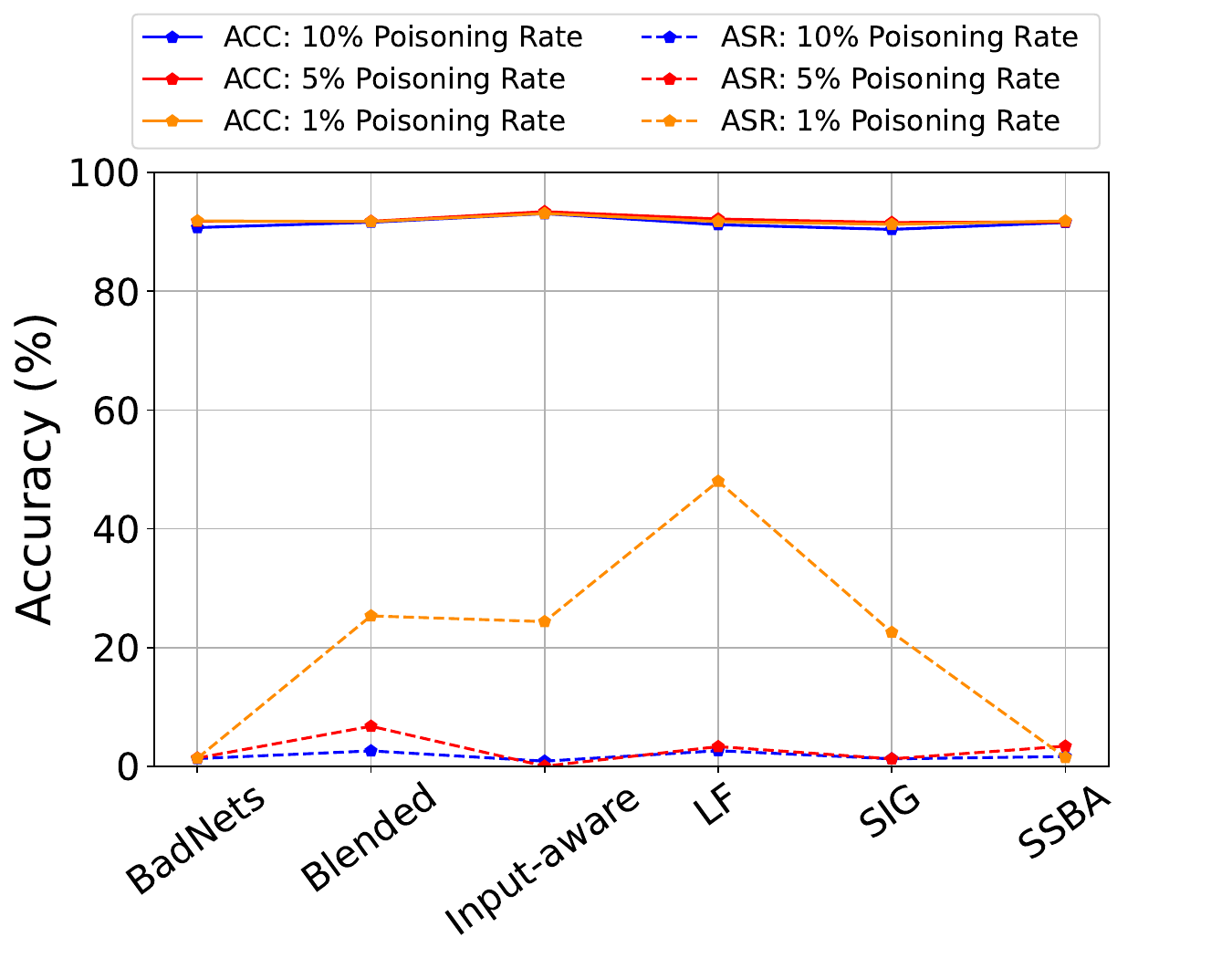}
    \vspace{-3mm}
    \caption{ACC on Different Poisoning Ratio.}
    \label{fig:Fig8_prate}
\vspace{-3mm}
\end{wrapfigure}
\section{Evaluations with Different Poisoning Ratio.}
\label{sec:perform_poison_ratio}
We further investigate the performance of \nameFramework on different poisoning ratios, \textit{e.g.}, 10\%, 5\%, and 1\%. Note that the larger poisoning ratio represents the stronger attack mode. We test the performance with six attacks on these three ratios.
Figure~\ref{fig:Fig8_prate} shows the performances of ACC and ASR. We can observe that \nameFramework successfully defends all the attacks on 10\% and 5\% with a low ASR and a high ACC while performing less effectively on 1\%. A possible reason is that the unlearning weight changes of backdoor neurons are less obvious in the weak attack mode compared to the strong attack mode with 10\% or 5\% poisoning ratios. 

\section{Evaluations with Different Clean Data Ratios}
\label{sec:perform_clean_ratio}
We are here to test the performance of \nameFramework under different clean data ratios. The performance on CIFAR-10 and PreAct-ResNet18 with 10\% poisoning ratio is illustrated in Table~\ref{tab:clean_ratio}. Except for the default 5\% clean data ratio, we also test the performance under 10\%, 1\%, and 0.5\%.
We can observe that a larger ratio of clean data can always bring better performance in ACC, \textit{e.g.}, $92.18\% > 91.70\% > 89.63\% > 86.42\%$ on average values of the four ratios in decreasing order, respectively. For the defense performance in ASR, all of them can successfully defend the attacks to under 10\%, and large clean data ratios can achieve promising DERs. Overall, \nameFramework is robust to the clean data ratio, with only 0.5\% clean data can also defend most tested defenses successfully.

\section{Evaluations with Different Learning Rates on Fine-tuning}
\label{sec:perform_ftlr}
We test the performance of our proposed method under different learning rates on fine-tuning. Table~\ref{tab:lr_ft} illustrates the results. Specifically, with the same settings on CIFAR-10 and PreAct-ResNet18 with 10\% poisoning ratio and 5\% clean data ratio, we test the performance when the learning rate is set to 0.1, 0.01, 0.001, and 0.0001, where 0.01 is our default setting in the previous experiments.
We can observe that the default learning rate of 0.01 is the most suitable one chosen for defense, which can achieve SOTA ASR and DER on average with comparable average ACC. Besides, setting the learning rate to 0.001 can perform the best in ACC, while making it fail on some strong attacks, \eg, Blended and LF. On the contrary, setting the learning rate to 0.1 will hurt the ACC greatly, which may imply that clean knowledge cannot be re-learned properly. 

\begin{wrapfigure}{R}{0.5\textwidth}
    \centering
    \includegraphics[width=0.48\linewidth]{ 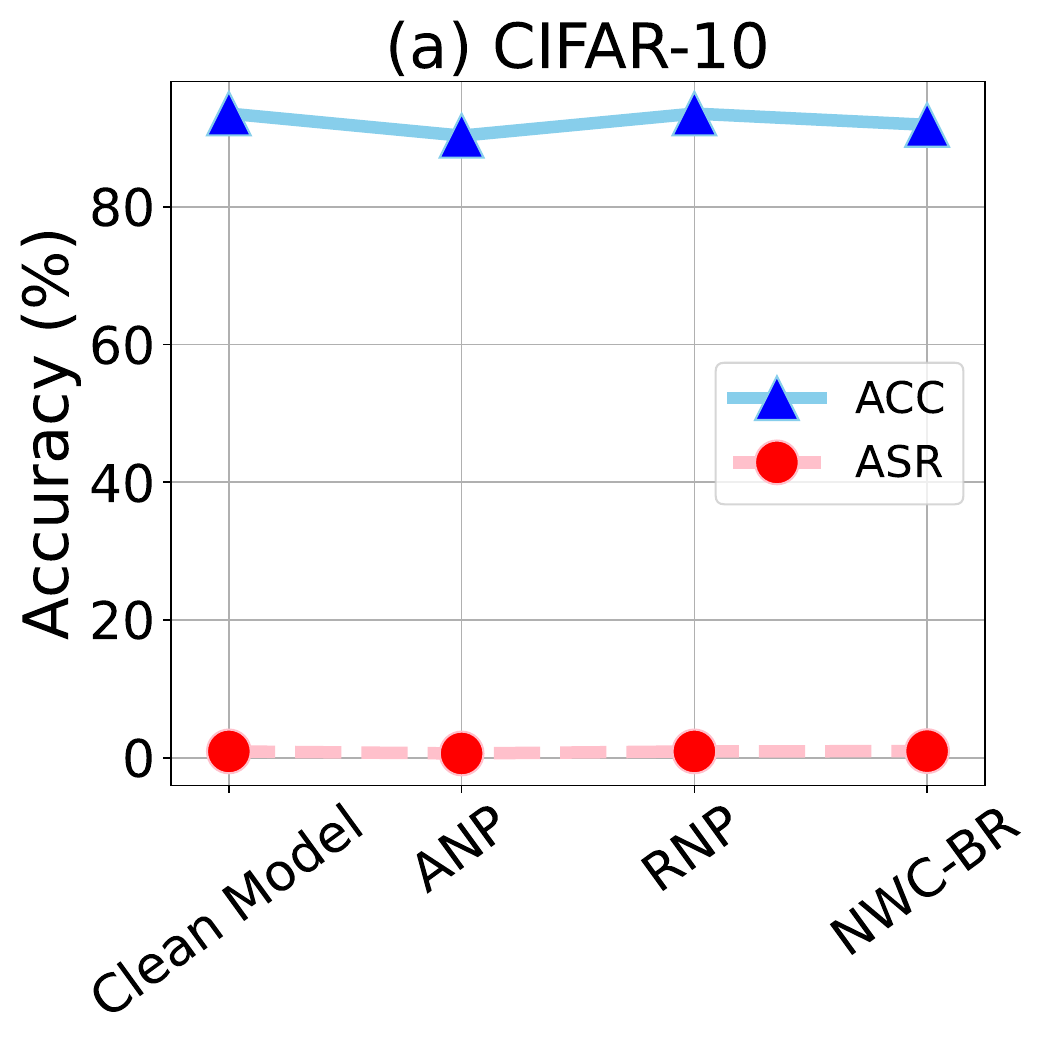}
    \includegraphics[width=0.48\linewidth]{ 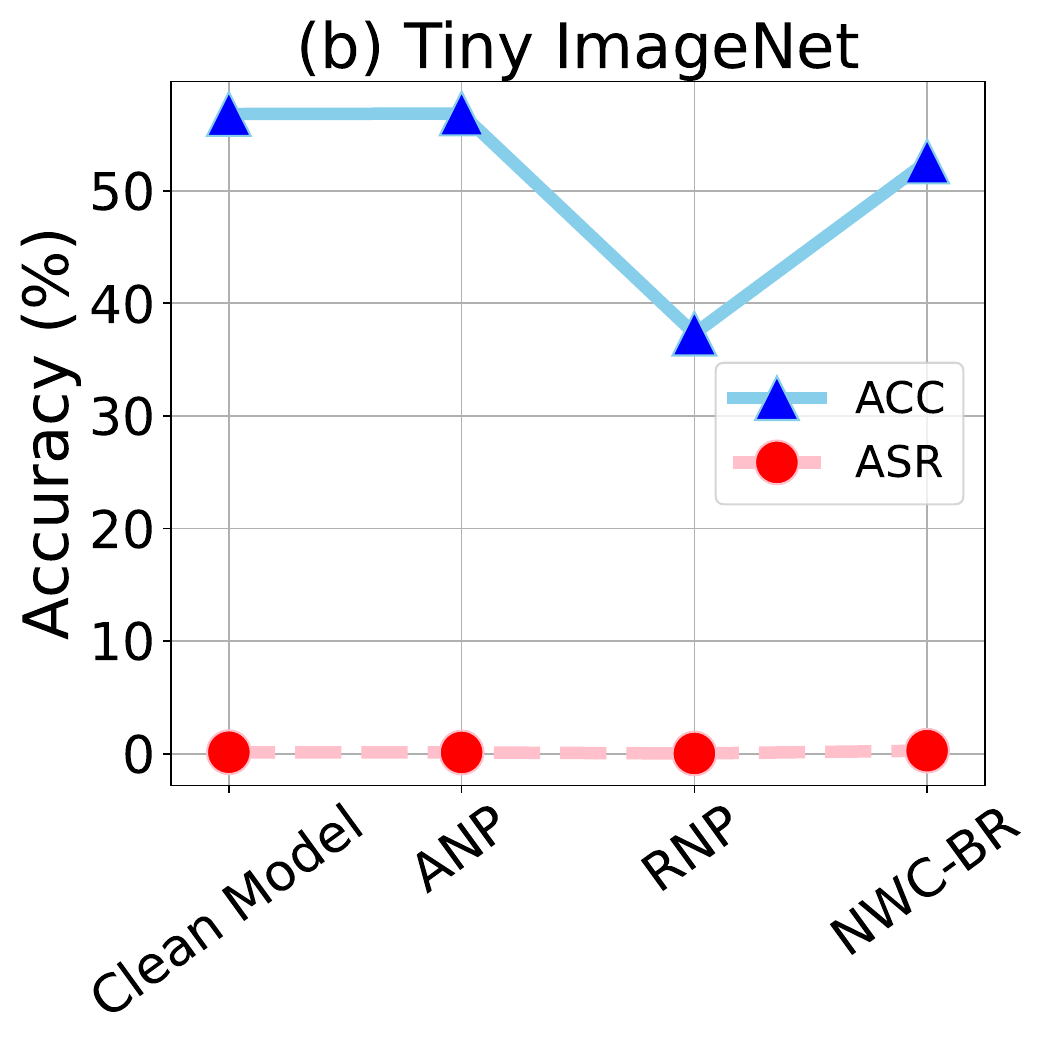}
    \caption{Performance of Defenses on Clean Model.}
    \label{fig:clean_model_test}
\end{wrapfigure}
\section{Evaluations on Clean Model}
\label{sec:clean_model}
To test whether our defense method will hurt the performance of the clean model if no backdoor attack occurs, we compare it with several defense methods on CIFAR-10 and Tiny ImageNet datasets on PreAct-ResNet18. Figure~\ref{fig:clean_model_test} illustrates the results, where the ACC and ASR on the original clean model, ANP, and RNP, are used for comparison. We can observe that most of the defense methods (including ours) will barely damage the clean model though there is no backdoor occurs, except for RNP on Tiny ImageNet. It exhibits the potential to largely employ our defense methods in real-world AI systems.

\begin{table*}[]
\caption{Performance with Different Clean Data Ratios on CIFAR-10 dataset with PreAct-ResNet18 (\%).}
\label{tab:clean_ratio}
\centering
\resizebox{\linewidth}{!}{
\begin{tabular}{c|ccc|ccc|ccc|ccc|ccc}
\hline
Clean Data Ratio & \multicolumn{3}{c|}{- (No Defense)} & \multicolumn{3}{c|}{10\% (\nameFramework)}               & \multicolumn{3}{c|}{5\% (\nameFramework)}              & \multicolumn{3}{c|}{1\% (\nameFramework)}          & \multicolumn{3}{c}{0.5\% (\nameFramework)} \\ \hline
Attacks       & ACC $\uparrow$        & ASR $\downarrow$         & DER $\uparrow$      & ACC $\uparrow$            & ASR $\downarrow$           & DER $\uparrow$            & ACC $\uparrow$            & ASR $\downarrow$           & DER $\uparrow$            & ACC $\uparrow$         & ASR $\downarrow$           & DER $\uparrow$            & ACC $\uparrow$     & ASR $\downarrow$             & DER $\uparrow$     \\ \hline
BadNets~\cite{gu2019badnets}       & 91.32      & 95.03       & -        & \textbf{91.44} & \textbf{0.82} & \textbf{97.11} & {\ul 90.72}    & 1.37          & {\ul 96.53}    & 87.42       & 2.00          & 94.57          & 81.13   & {\ul 1.18}      & 91.83   \\
Blended~\cite{chen2017targeted}       & 93.47      & 99.92       & -        & \textbf{91.76} & 5.79          & 96.21          & {\ul 91.61}    & {\ul 2.61}    & \textbf{97.73} & 89.36       & \textbf{1.32} & {\ul 97.25}    & 86.68   & 4.80            & 94.17   \\
Input-aware~\cite{nguyen2020input}   & 90.67      & 98.26       & -        & \textbf{93.65} & {\ul 1.56}    & \textbf{98.35} & {\ul 93.06}    & 1.94          & {\ul 98.16}    & 91.54       & 2.62          & 97.82          & 89.51   & \textbf{1.43}   & 97.83   \\
LF~\cite{zeng2021rethinking}            & 93.19      & 99.28       & -        & \textbf{92.24} & {\ul 4.62}    & {\ul 96.85}    & {\ul 91.20}    & \textbf{2.64} & \textbf{97.32} & 89.58       & 8.41          & 93.63          & 88.49   & 5.00            & 94.79   \\
SIG~\cite{barni2019new}           & 84.48      & 98.27       & -        & \textbf{90.41} & {\ul 1.01}    & {\ul 98.63}    & \textbf{90.41} & 1.27          & 98.50          & {\ul 88.53} & \textbf{0.96} & \textbf{98.65} & 83.55   & 2.44            & 97.45   \\
SSBA~\cite{li2021invisible}          & 92.88      & 97.86       & -        & \textbf{91.94} & 2.14          & {\ul 97.39}    & {\ul 91.57}    & {\ul 1.66}    & \textbf{97.44} & 89.24       & 1.71          & 96.25          & 84.46   & \textbf{0.70}   & 94.37   \\
Trojan~\cite{liu2018trojaning}        & 93.42      & 100.00      & -        & \textbf{92.37} & 7.00          & 95.98          & {\ul 91.76}    & {\ul 5.06}    & \textbf{96.64} & 90.18       & \textbf{4.66} & {\ul 96.05}    & 87.98   & 7.21            & 93.68   \\
WaNet~\cite{nguyen2021wanet}         & 91.25      & 89.73       & -        & \textbf{93.66} & 0.98          & {\ul 94.38}    & {\ul 93.26}    & {\ul 0.88}    & \textbf{94.43} & 91.19       & 1.24          & 94.22          & 89.57   & \textbf{0.69}   & 93.68   \\ \hline
Average       & 91.34      & 97.29       & -        & \textbf{92.18} & 2.99          & {\ul 96.86}    & {\ul 91.70}    & \textbf{2.18} & \textbf{97.09} & 89.63       & {\ul 2.87}    & 96.05          & 86.42   & 2.93            & 94.72   \\ \hline
\end{tabular}}
\end{table*}

\begin{table*}[]
\caption{Performance with Different Learning Rates of Fine-tuning on CIFAR-10 dataset with PreAct-ResNet18 (\%).}
\label{tab:lr_ft}
\centering
\resizebox{\linewidth}{!}{
\begin{tabular}{c|ccc|ccc|ccc|ccc|ccc}
\hline
Learning Rate & \multicolumn{3}{c|}{- (No Defense)} & \multicolumn{3}{c|}{0.1 (\nameFramework)} & \multicolumn{3}{c|}{0.01 (\nameFramework)}           & \multicolumn{3}{c|}{0.001 (\nameFramework)}             & \multicolumn{3}{c}{0.0001 (\nameFramework)}         \\ \hline
Attacks       & ACC $\uparrow$        & ASR $\downarrow$         & DER $\uparrow$      & ACC $\uparrow$     & ASR $\downarrow$            & DER $\uparrow$    & ACC $\uparrow$         & ASR $\downarrow$           & DER $\uparrow$            & ACC $\uparrow$            & ASR $\downarrow$           & DER $\uparrow$            & ACC $\uparrow$         & ASR $\downarrow$           & DER $\uparrow$            \\ \hline
BadNets~\cite{gu2019badnets}       & 91.32      & 95.03       & -        & 78.11   & 4.19           & 88.82  & 90.72          & \textbf{1.37} & \textbf{96.53} & \textbf{91.62} & {\ul 2.10}  & {\ul 96.47} & {\ul 91.44}    & 4.87          & 95.08          \\
Blended~\cite{chen2017targeted}       & 93.47      & 99.92       & -        & 74.79   & \textbf{0.57}  & 90.34  & 91.61          & {\ul 2.61}    & \textbf{97.73} & {\ul 92.85}    & 15.19       & {\ul 92.06} & \textbf{92.88} & 29.33         & 85.00          \\
Input-aware~\cite{nguyen2020input}   & 90.67      & 98.26       & -        & 88.70   & 3.00           & 96.64  & {\ul 93.06}    & {\ul 1.94}    & {\ul 98.16}    & \textbf{93.22} & 2.05        & 98.10       & 91.46          & \textbf{0.64} & \textbf{98.81} \\
LF~\cite{zeng2021rethinking}            & 93.19      & 99.28       & -        & 79.38   & {\ul 3.50}     & 90.98  & 91.20          & \textbf{2.64} & \textbf{97.32} & \textbf{92.90} & 11.98       & {\ul 93.50} & {\ul 92.79}    & 37.03         & 80.92          \\
SIG~\cite{barni2019new}           & 84.48      & 98.27       & -        & 73.15   & {\ul 3.68}     & 91.63  & {\ul 90.41}    & \textbf{1.27} & \textbf{98.50} & \textbf{90.87} & 4.81        & {\ul 96.73} & 89.23          & 14.47         & 91.90          \\
SSBA~\cite{li2021invisible}          & 92.88      & 97.86       & -        & 79.90   & {\ul 4.21}     & 90.33  & 91.57          & \textbf{1.66} & \textbf{97.44} & {\ul 92.64}    & 9.26        & {\ul 94.18} & \textbf{92.72} & 15.27         & 91.21          \\
Trojan~\cite{liu2018trojaning}        & 93.42      & 100.00      & -        & 73.83   & 32.87          & 73.77  & 91.76          & \textbf{5.06} & \textbf{96.64} & \textbf{92.75} & {\ul 10.76} & {\ul 94.29} & {\ul 92.73}    & 21.07         & 89.12          \\
WaNet~\cite{nguyen2021wanet}         & 91.25      & 89.73       & -        & 87.61   & 1.48           & 92.31  & \textbf{93.26} & 0.88          & 94.43          & {\ul 93.11}    & {\ul 0.79}  & {\ul 94.47} & 91.36          & \textbf{0.31} & \textbf{94.71} \\ \hline
Average       & 91.34      & 97.29       & -        & 79.43   & {\ul 6.69}     & 89.35  & 91.70          & \textbf{2.18} & \textbf{97.09} & \textbf{92.50} & 7.12        & {\ul 94.97} & {\ul 91.83}    & 15.37         & 90.84         \\ \hline
\end{tabular}}
\end{table*}


\end{document}